\documentstyle[sprocl,epsfig]{article}

\newcommand{\sqrts}{\mbox{$\sqrt {s}$}}
\newcommand{\ee}{\mbox{$\mathrm{e}^{+}\mathrm{e}^{-}$}}
\newcommand{\ra}{\mbox{$\rightarrow$}}
\newcommand{\mHrec}{\mbox{$m_{\mathrm{H}}^{\mathrm{rec}}$}}

\begin{document}
\begin{titlepage}
\def\thefootnote{\fnsymbol{footnote}}       

\begin{center}
\mbox{ } 

\end{center}
\begin{flushright}
\Large
\mbox{\hspace{10.2cm} hep-ph/0402231} \\
\mbox{\hspace{12.0cm} February 2004}
\end{flushright}
\begin{center}
\vskip 1.0cm
{\Huge\bf
Highlights of Higgs Physics at LEP}
\vskip 1cm
{\LARGE\bf Andr\'e Sopczak}\\
\smallskip
\Large Lancaster University

\vskip 2.5cm
\centerline{\Large \bf Abstract}
\end{center}

\vskip 3.5cm
\hspace*{-1cm}
\begin{picture}(0.001,0.001)(0,0)
\put(,0){
\begin{minipage}{16cm}
\Large
\renewcommand{\baselinestretch} {1.2}
Final results from the combined data of the four LEP experiments 
ALEPH, DELPHI, L3 and OPAL on Standard Model (SM) Higgs boson 
searches are presented.
New preliminary results of searches in extended models are reviewed.
\renewcommand{\baselinestretch} {1.}

\normalsize
\vspace{10cm}
\begin{center}
{\sl \large
Presented at the Fourth International Conference on Non-Accelerator
New Physics, NANP--03, Dubna, Russia, 2003
\vspace{-6cm}
}
\end{center}
\end{minipage}
}
\end{picture}
\vfill

\end{titlepage}


\newpage
\thispagestyle{empty}
\mbox{ }
\newpage
\setcounter{page}{1}

\large
\title{\Large Highlights of Higgs Physics at LEP}

\author{\large Andr\'e Sopczak}

\address{\large Lancaster University \\E-mail: andre.sopczak@cern.ch} 


\maketitle\abstracts{\large
Final results from the combined data of the four LEP experiments 
ALEPH, DELPHI, L3 and OPAL on Standard Model (SM) Higgs boson 
searches are presented.
New preliminary results of searches in extended models are reviewed.
}

The LEP experiments took data between August 1989 and November 2000
at centre-of-mass energies first around the Z resonance (LEP-1) and later
up to 209 GeV (LEP-2). In 2000 most data was taken around 206 GeV.
The LEP accelerator operated very successfully and 
a total luminosity of ${\cal L} = 2461$ pb$^{-1}$ was accumulated
at LEP-2 energies.
Data-taking ended on 3 November 2000, although some data excess was 
observed in searches for the SM Higgs boson with 115~GeV mass.
In this report several different research lines are addressed:
1) the Standard Model Higgs boson: candidates, confidence levels, 
mass limit, coupling limits;
2) the Minimal Supersymmetric extension of the SM (MSSM):
dedicated searches, three-neutral-Higgs-boson hypothesis,
benchmark and general scan mass limits;
3) CP-violating models;
4) invisible Higgs boson decays;
5) flavour-independent hadronic Higgs boson decays;
6) neutral Higgs bosons in the general 2-doublet Higgs model;
7) Yukawa Higgs boson processes $\rm b\bar b h$ and $\rm b\bar b A$;
8) singly-charged Higgs bosons;
9) doubly-charged Higgs bosons;
10) fermiophobic Higgs boson decays $\rm h\ra WW, ZZ, \gamma\gamma$.

The results from Standard Model Higgs boson searches are final~\cite{sm},
and the results of searches in extended models
are mostly preliminary~\cite{summer2003}.
Limits are given at 95\% CL.

\section{Standard Model Higgs Boson}

\subsection{Combined Test Statistics and Candidates}
\vspace{-0.2cm}

\begin{figure}[htb]
\vspace*{-0.6cm}
\begin{center}
\begin{minipage}{0.43\textwidth}
\includegraphics[scale=0.38]{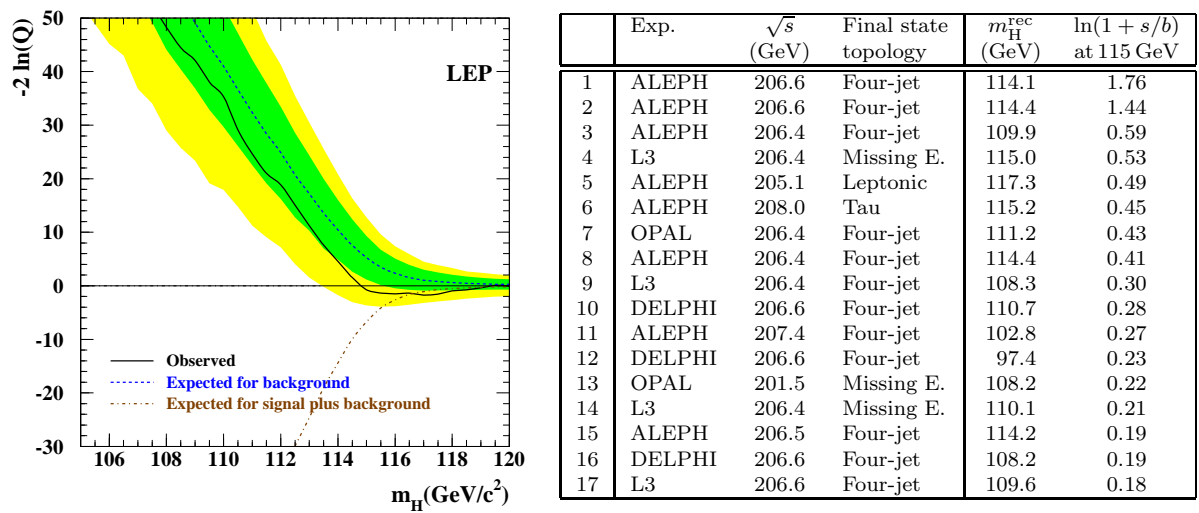}
\end{minipage}\hfill
\begin{minipage}{0.53\textwidth}
\footnotesize
\begin{tabular}{|c|lcl|cc|}
\hline
   & Exp.&  \sqrts   &  Final state      &   $\mHrec$ &  $\ln(1+s/b)$ \\
   &     &  (GeV)    &  topology         &   (GeV)    &  \hspace*{-2mm}at\,115\,GeV\hspace*{-2mm}    \\
\hline\hline
1&  ALEPH  &  206.6  &  Four-jet         &   114.1             & 1.76 \\
2&  ALEPH  &  206.6  &  Four-jet         &   114.4             & 1.44 \\
3&  ALEPH  &  206.4  &  Four-jet         &   109.9             & 0.59 \\
4&  L3     &  206.4  &  Missing E.       &   115.0             & 0.53 \\
5&  ALEPH  &  205.1  &  Leptonic         &   117.3             & 0.49 \\
6&  ALEPH  &  208.0  &  Tau              &   115.2             & 0.45 \\
7&  OPAL   &  206.4  &  Four-jet         &   111.2             & 0.43 \\
8&  ALEPH  &  206.4  &  Four-jet         &   114.4             & 0.41 \\
9&  L3     &  206.4  &  Four-jet         &   108.3             & 0.30 \\
10& DELPHI &  206.6  &  Four-jet         &   110.7             & 0.28 \\
11& ALEPH  &  207.4  &  Four-jet         &   102.8             & 0.27 \\
12& DELPHI &  206.6  &  Four-jet         & $\phantom{0}$97.4   & 0.23 \\ 
13& OPAL   &  201.5  &  Missing E.       &   108.2             & 0.22 \\
14& L3     &  206.4  &  Missing E.       &   110.1             & 0.21 \\
15& ALEPH  &  206.5  &  Four-jet         &   114.2             & 0.19 \\
16& DELPHI &  206.6  &  Four-jet         &   108.2             & 0.19 \\
17& L3     &  206.6  &  Four-jet         &   109.6             & 0.18 \\
\hline
\end{tabular}
\end{minipage}
\vspace*{-0.6cm}
\caption[]{
SM Higgs boson. Left: test statistics for the likelihood ratio
$Q={\cal L}_{\rm signal+background}/{\cal L}_{\rm background}$.
The $1\sigma$ and $2\sigma$ error bands are indicated (shaded area).
Right: final candidates. The signal $s$ and background $b$ estimates
are used to construct an event weight $\ln(1+s/b)$.   
\label{fig:noexcess}}
\end{center}
\vspace*{-0.3cm}
\end{figure}

Figure~\ref{fig:noexcess} shows that the observed SM excess is less 
than $2\sigma$ for combined LEP data, and lists the final candidates. 
The excess is reduced compared to previous reports. A short summary of
the development of the data excess was given previously~\cite{nanp01}.

\clearpage
\subsection{Test Statistics for Each Experiment}
\begin{figure}[htb]
\vspace*{-0.5cm}
\begin{center}
\includegraphics[scale=0.3]{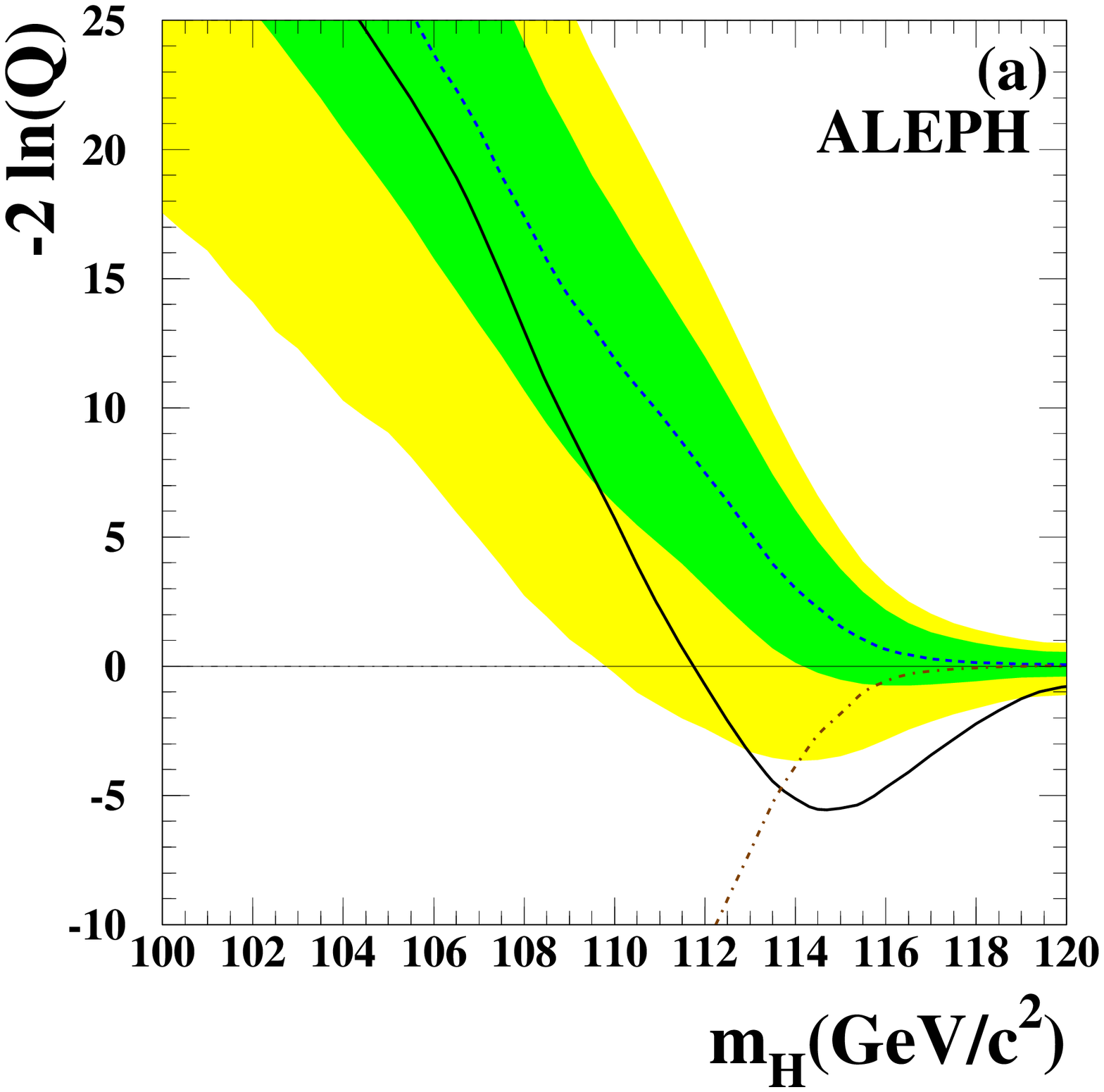}\hfill
\includegraphics[scale=0.3]{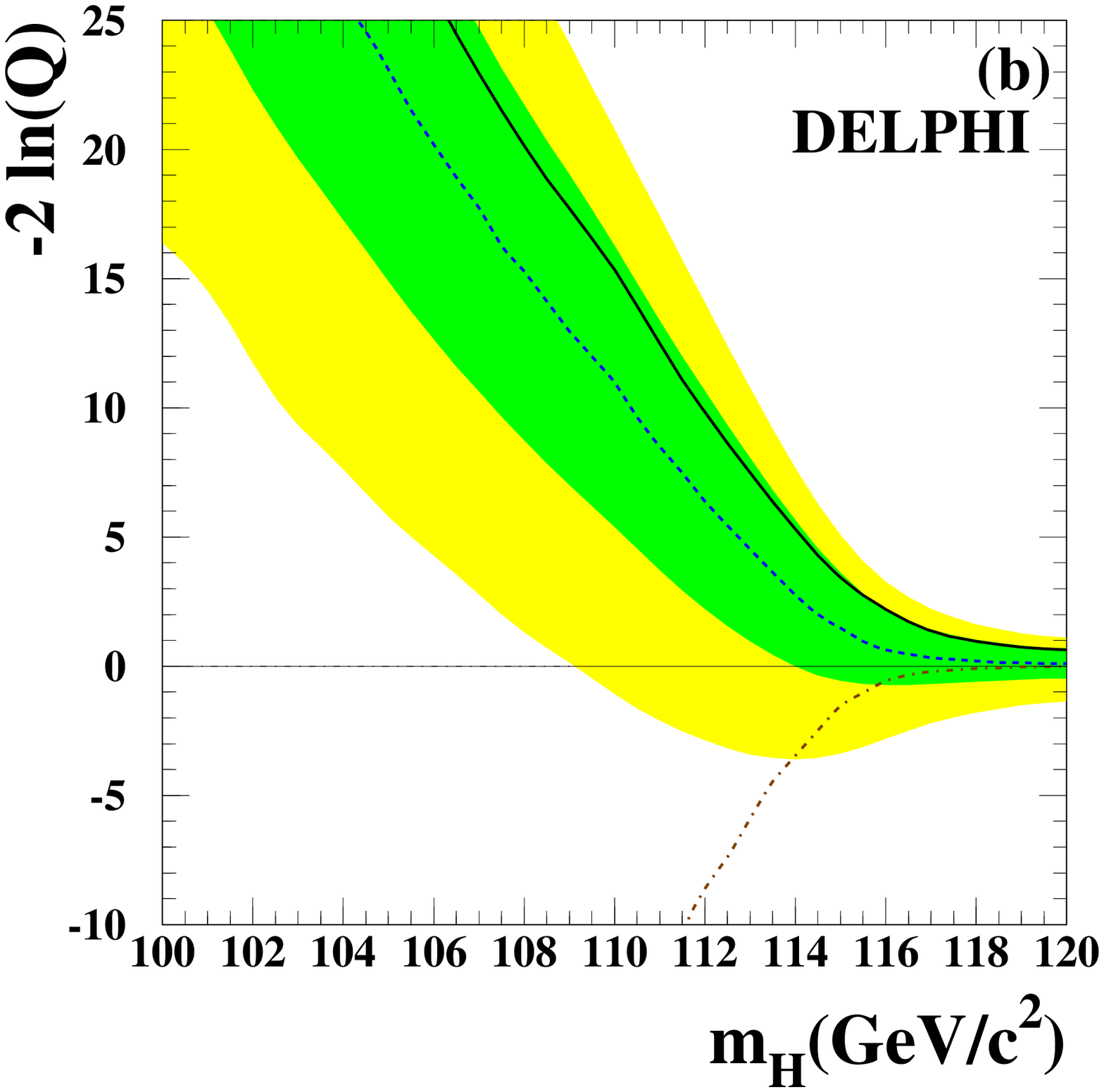}
\includegraphics[scale=0.3]{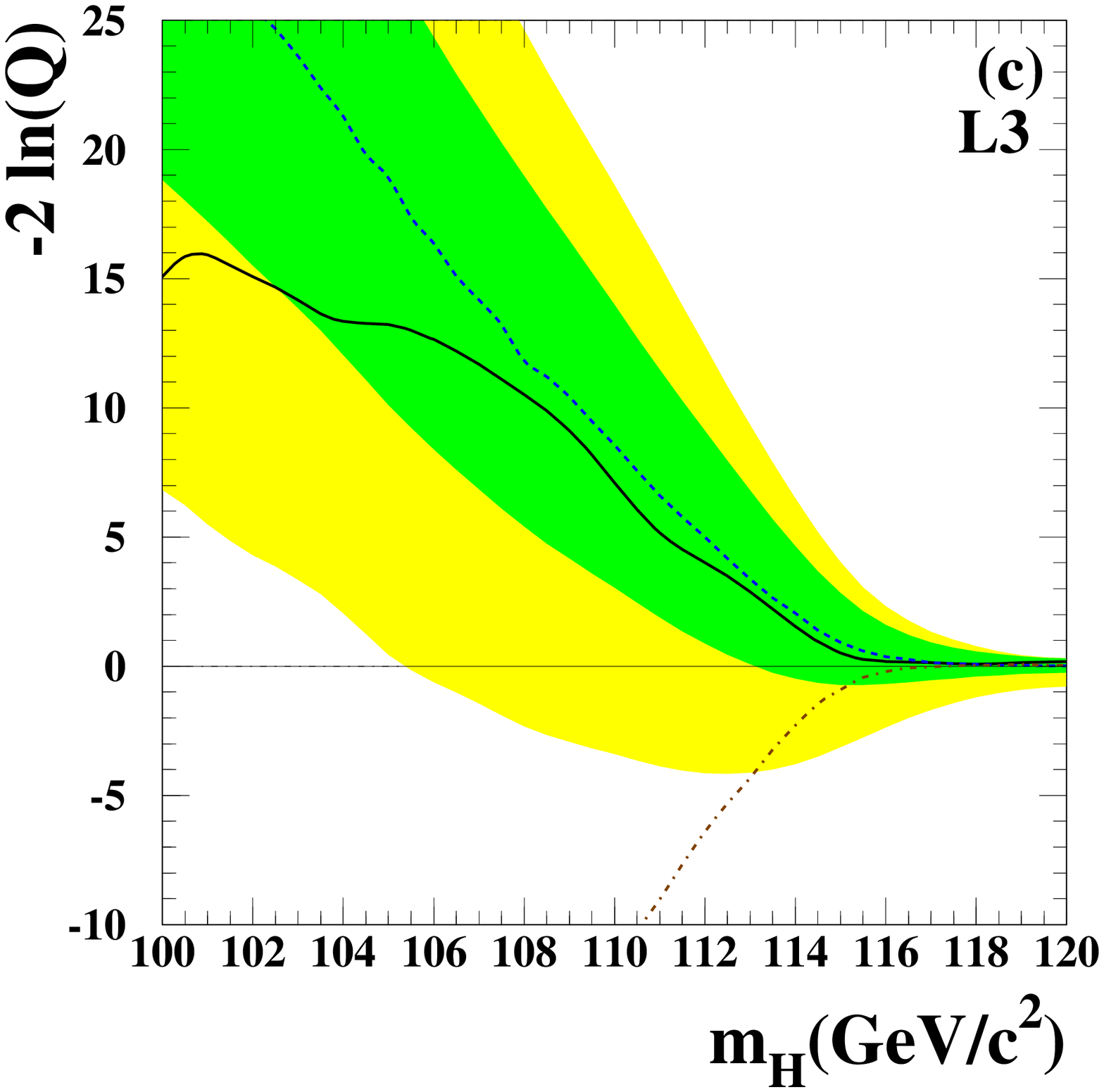}\hfill
\includegraphics[scale=0.3]{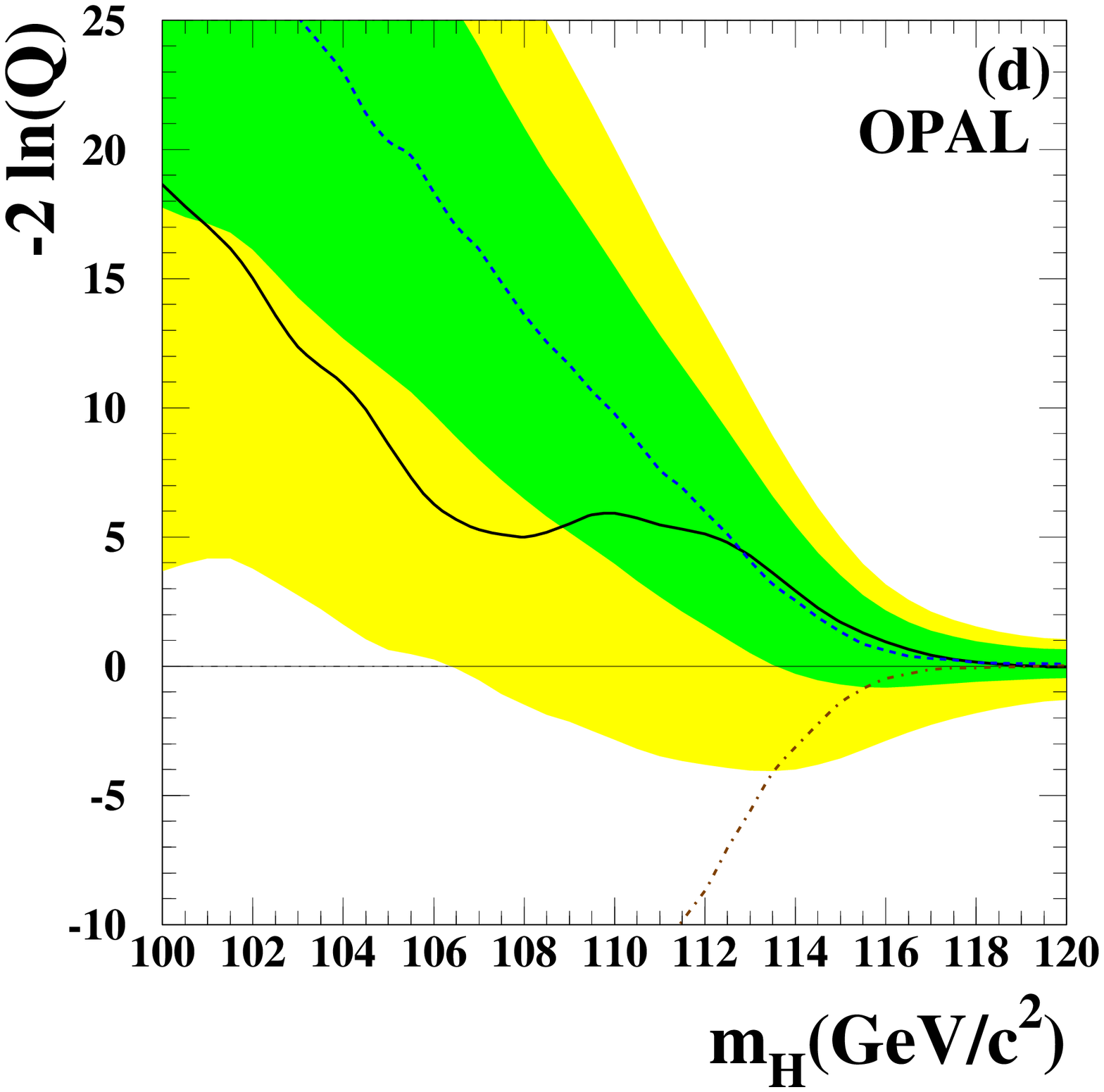}
\vspace*{-0.5cm}
\caption[]{SM Higgs boson: test statistics for each experiment.
As described in Fig.~\ref{fig:noexcess}, 
solid line: data;
dotted lines: expectation for background and for signal plus background.
\label{fig:sm_each}}
\end{center}
\end{figure}

\vspace*{-0.2cm}
Figure~\ref{fig:sm_each} shows the test statistics for each experiment separately.

\subsection{Test Statistics for the Four-Jet and Other Channels}
\vspace*{-0.2cm}

\begin{figure}[htb]
\vspace*{-0.5cm}
\begin{center}
\includegraphics[scale=0.3]{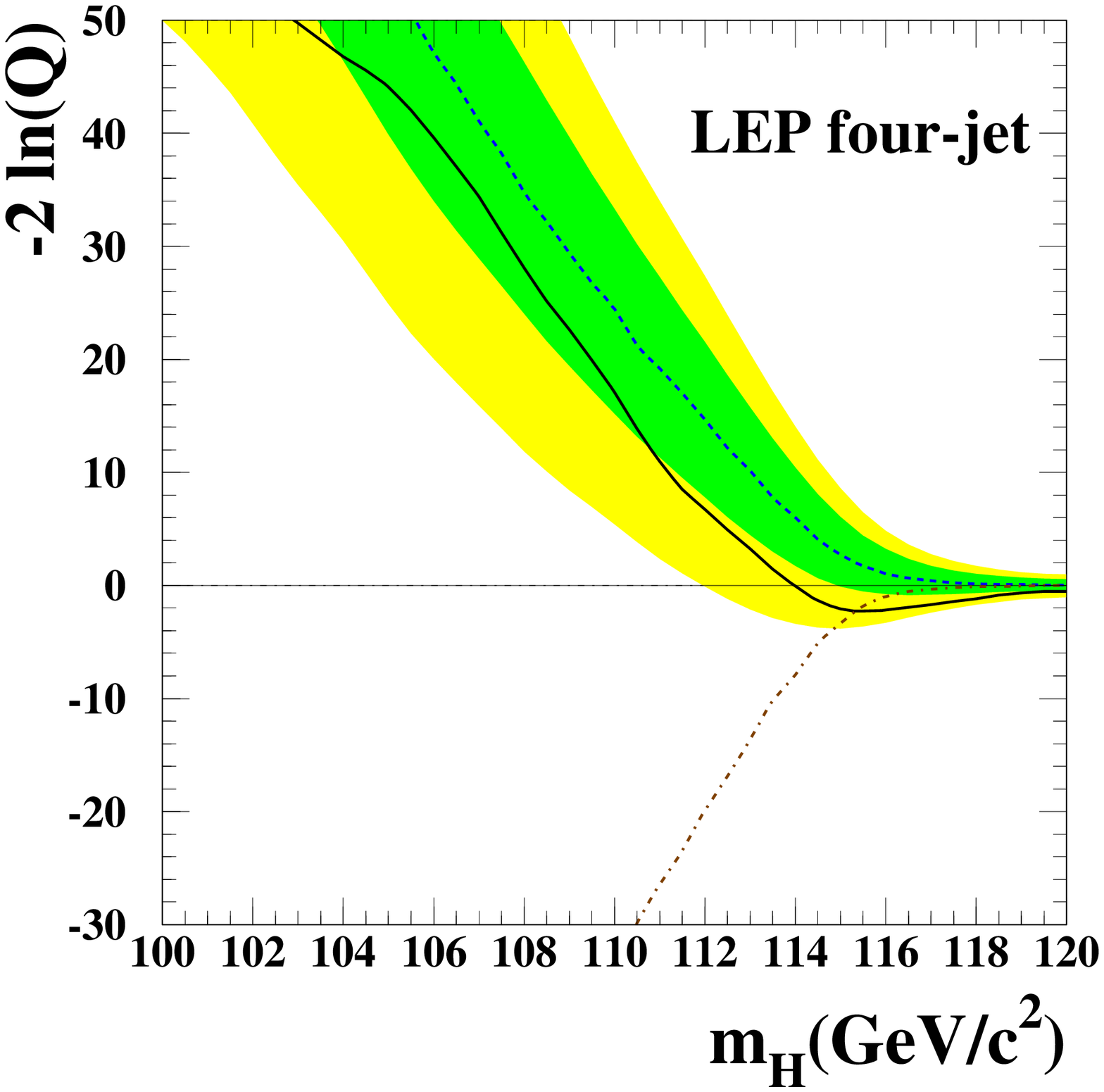}\hfill
\includegraphics[scale=0.3]{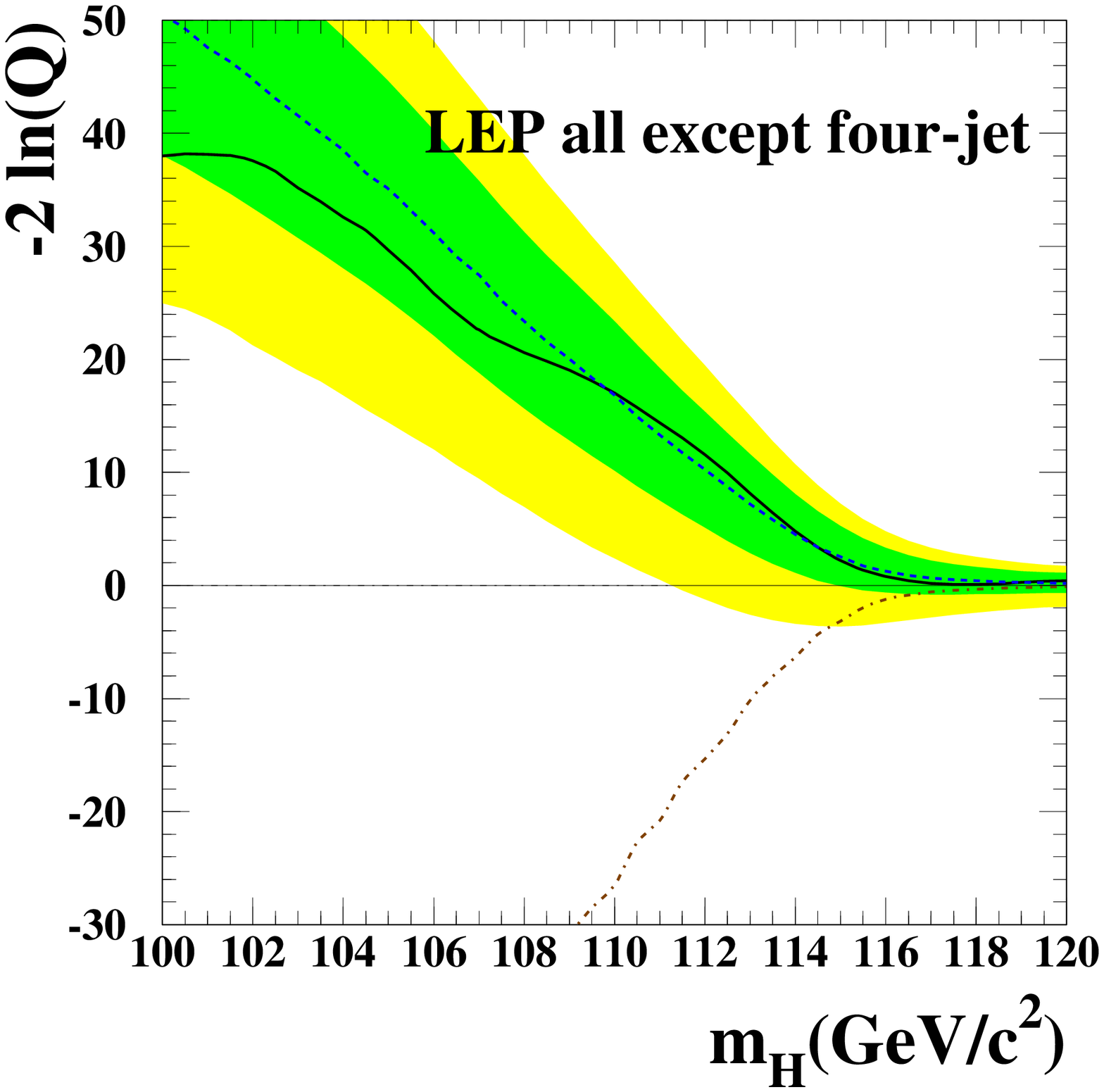}
\vspace*{-0.5cm}
\caption[]{SM Higgs boson: test statistics for the four-jet channel
and all other search channels combined.
As described in Fig.~\ref{fig:noexcess}, solid line: data; 
dotted lines: expectation for background and for signal plus background.
\label{fig:sm_4jet}}
\end{center}
\vspace*{-0.7cm}
\end{figure}

Figure~\ref{fig:sm_4jet} shows the test statistics for the four-jet channel
and all other search channels combined.

\clearpage
\subsection{Significance of Candidates for Each Experiment}
\vspace*{-0.2cm}

\begin{figure}[htb]
\vspace*{-0.5cm}
\begin{center}
\includegraphics[scale=0.3]{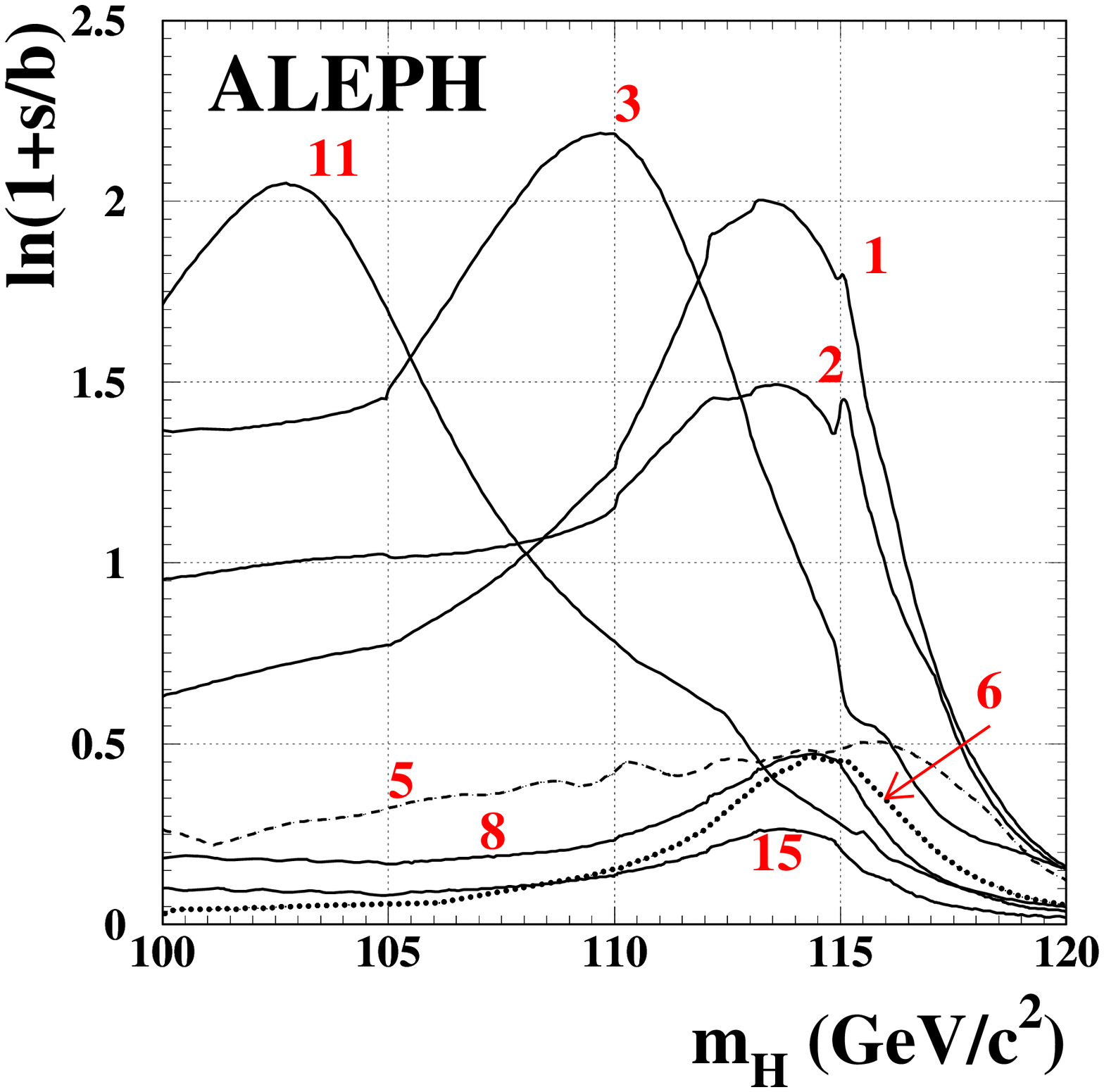}\hfill
\includegraphics[scale=0.3]{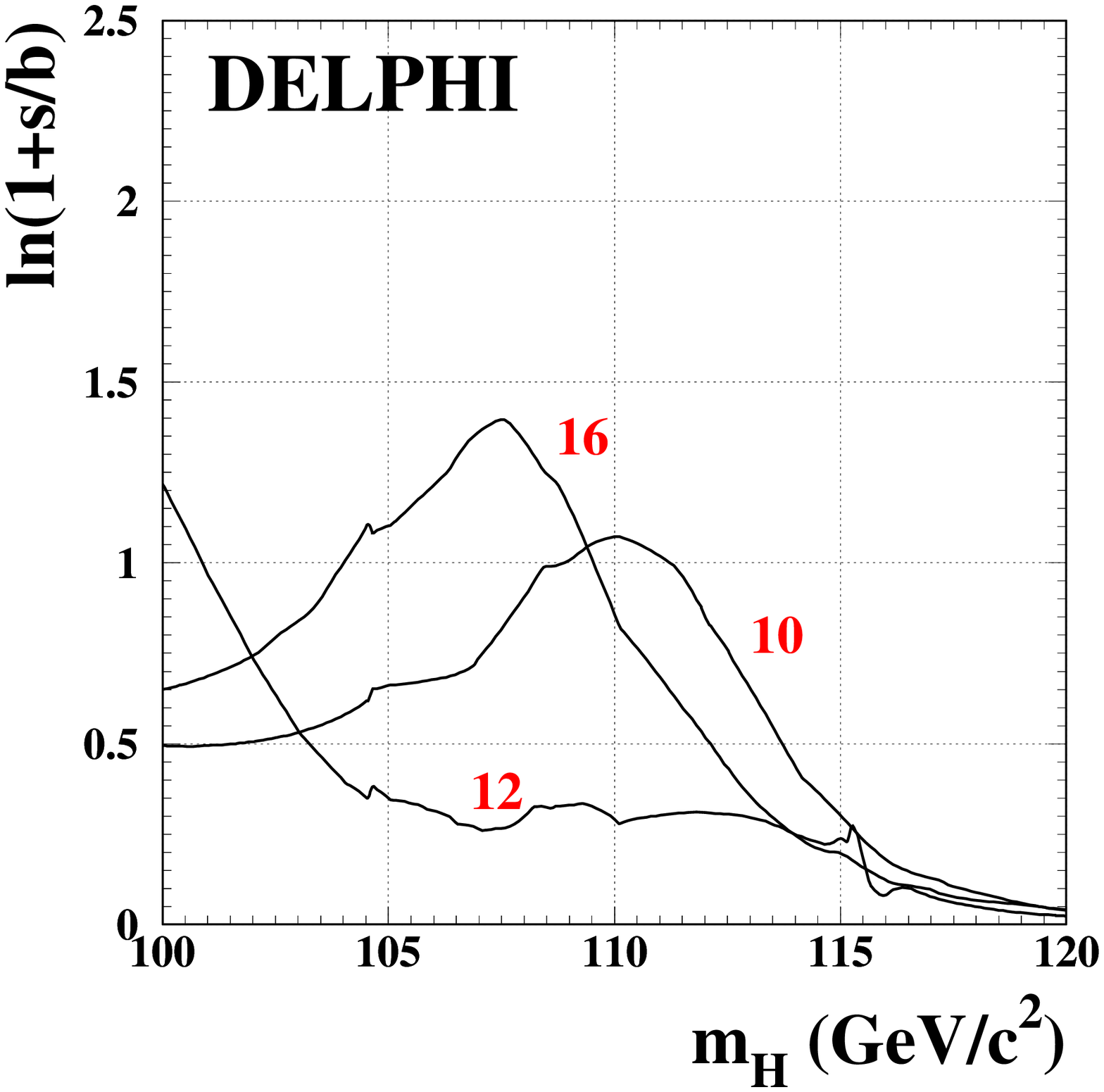}
\includegraphics[scale=0.3]{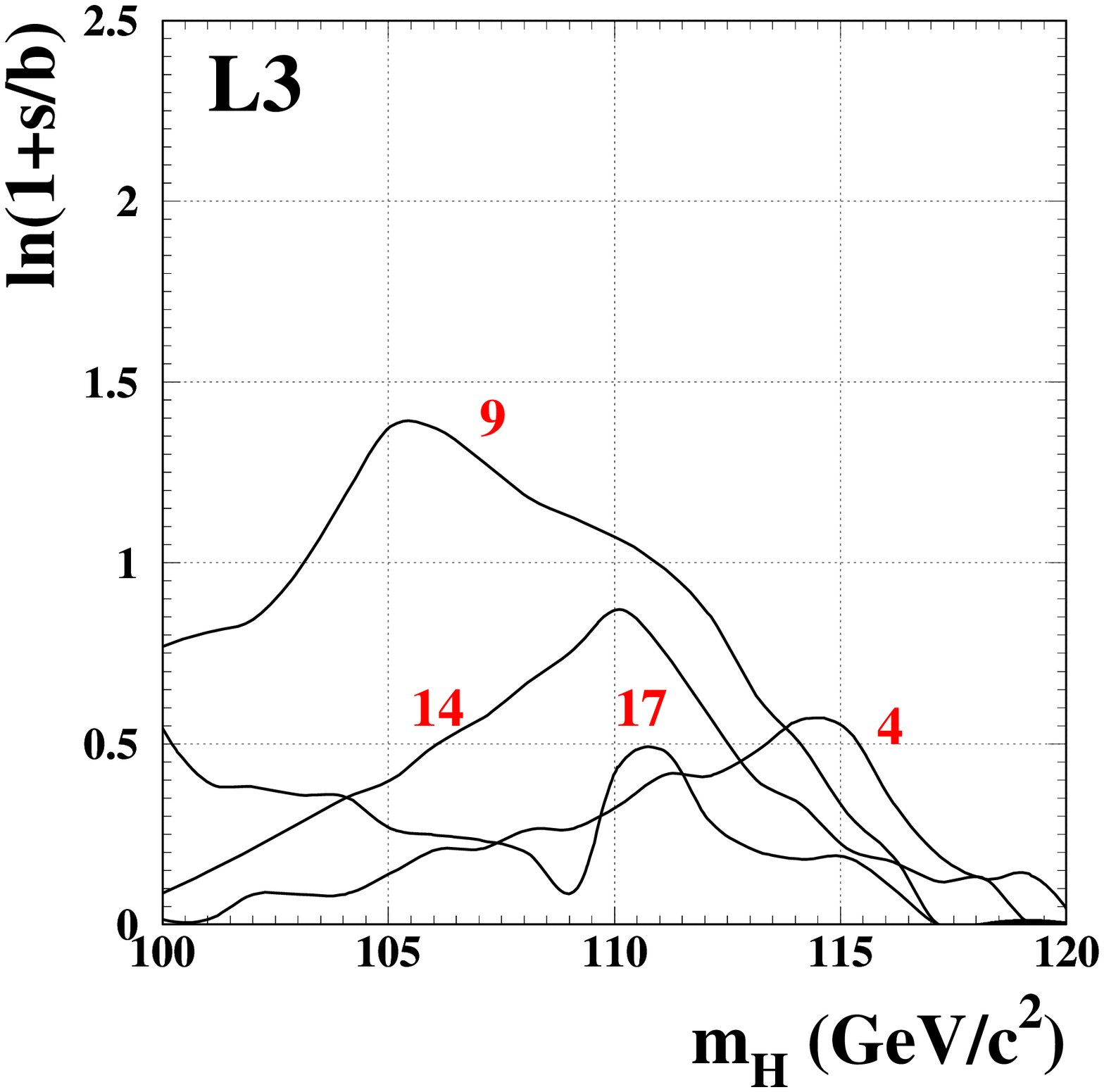}\hfill
\includegraphics[scale=0.3]{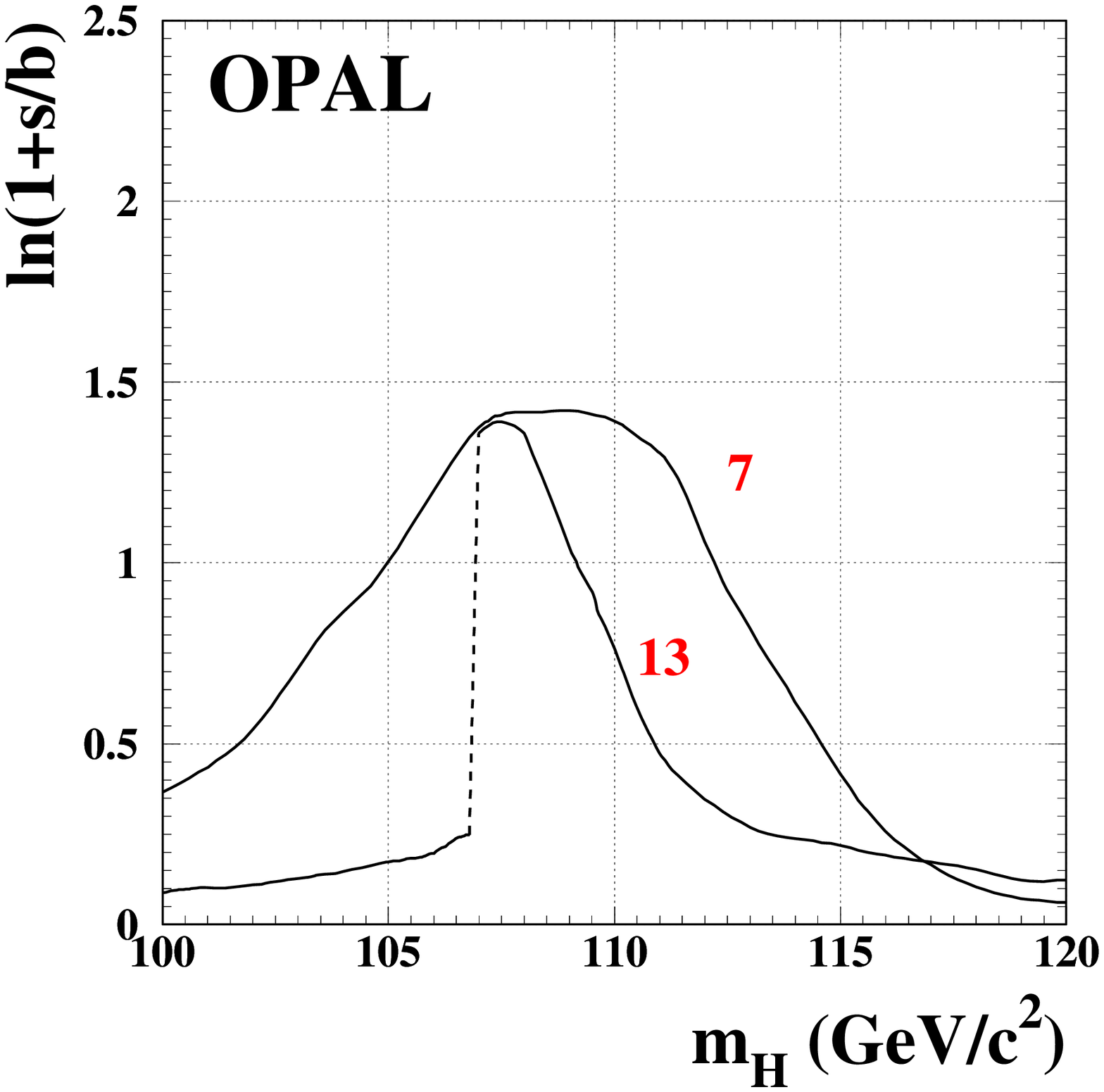}
\vspace*{-0.5cm}
\caption[]{SM Higgs boson: significance of candidates as a function of 
the Higgs boson mass.
\label{fig:cand}}
\end{center}
\end{figure}

Figure~\ref{fig:cand} shows the significance of the SM candidates as listed in 
Fig.~\ref{fig:noexcess} for each experiment separately as a function of the 
Higgs boson mass.

\subsection{Combined Significance of Candidates}
\begin{figure}[htb]
\vspace*{-0.5cm}
\begin{center}
\includegraphics[scale=0.3]{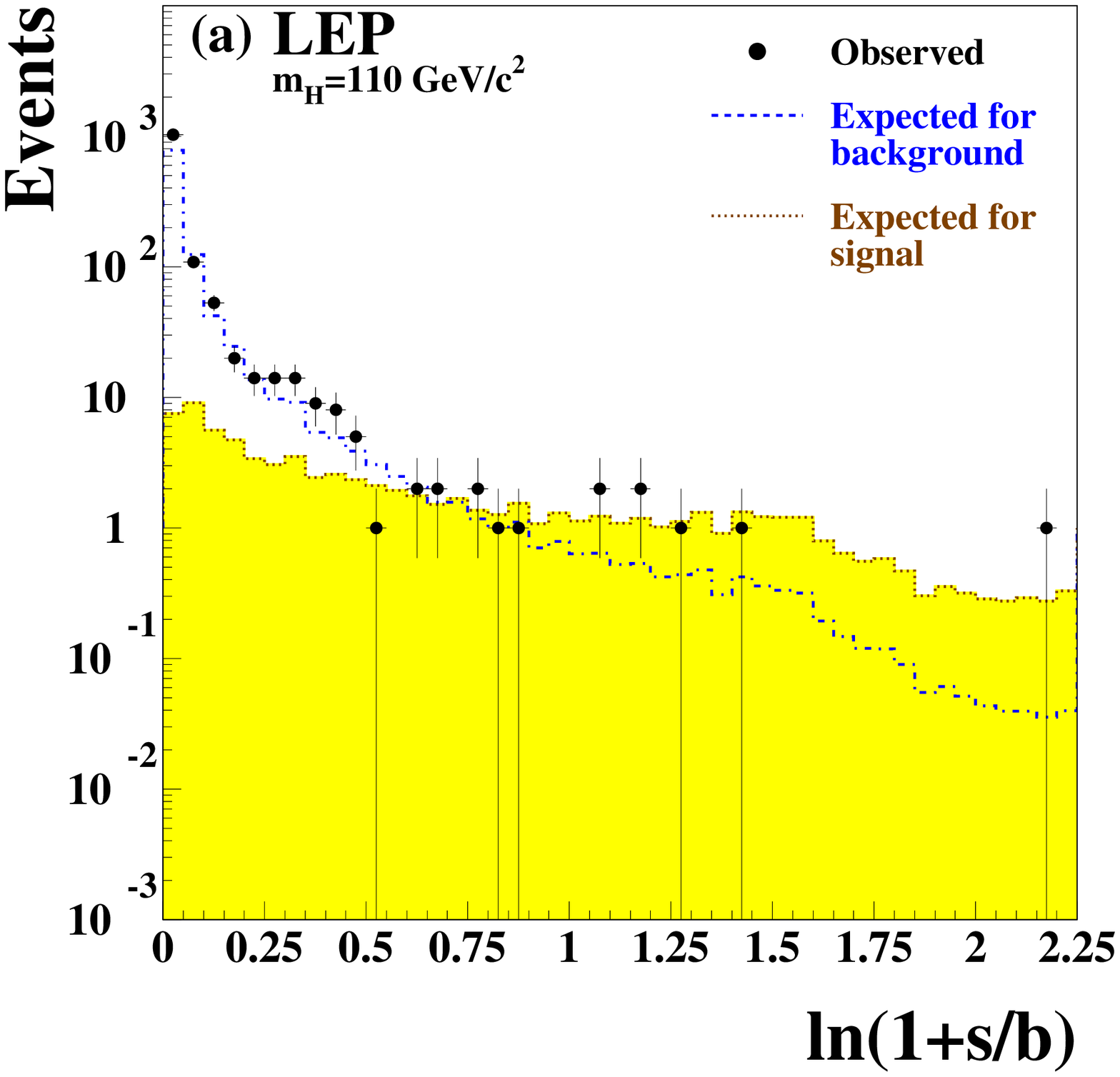}\hfill
\includegraphics[scale=0.3]{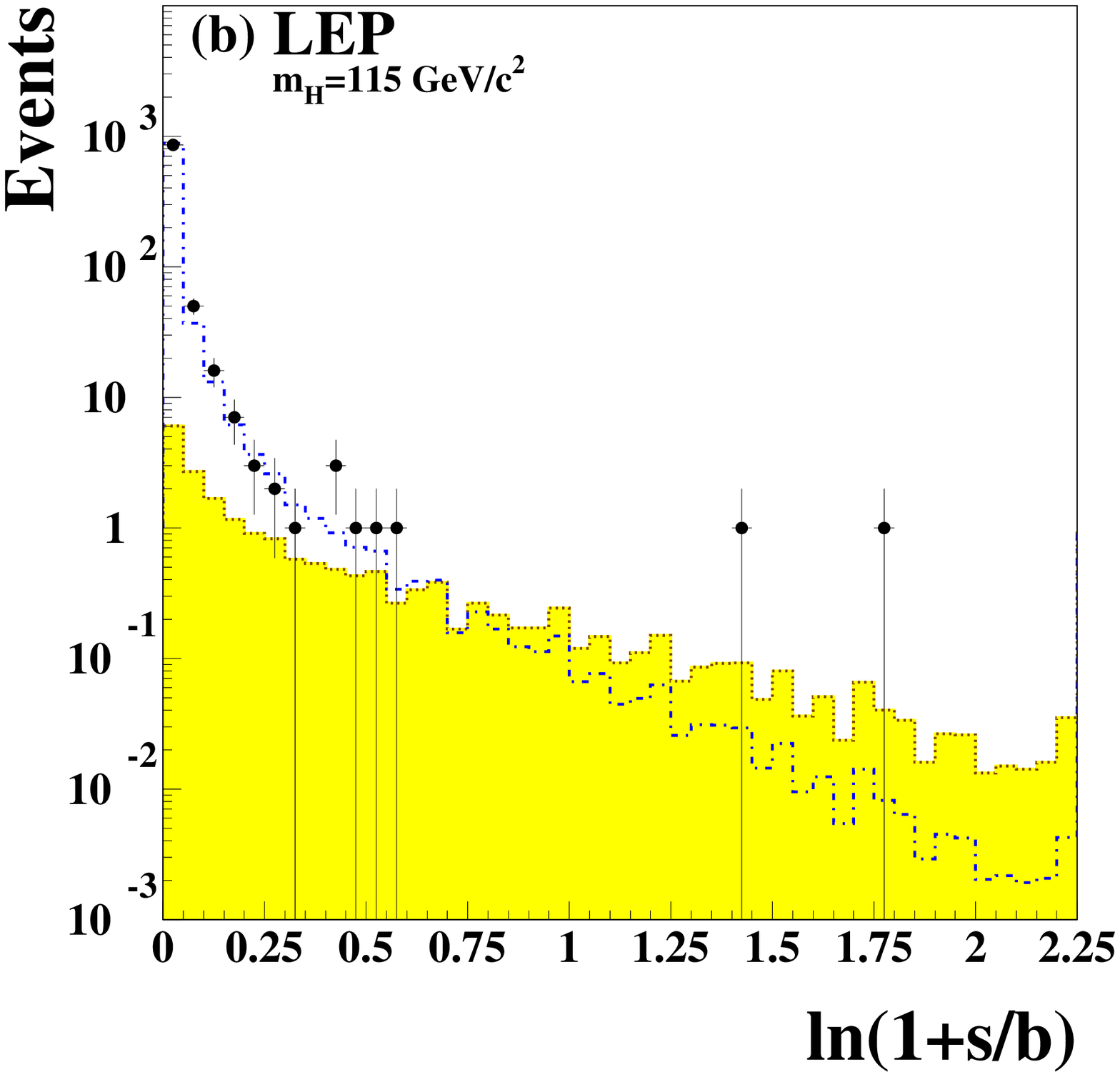}
\vspace*{-0.5cm}
\caption[]{SM Higgs boson: significance of candidate events 
           for a 100~GeV and 115~GeV Higgs boson hypothesis. 
\label{fig:sm_weight}}
\end{center}
\vspace*{-0.5cm}
\end{figure}

\vspace*{-0.2cm}
Figure~\ref{fig:sm_weight} shows the significance of the candidate events
for a 110~GeV and 115~GeV Higgs boson hypothesis.

\clearpage
\subsection{Reconstructed Candidate Masses}
\vspace*{-0.2cm}

\begin{figure}[htb]
\vspace*{-0.5cm}
\begin{center}
\includegraphics[scale=0.3]{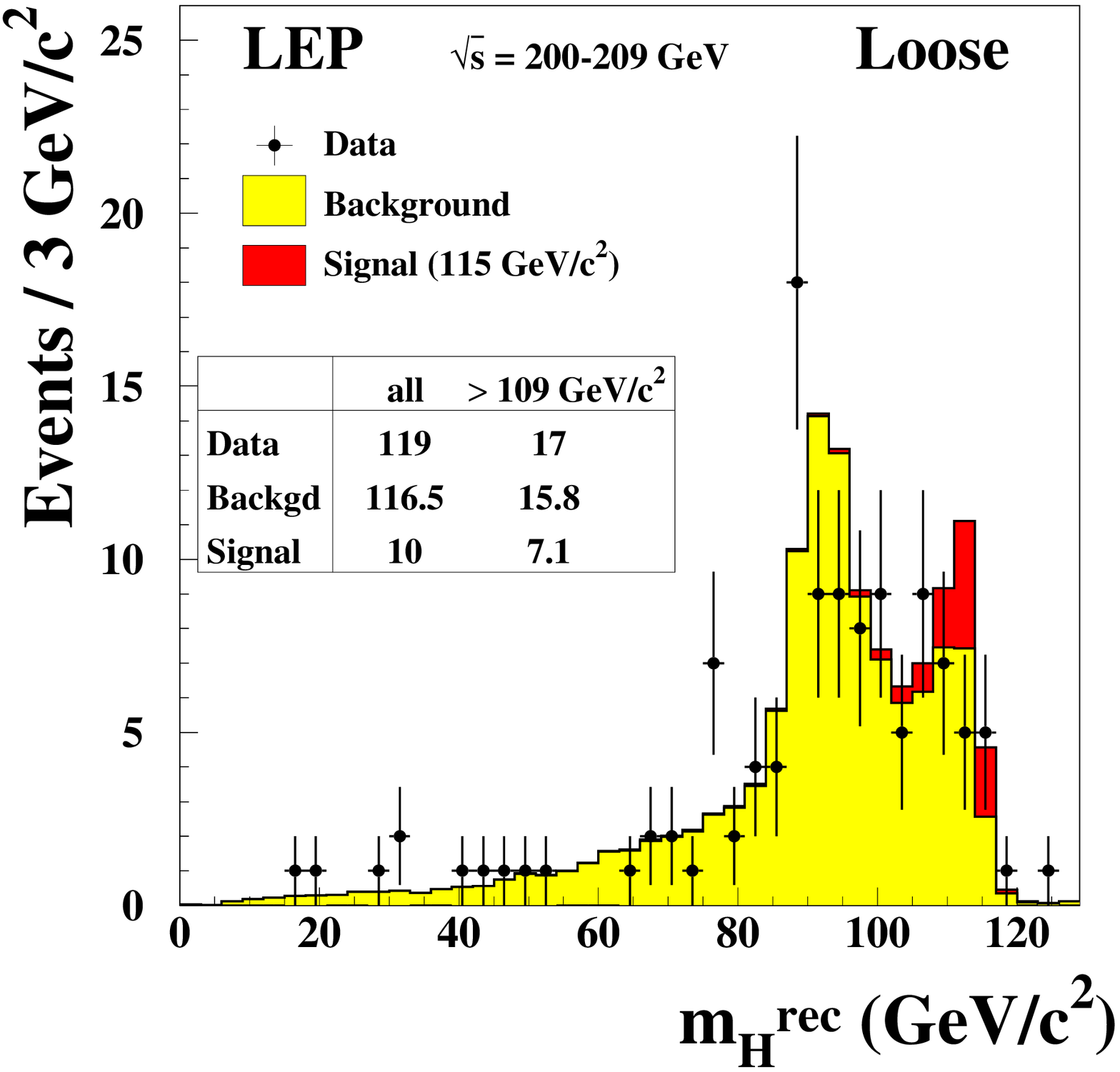}\hfill
\includegraphics[scale=0.3]{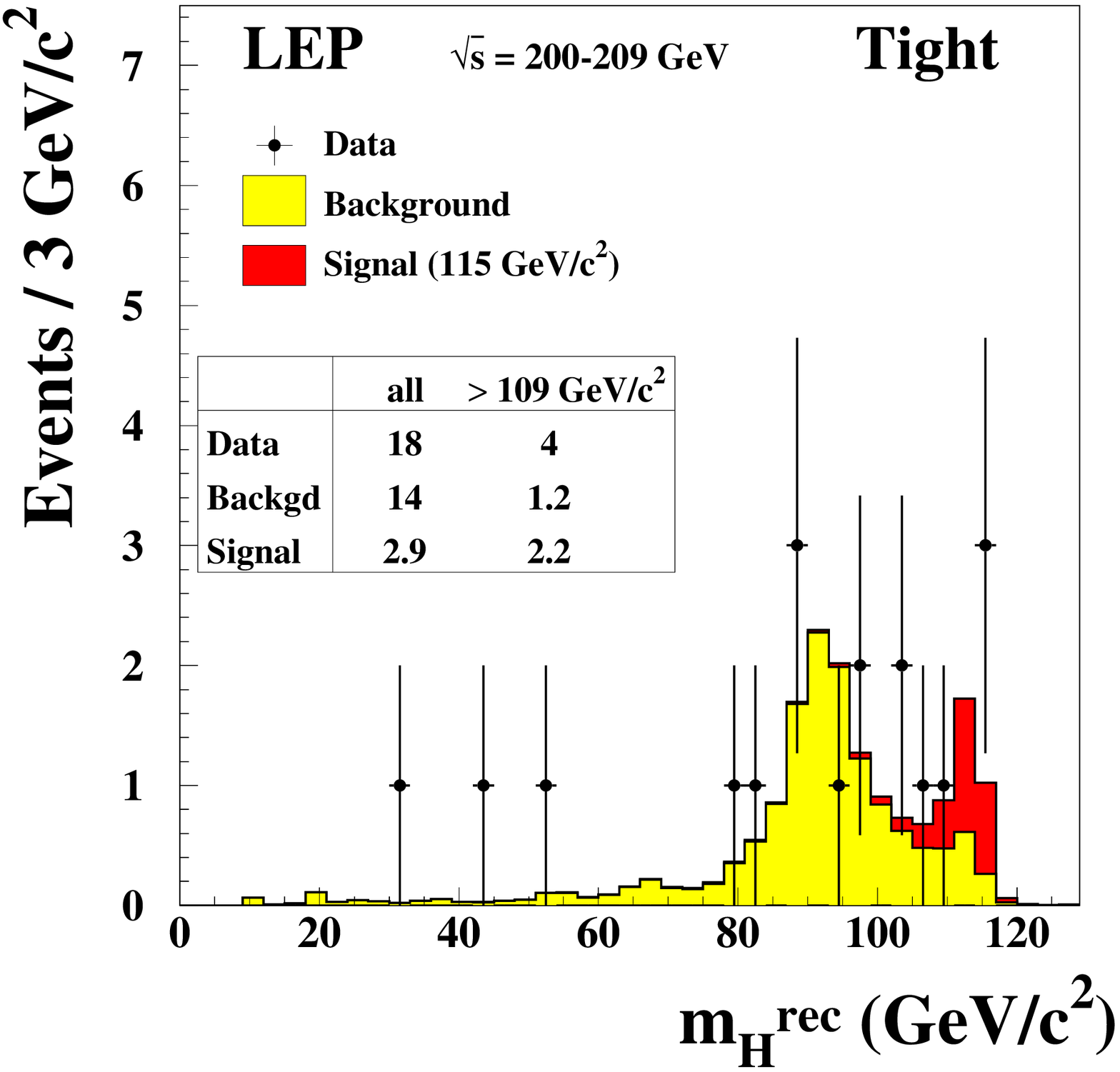}
\vspace*{-0.5cm}
\caption[]{SM Higgs boson: reconstructed mass of Higgs boson candidates
           for loose and tight selection cuts.
\label{fig:sm_mass}}
\end{center}
\end{figure}

Figure~\ref{fig:sm_mass} shows the reconstructed mass 
of the Higgs boson candidates with loose and tight selection cuts
for a 115~GeV Higgs boson hypothesis.

\subsection{Background Confidence Levels for Each Experiment}
\vspace*{-0.2cm}

\begin{figure}[htb]
\vspace*{-0.5cm}
\begin{center}
\includegraphics[scale=0.3]{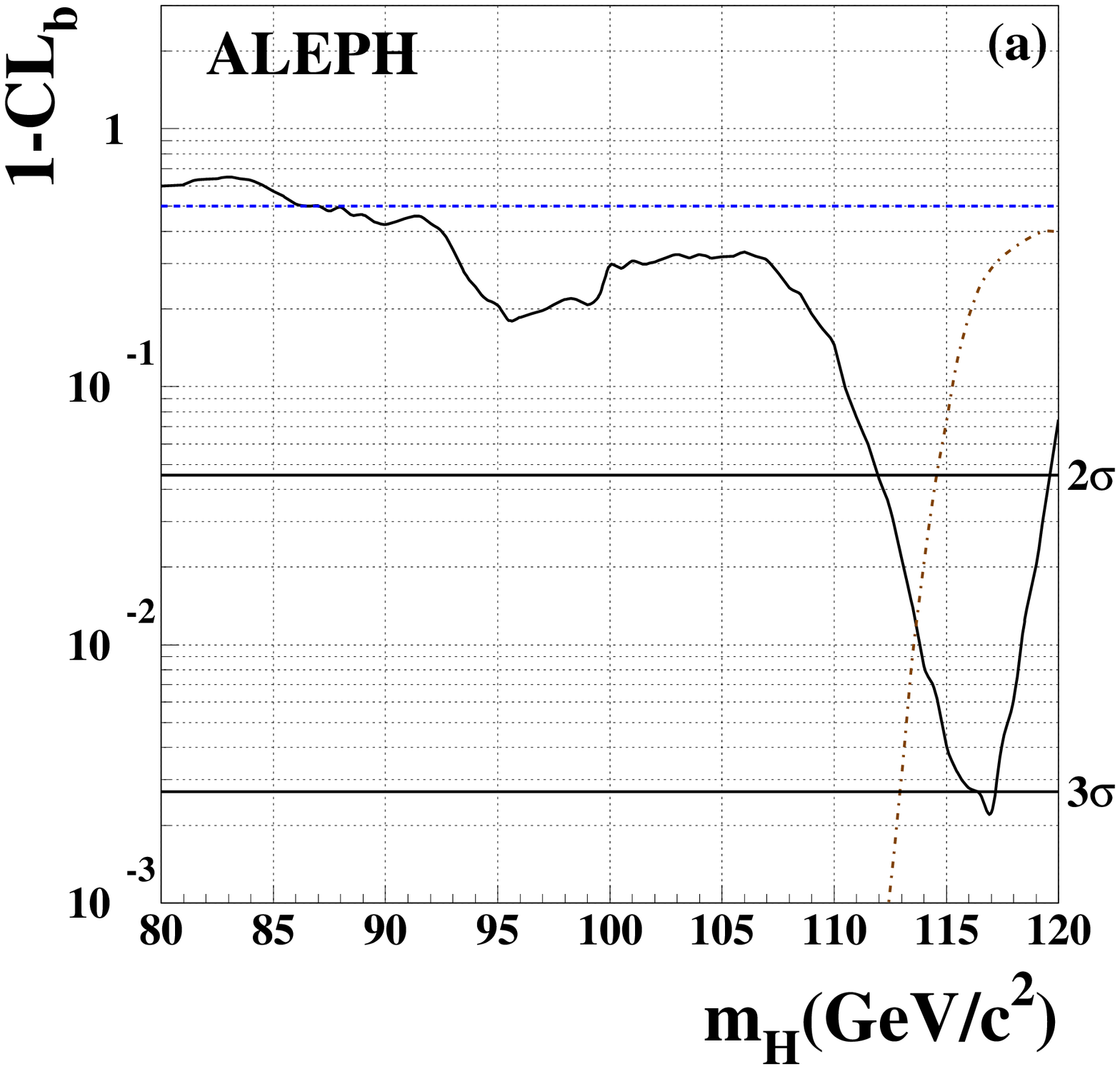}\hfill
\includegraphics[scale=0.3]{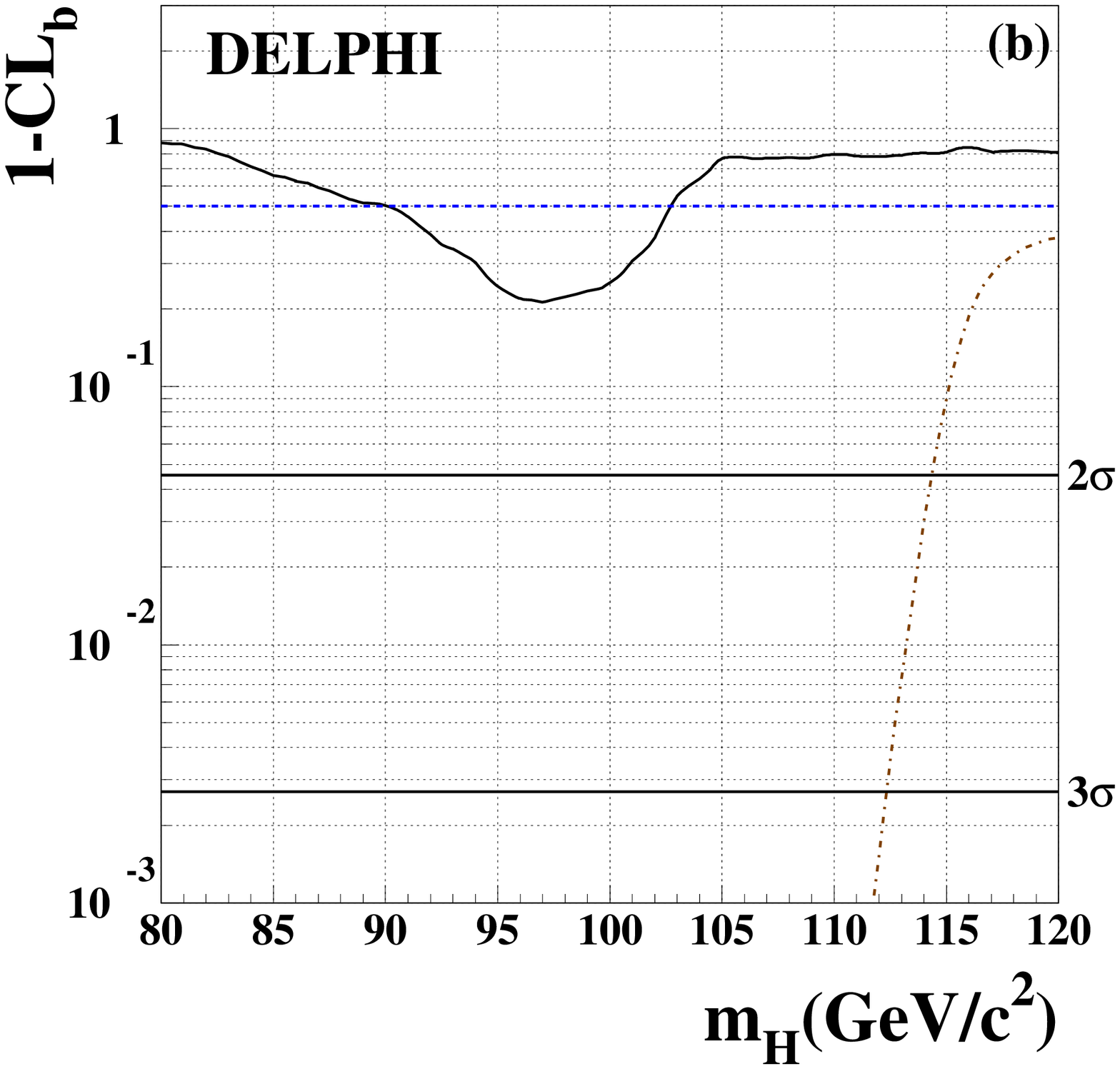}
\includegraphics[scale=0.3]{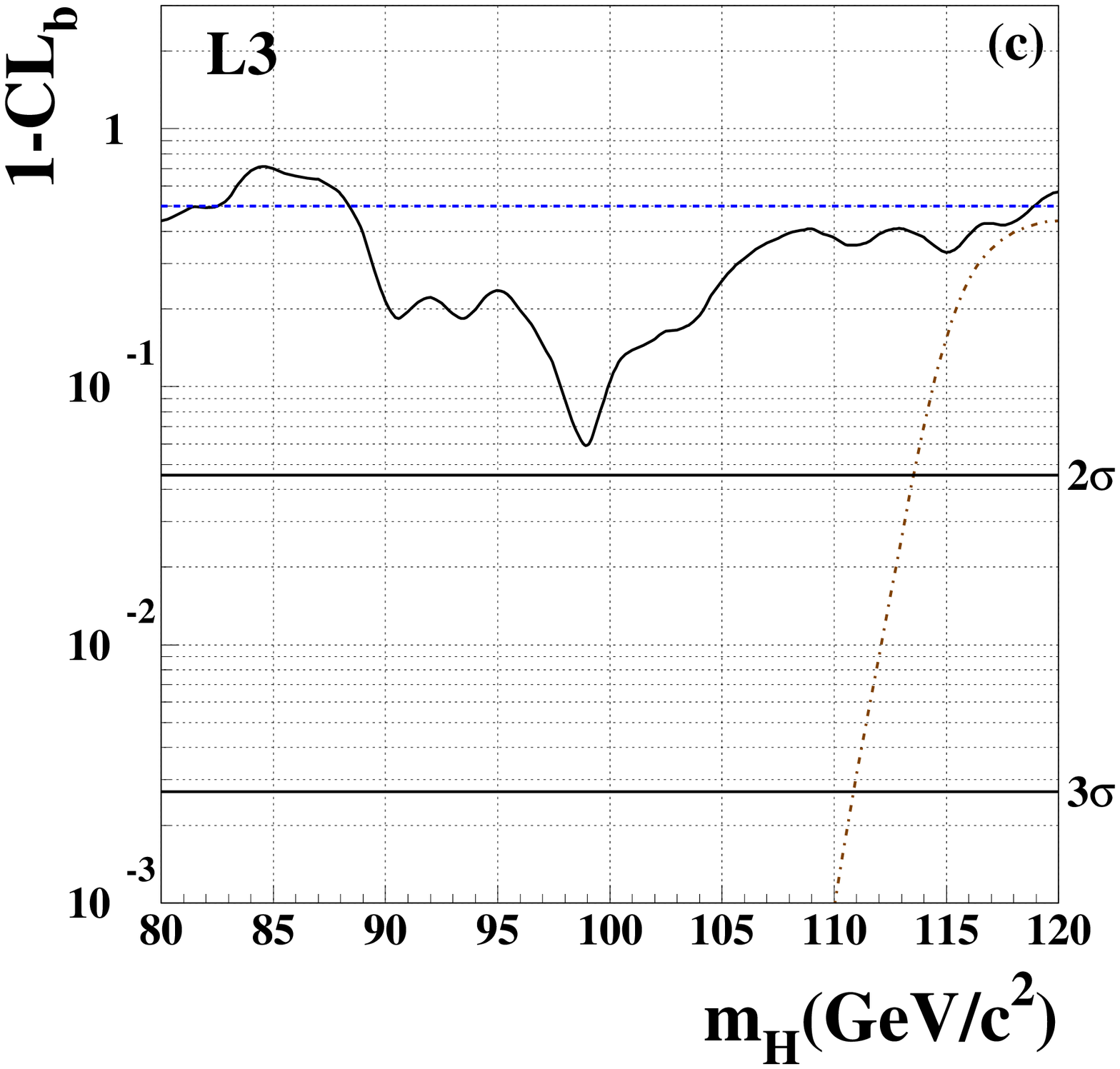}\hfill
\includegraphics[scale=0.3]{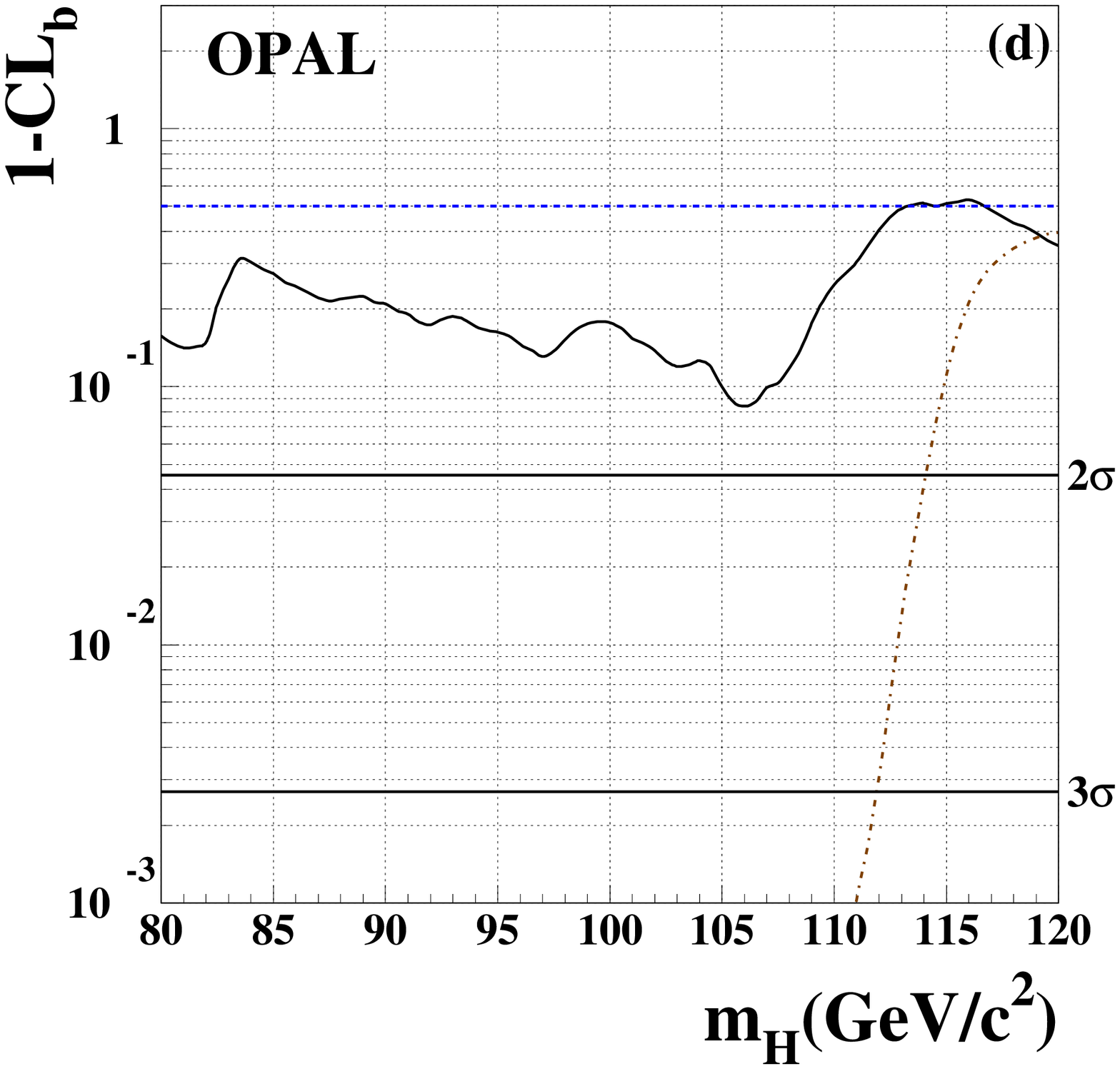}
\vspace*{-0.5cm}
\caption[]{SM Higgs boson: confidence levels for background-only
hypotheses for each experiment.
The value $1-CL_{\rm b}$ expresses the incompatibility of the observation
with the background-only hypothesis.
\label{fig:bgonly}}
\end{center}
\vspace*{-0.2cm}
\end{figure}

Figure~\ref{fig:bgonly} shows the confidence levels for background-only
hypotheses for each experiment separately.

\clearpage
\subsection{Background Confidence Levels: About $2\sigma$ Deviations}
\vspace*{-0.2cm}

\begin{figure}[htb]
\vspace*{-0.8cm}
\begin{center}
\includegraphics[scale=0.3]{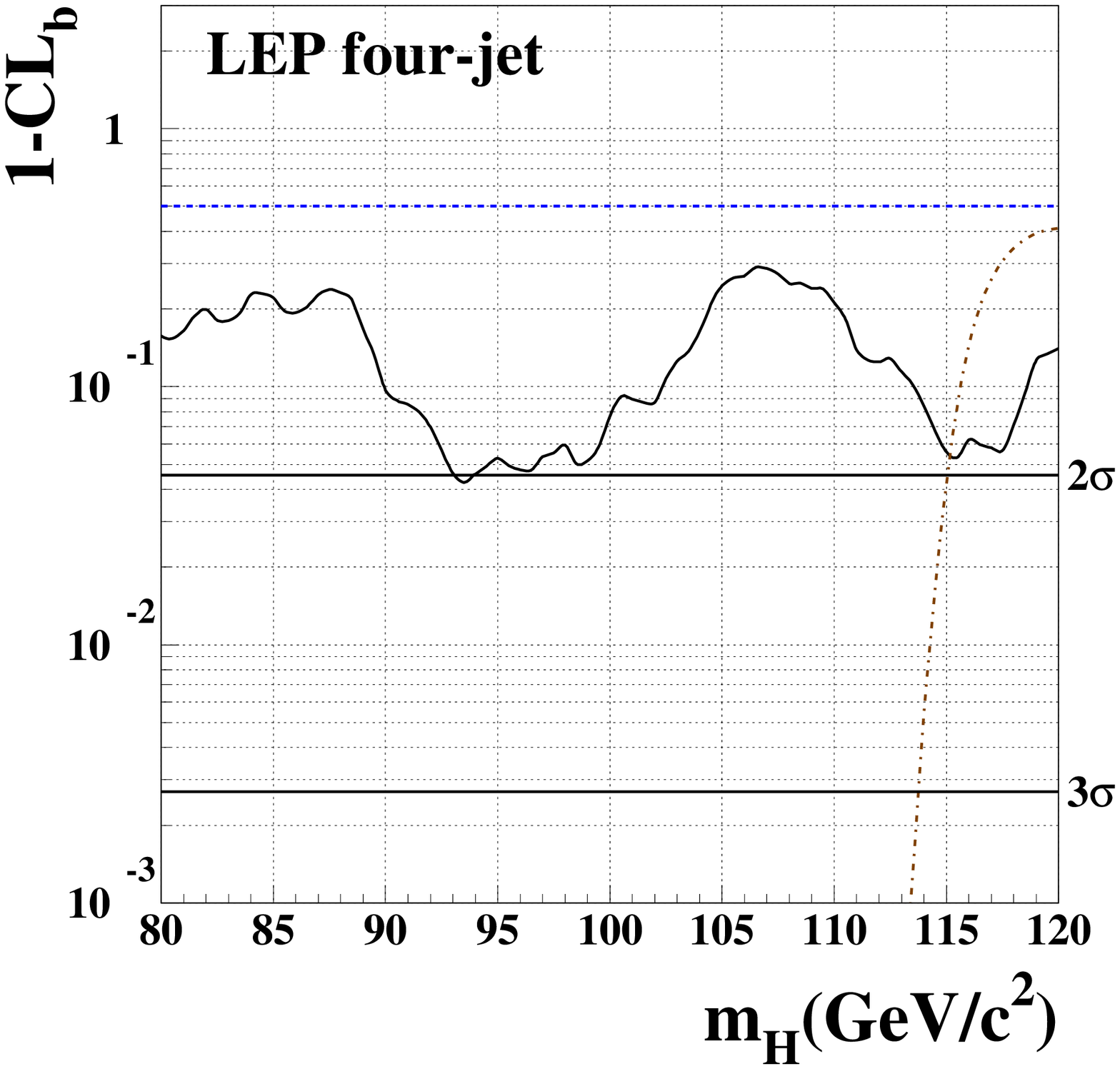}\hfill
\includegraphics[scale=0.3]{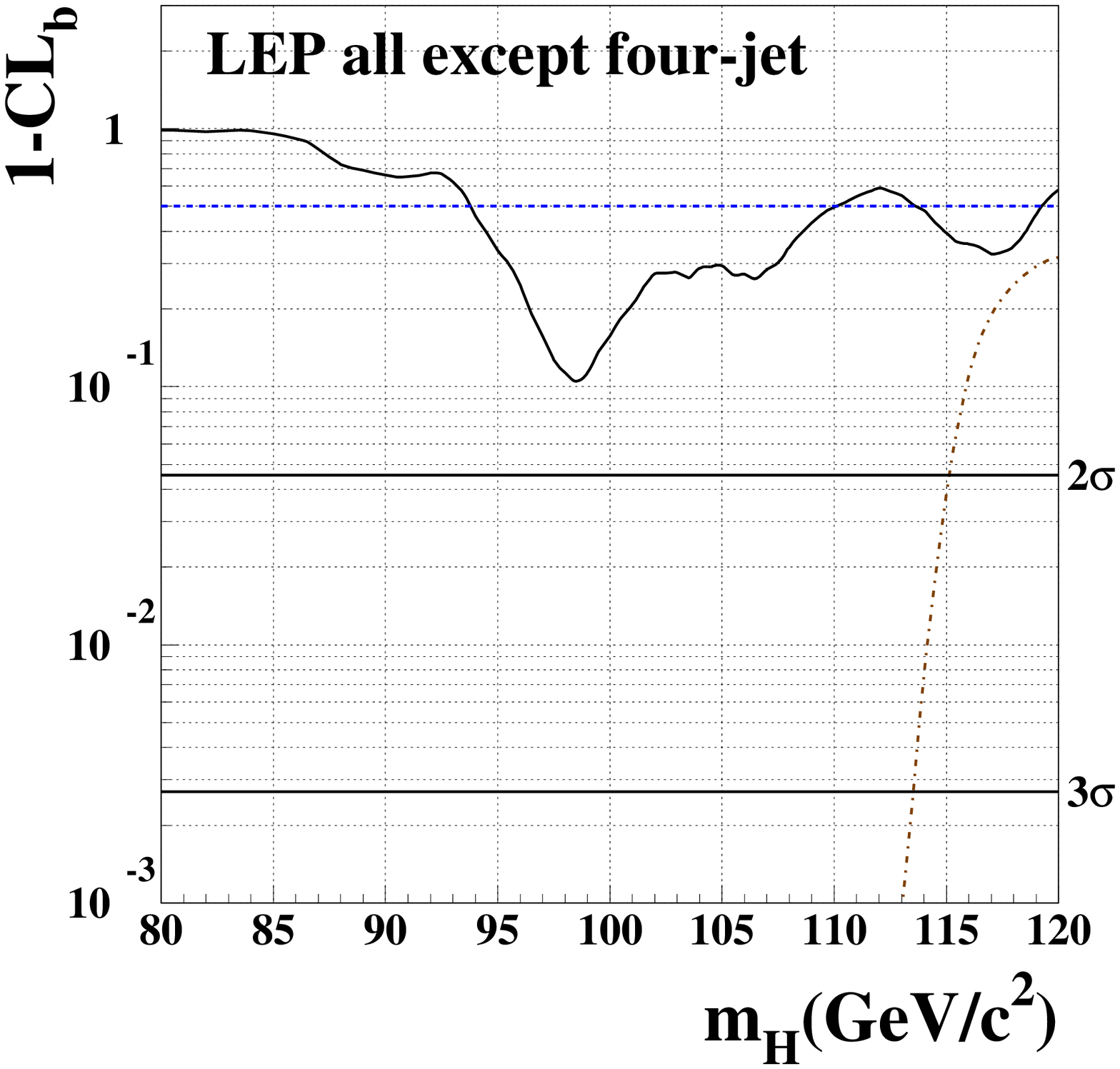}
\vspace*{-0.5cm}
\caption[]{SM Higgs boson: a small data excess at 98 GeV in the four-jet and 
also in all other channels combined. In addition,
a small excess at 115 GeV in the four-jet channel is observed.
\label{fig:98excess}}
\end{center}
\vspace*{-0.6cm}
\end{figure}

Figure~\ref{fig:98excess} shows a small data excess at 98 GeV in the four-jet 
channel and also in all other channels\,combined.\,In\,addition,\,a\,small\,excess\,at\,115\,GeV\,in\,the\,four-jet\,channel\,is\,observed.

\subsection{Probability Densities for Signal and Background 
            Hypotheses}
\vspace*{-0.2cm}

\begin{figure}[htb]
\vspace*{-0.8cm}
\begin{minipage}{0.48\textwidth}
\begin{center}
\includegraphics[scale=0.3]{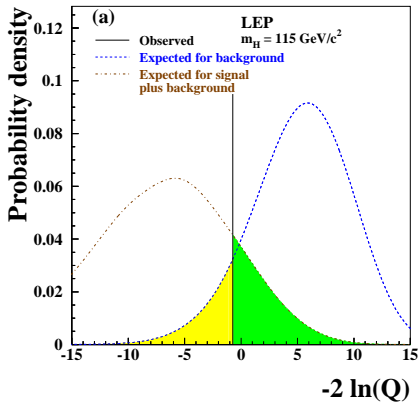}
\end{center}
\end{minipage}
\begin{minipage}{0.48\textwidth}
\vspace*{-0.5cm}
\caption[]{SM Higgs boson: probability densities of the 
test statistics $-2\ln Q$ for background-only and signal plus background 
hypotheses, and the observed value of $-2\ln Q$.
\label{fig:sm_density}}
\end{minipage}
\vspace*{-0.9cm}
\end{figure}

Figure~\ref{fig:sm_density} shows the probability densities of the 
test statistics $-2\ln Q$
for background-only and signal plus background hypotheses
for a 115~GeV Higgs boson, and the observed value of $-2\ln Q$.

\subsection{Mass and Coupling Limits}
\vspace*{-0.2cm}

\begin{figure}[htb]
\vspace*{-0.7cm}
\begin{center}
\includegraphics[scale=0.33]{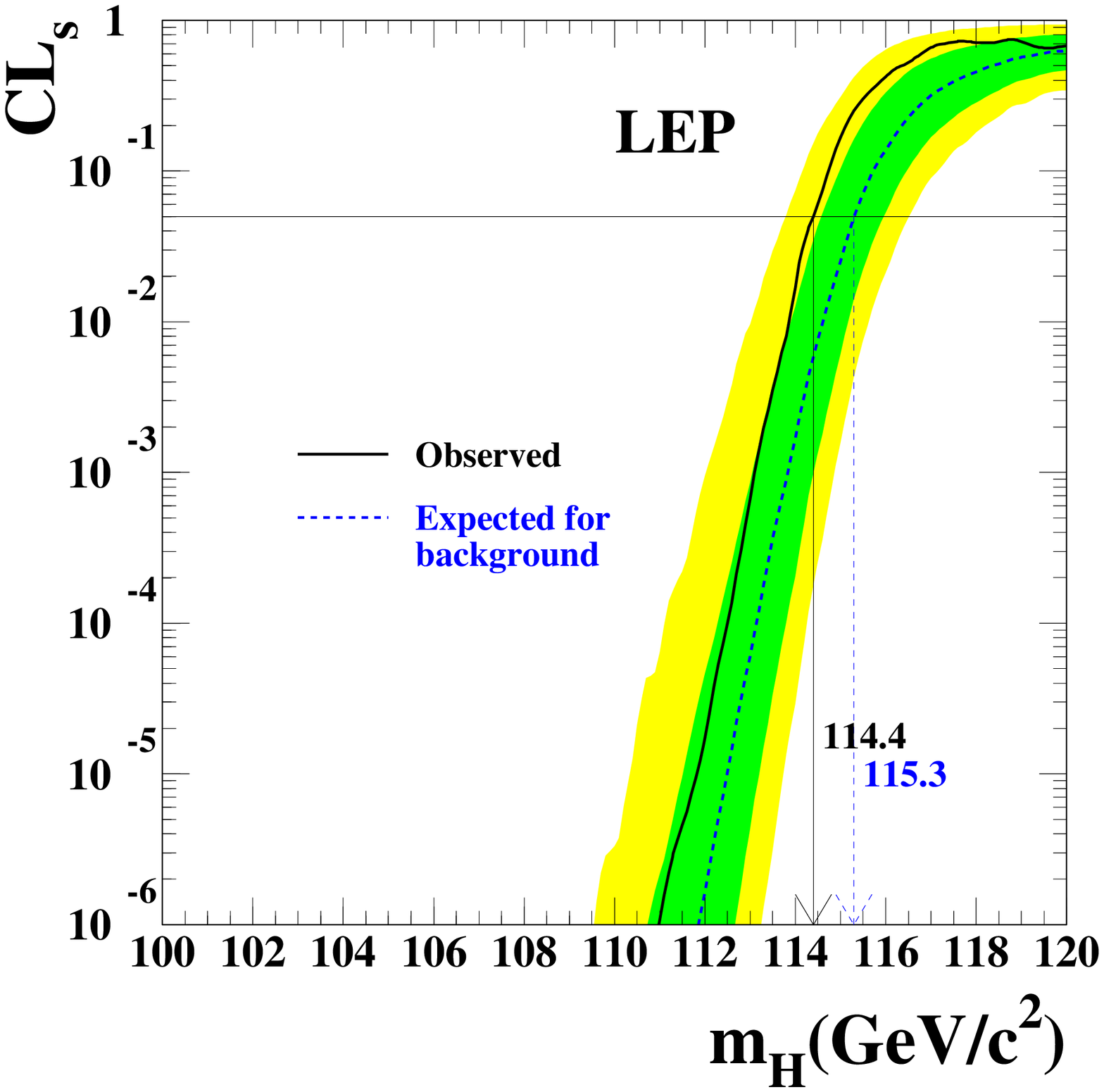}\hfill
\includegraphics[scale=0.33]{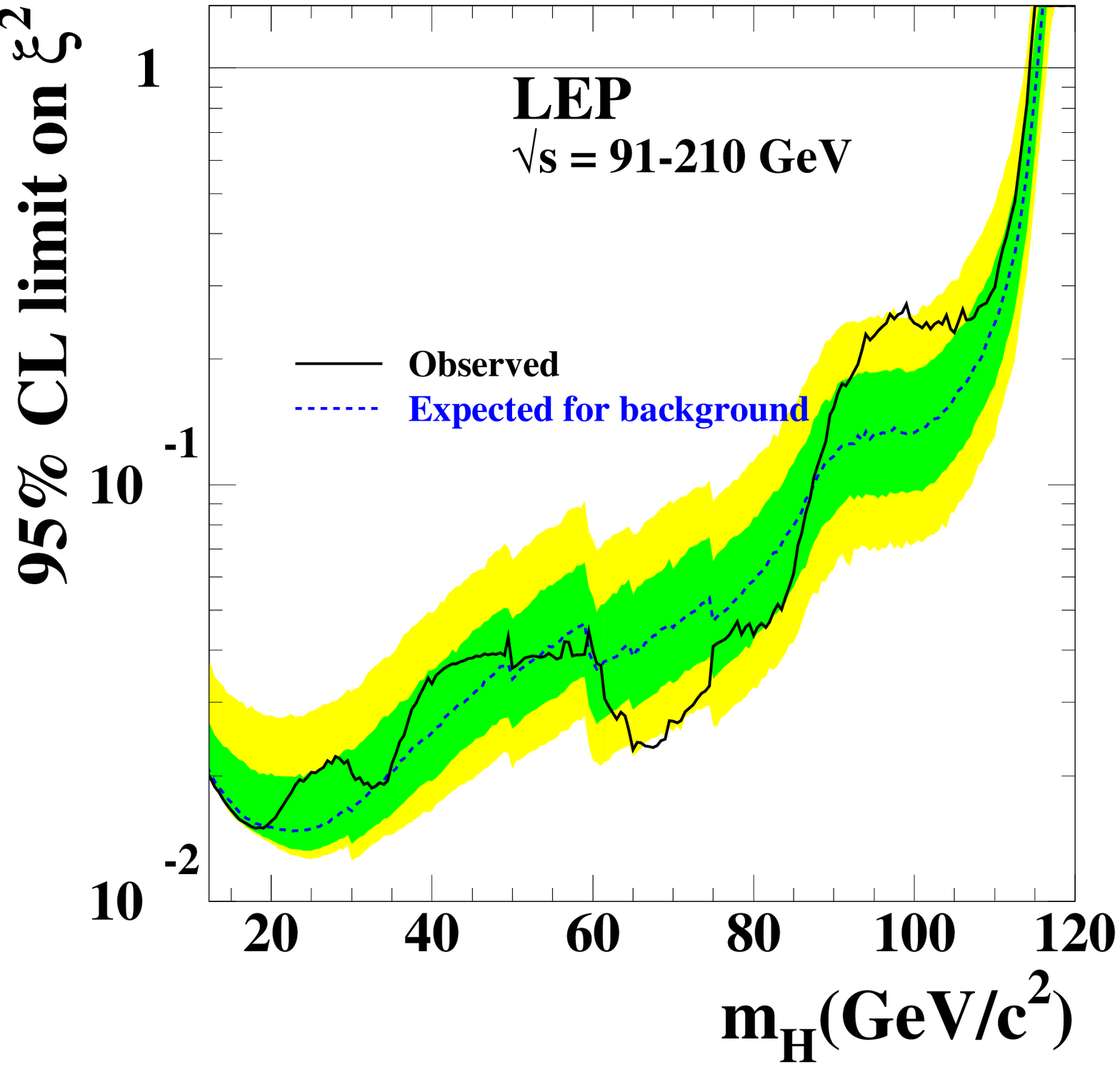}
\end{center}
\vspace*{-0.6cm}
\caption[]{SM Higgs boson. Left: mass limit. Right: 
coupling limits assuming the Higgs boson decays with SM branching
fractions and a SM production rate reduced by the factor 
$\xi^2 = (g_{\rm HZZ}/g_{\rm HZZ}^{\rm SM})^2$.
\label{fig:limits}}
\vspace*{-0.6cm}
\end{figure}

Figure~\ref{fig:limits} shows the SM mass limit,
and coupling limits assuming the Higgs boson decays with SM branching
fractions and a SM production rate reduced by the factor 
$\xi^2 = (g_{\rm HZZ}/g_{\rm HZZ}^{\rm SM})^2$.

\clearpage
\subsection{Coupling Limits: b-Quark and $\tau$-Lepton Decay Modes}
\vspace*{-0.2cm}

\begin{figure}[htb]
\vspace*{-0.7cm}
\begin{center}
\includegraphics[scale=0.3]{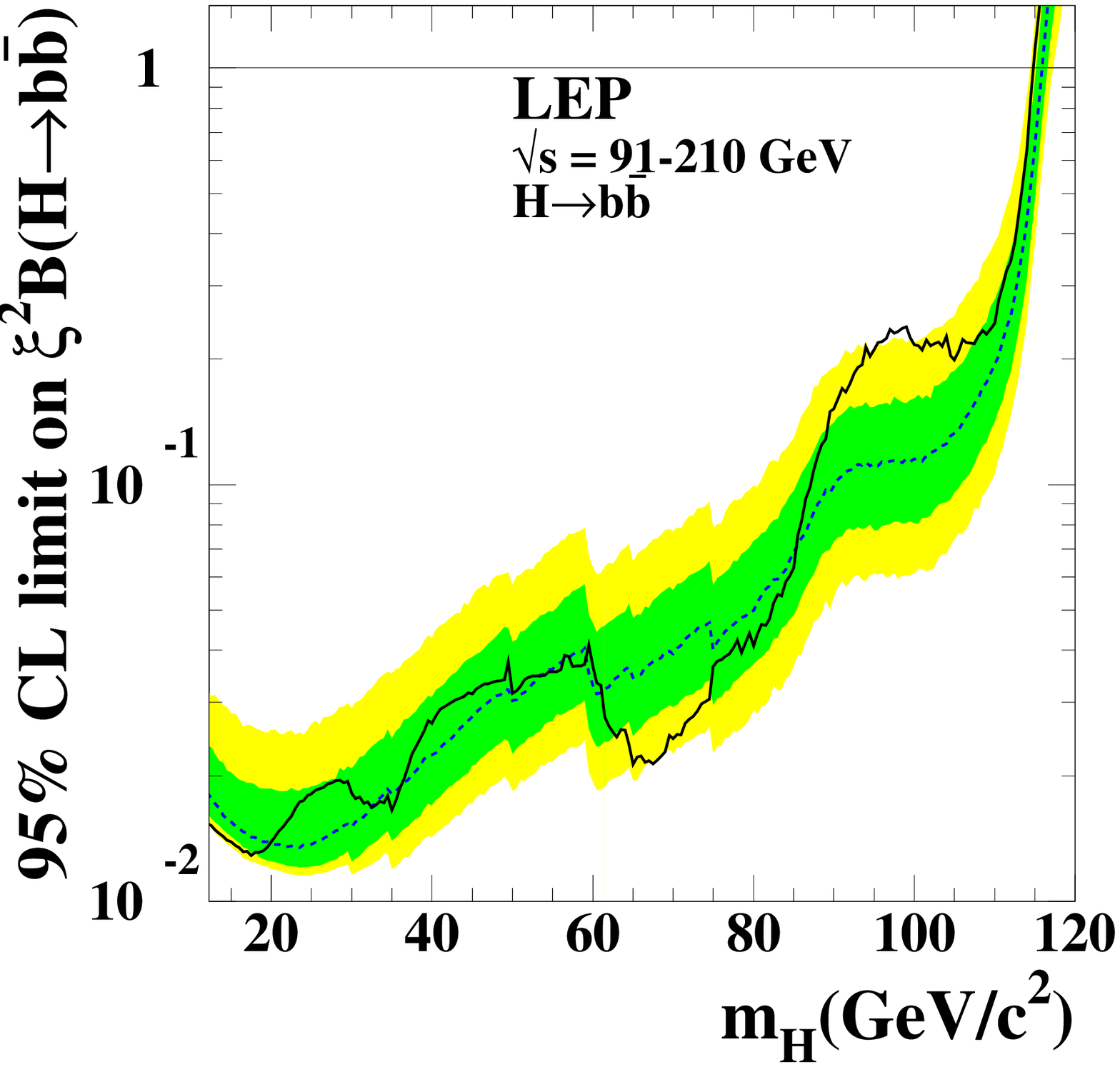}\hfill
\includegraphics[scale=0.3]{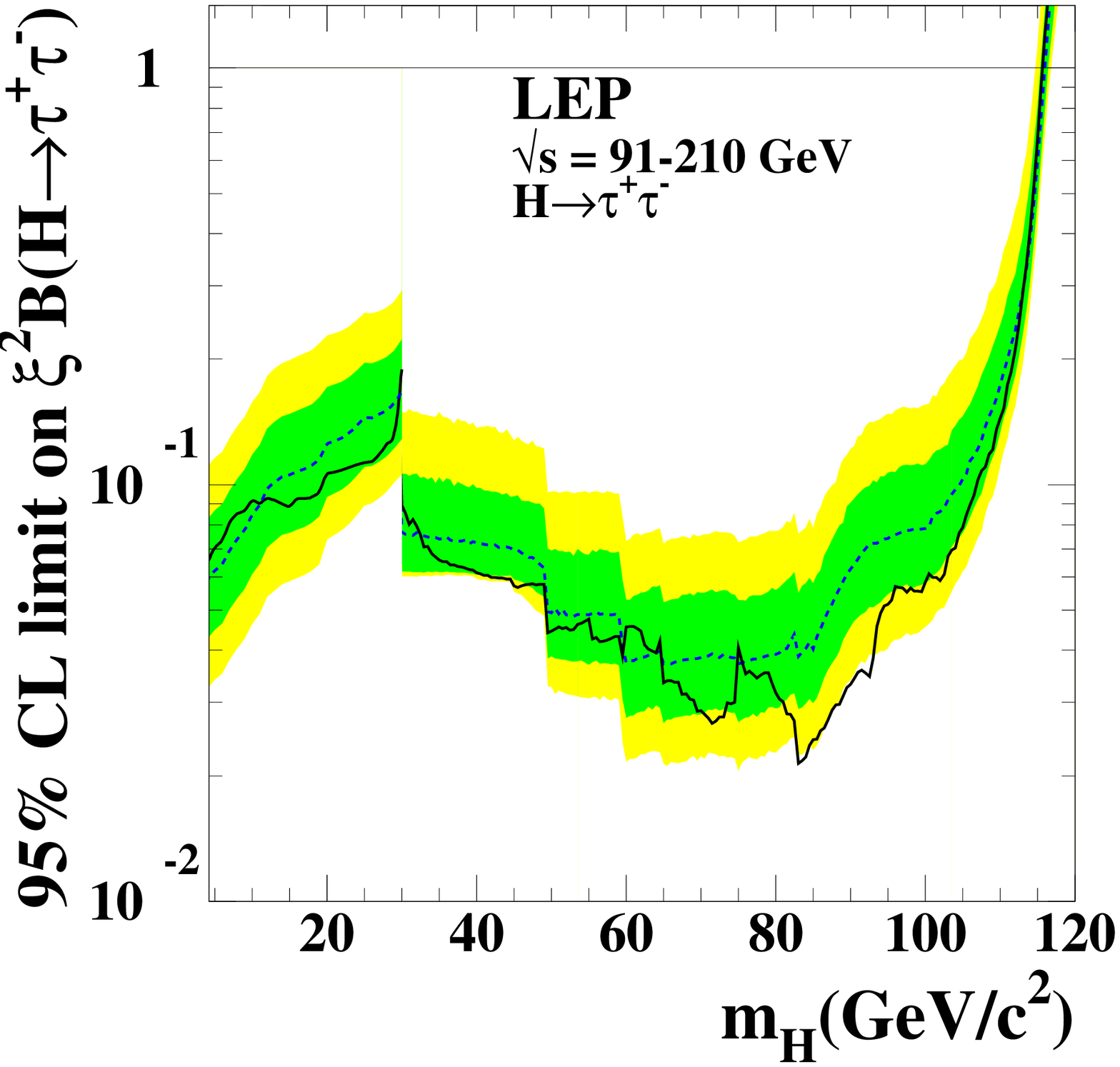}
\end{center}
\vspace*{-0.5cm}
\caption[]{SM Higgs boson: coupling limits for b-quark 
           and $\tau$-lepton decay modes.
\label{fig:coupling}}
\vspace*{-0.5cm}
\end{figure}

Figure~\ref{fig:coupling} shows coupling limits for b-quark 
and $\tau$-lepton decay modes.

\section{Minimal Supersymmetric Extension of the SM (MSSM)}

\subsection{Benchmark Limits and Dedicated Low $m_{\rm A}$ Searches}
\vspace*{-0.2cm}

\begin{figure}[htb]
\vspace*{-0.5cm}
\begin{center}
\includegraphics[scale=0.3]{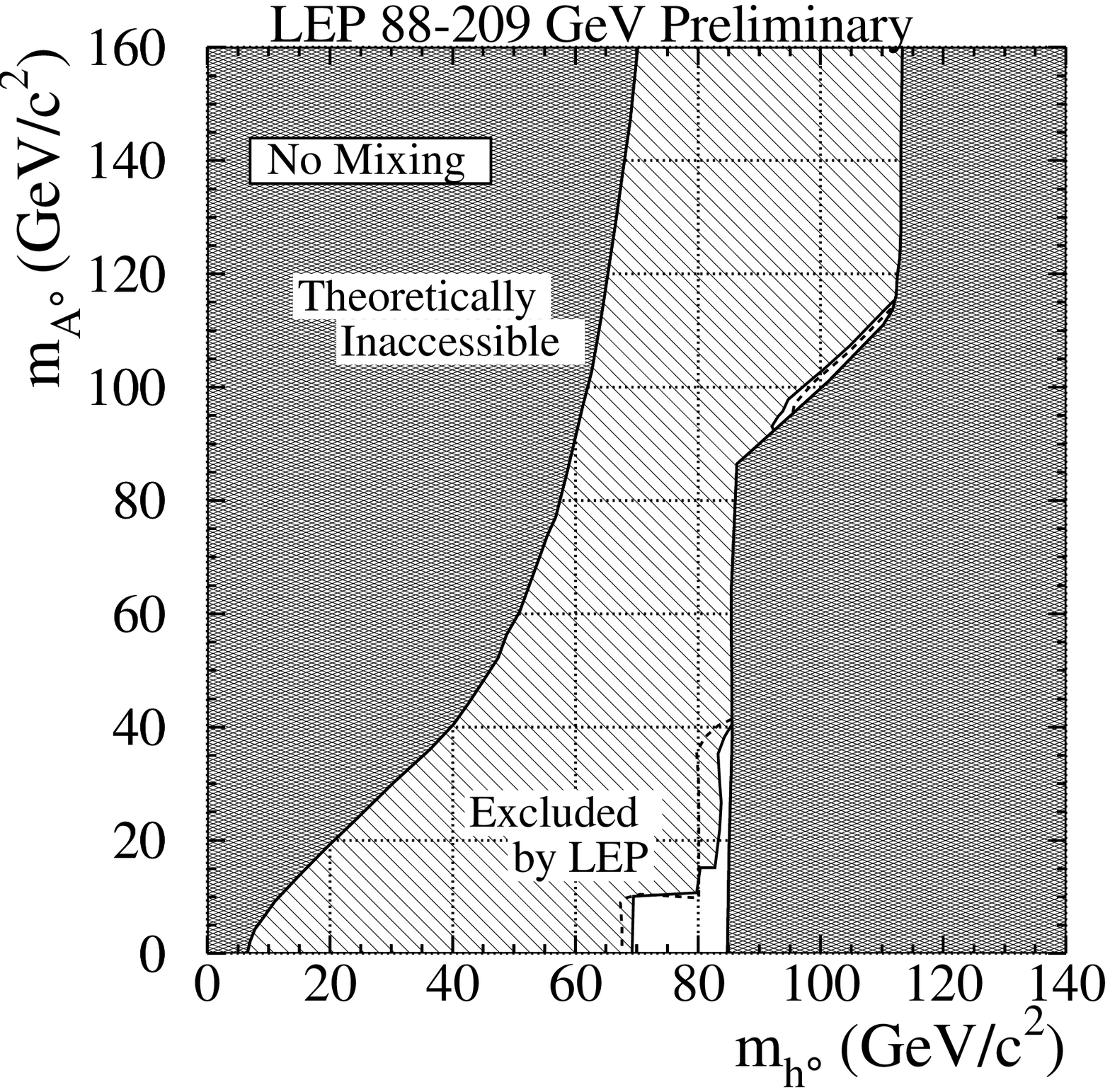}\hfill
\includegraphics[scale=0.3]{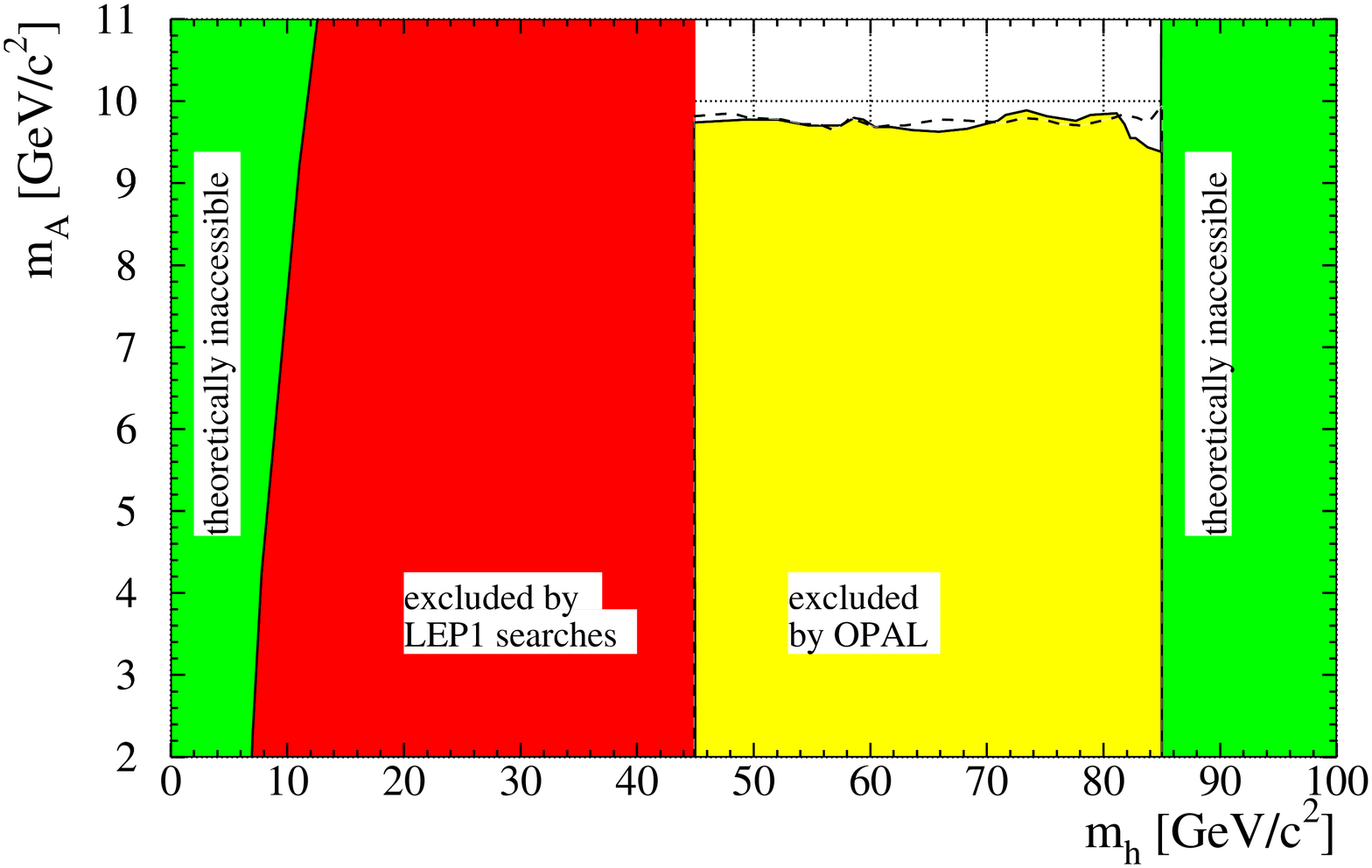}
\end{center}
\vspace*{-0.5cm}
\caption[]{MSSM. Left: unexcluded mass region for a light A boson
           in the no-mixing scalar top benchmark scenario.
                 Right: excluded mass region by dedicated searches
                 for a light A boson.
\label{fig:light}}
\vspace*{-0.5cm}
\end{figure}

Figure~\ref{fig:light} shows a previously small unexcluded mass region
for light A masses in the no-mixing scalar top benchmark scenario.
This region is mostly excluded by new dedicated searches for 
a light A boson (right plot).

\subsection{Benchmark Limits and Dedicated $\rm h\ra AA$ Searches}
\vspace*{-0.2cm}

\begin{figure}[htb]
\vspace*{-0.1cm}
\begin{minipage}{0.48\textwidth}
\begin{center}
\includegraphics[scale=0.4]{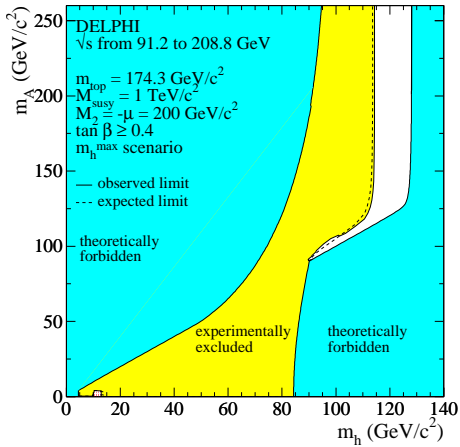}
\end{center}
\end{minipage}
\begin{minipage}{0.48\textwidth}
\vspace*{-0.5cm}
\caption[]{MSSM: excluded mass region for the maximum h-mass benchmark scenario, 
           including results from dedicated searches for $\rm h\ra AA$.
\label{fig:haa}}
\end{minipage}
\vspace*{-0.8cm}
\end{figure}

Figure~\ref{fig:haa} shows mass limits for the maximum h-mass benchmark 
scenario,
including results from dedicated searches for the reaction $\rm h\ra AA$.

\clearpage
\subsection{Three-Neutral-Higgs-Boson Hypothesis}
\begin{figure}[htb]
\vspace*{-0.8cm}
\begin{center}
\includegraphics[scale=0.3]{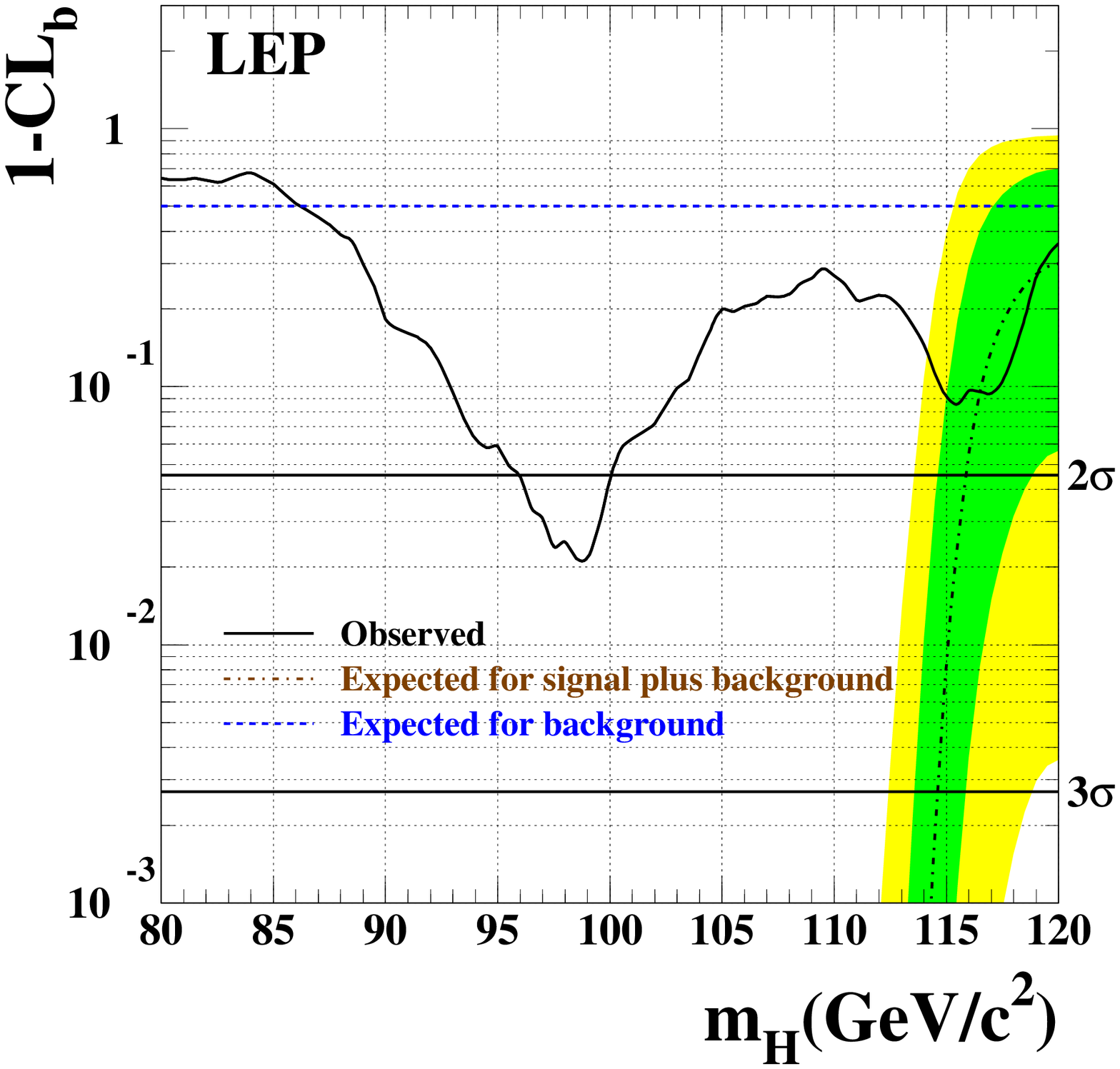}\hfill
\includegraphics[scale=0.37]{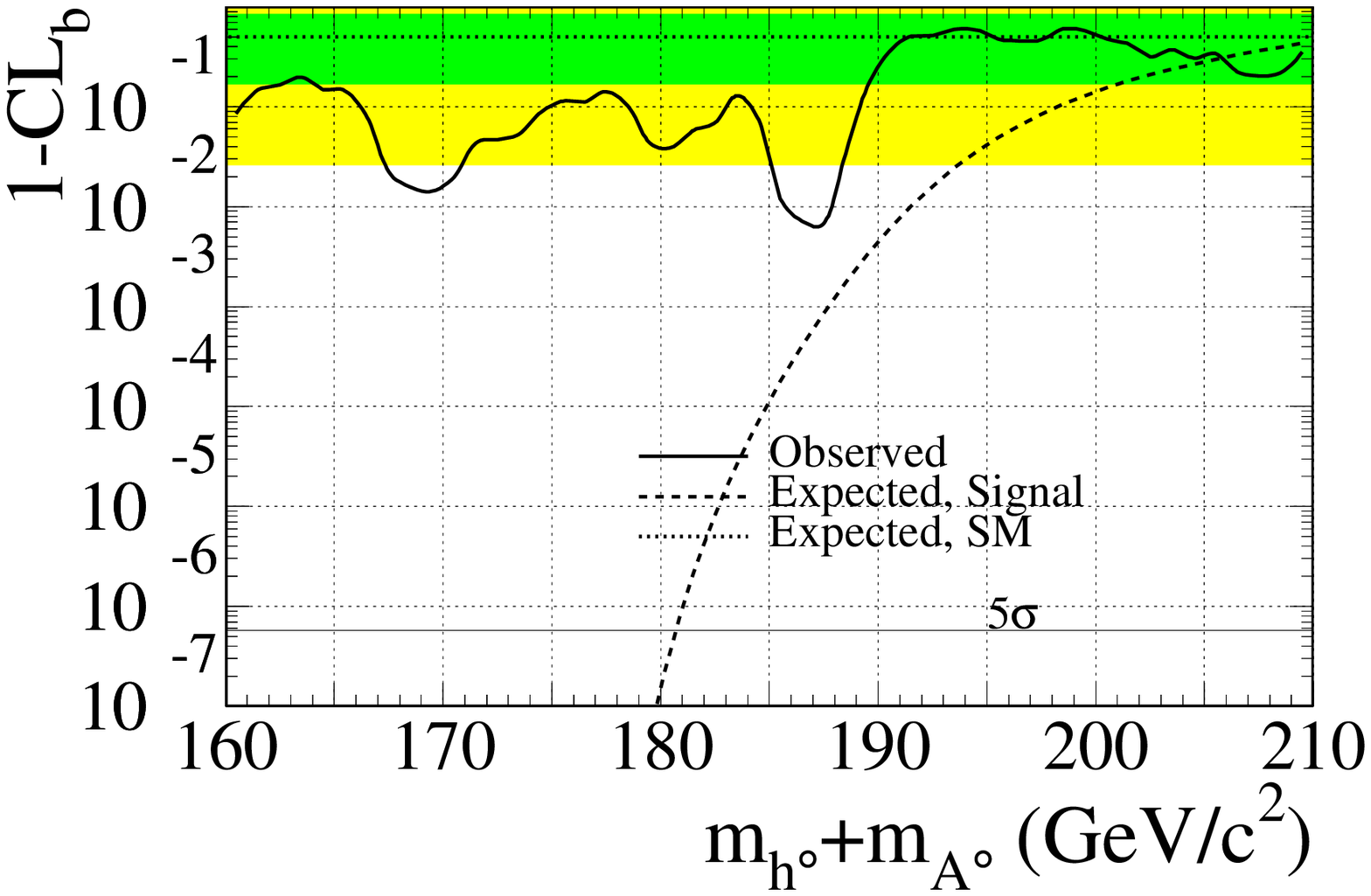}
\end{center}
\vspace*{-0.6cm}
\caption[]{MSSM. Left: small data excess at 99 GeV and 116 GeV in hZ/HZ searches.
                 Right: small data excess at 
                 $m_{\rm h}+m_{\rm A}=187$~GeV in hA searches.
\label{fig:three}}
\vspace*{-0.5cm}
\end{figure}

\vspace*{-0.2cm}
The hypothesis of three-neutral-Higgs-boson production, via hZ, HZ and hA
is compatible with the data excess seen in Fig.~\ref{fig:three}.
For the reported MSSM parameters~\cite{as2000} reduced hZ production 
near 100 GeV and HZ production near 115 GeV is compatible with the 
data (left plot). 
For $m_{\rm h} \approx m_{\rm A}$, hA production is also compatible 
with the data (right plot).

\subsection{MSSM Parameter Scan}
\begin{figure}[htb]
\vspace*{-0.7cm}
\begin{center}
\includegraphics[scale=0.3]{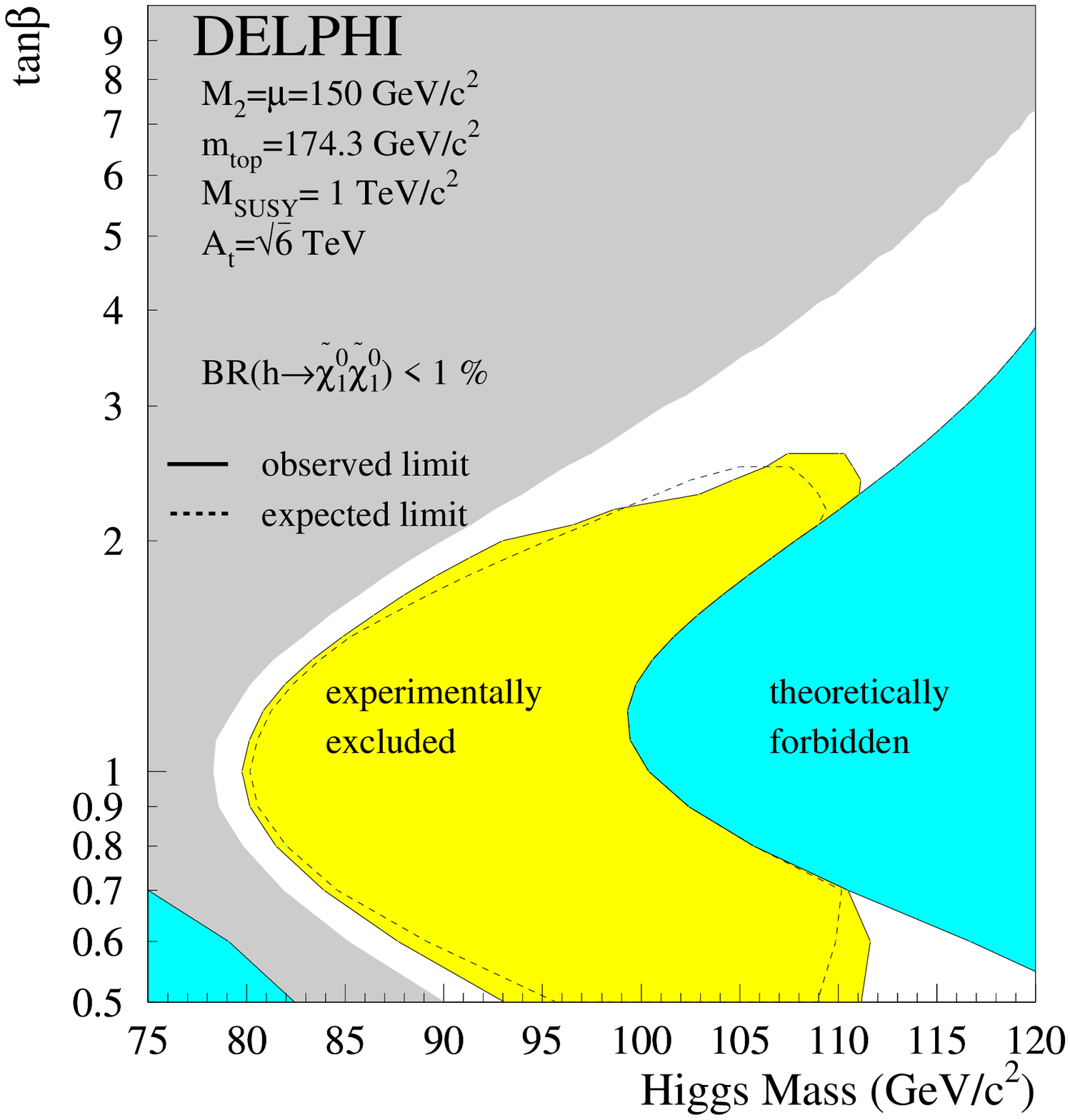}\hfill
\includegraphics[scale=0.3]{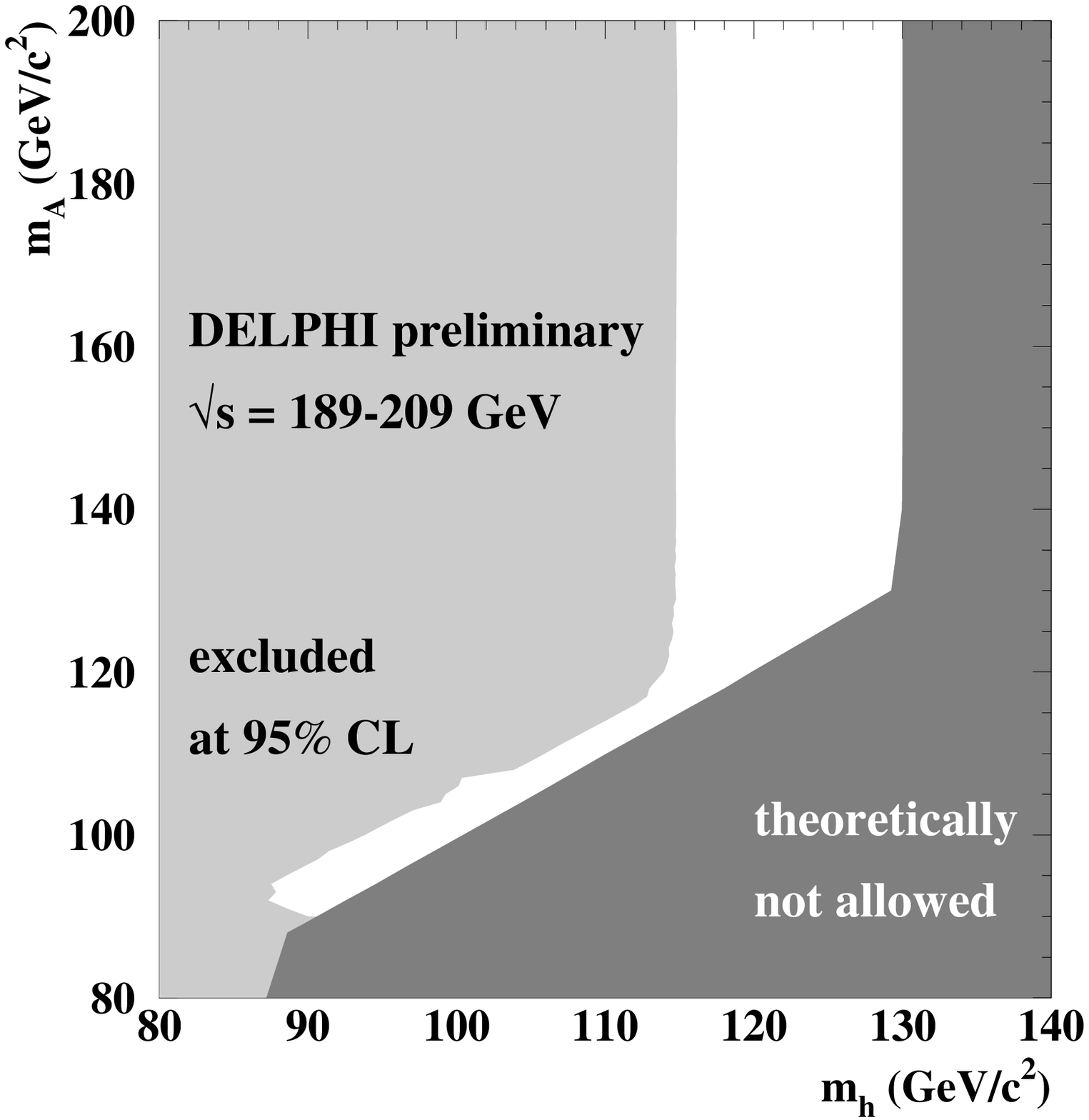}
\end{center}
\vspace*{-0.5cm}
\caption[]{MSSM. Left: mass limits from searches for invisibly-decaying Higgs bosons. 
                 Right: mass limits from a general MSSM parameter scan.
\label{fig:mssmscan}}
\vspace*{-0.5cm}
\end{figure}

\vspace*{-0.2cm}
Mass limits in the MSSM depend on invisibly-decaying Higgs boson searches,
in particular for general MSSM parameter scans as shown in Fig.~\ref{fig:mssmscan}.

\section{CP-Violating Models}
\vspace*{-0.2cm}

\begin{figure}[htb]
\vspace*{-0.3cm}
\begin{minipage}{0.56\textwidth}
\begin{center}
\includegraphics[scale=0.4]{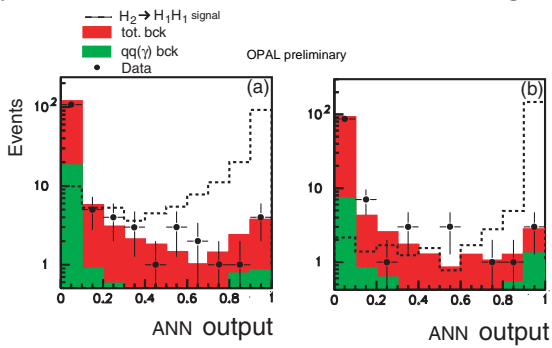}\\
\end{center}
\end{minipage}
\begin{minipage}{0.40\textwidth}
\vspace*{-0.8cm}
\caption[]{CP-violation models.
Artificial Neural Network (ANN) output distributions for the reactions
$\rm \ee\ra H_2 Z \ra b\bar b \nu\bar \nu$ and 
$\rm \ee\ra H_2 Z \ra H_1 H_1 Z \ra b\bar b b\bar b \nu\bar \nu$
for different data sub-samples. No indication of a signal is observed.
\label{fig:cpvio}}
\end{minipage}
\vspace*{-0.7cm}
\end{figure}

\begin{figure}[htb]
\vspace*{0.2cm}
\begin{center}
\includegraphics[scale=0.3]{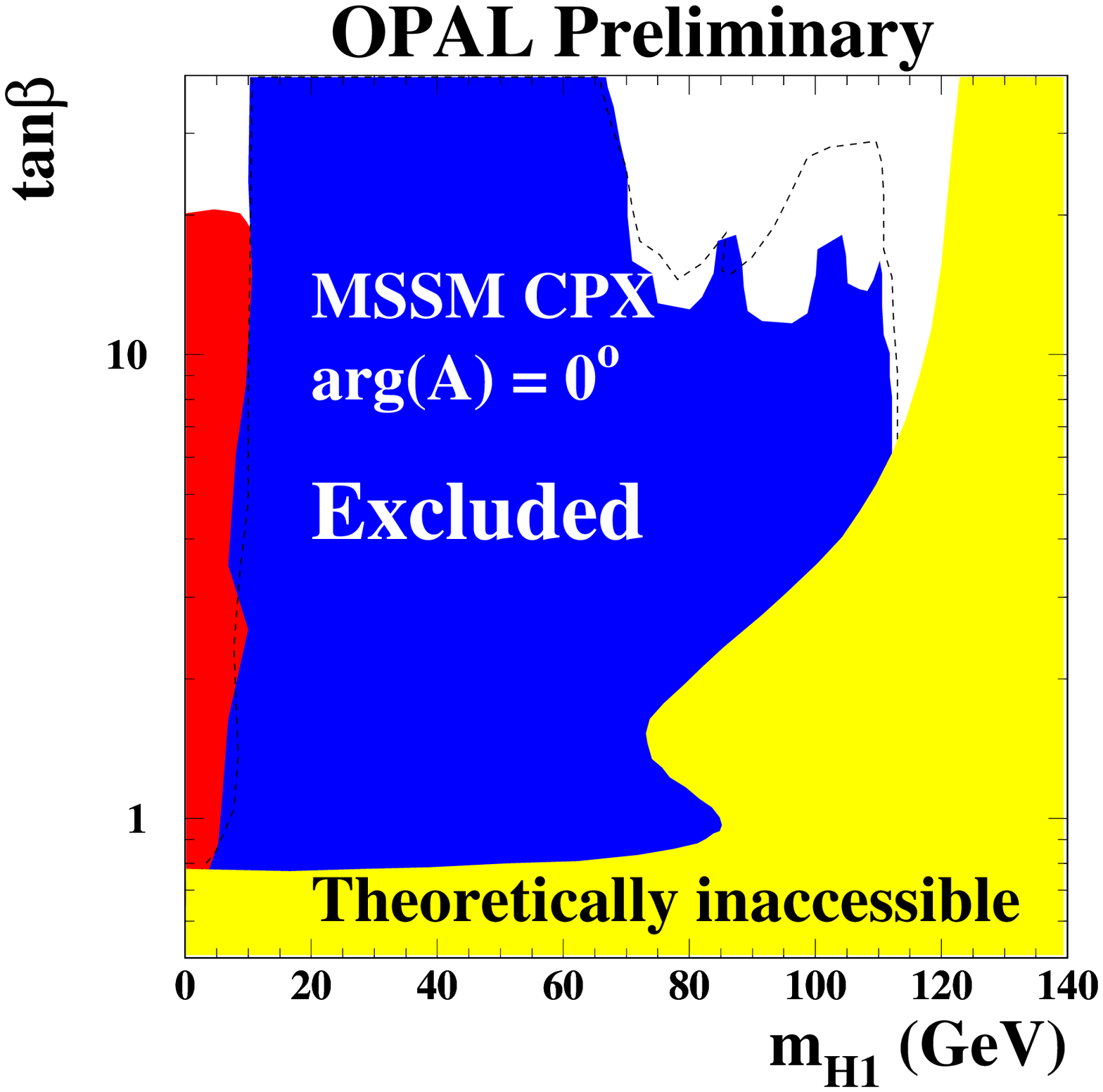}\hfill
\includegraphics[scale=0.3]{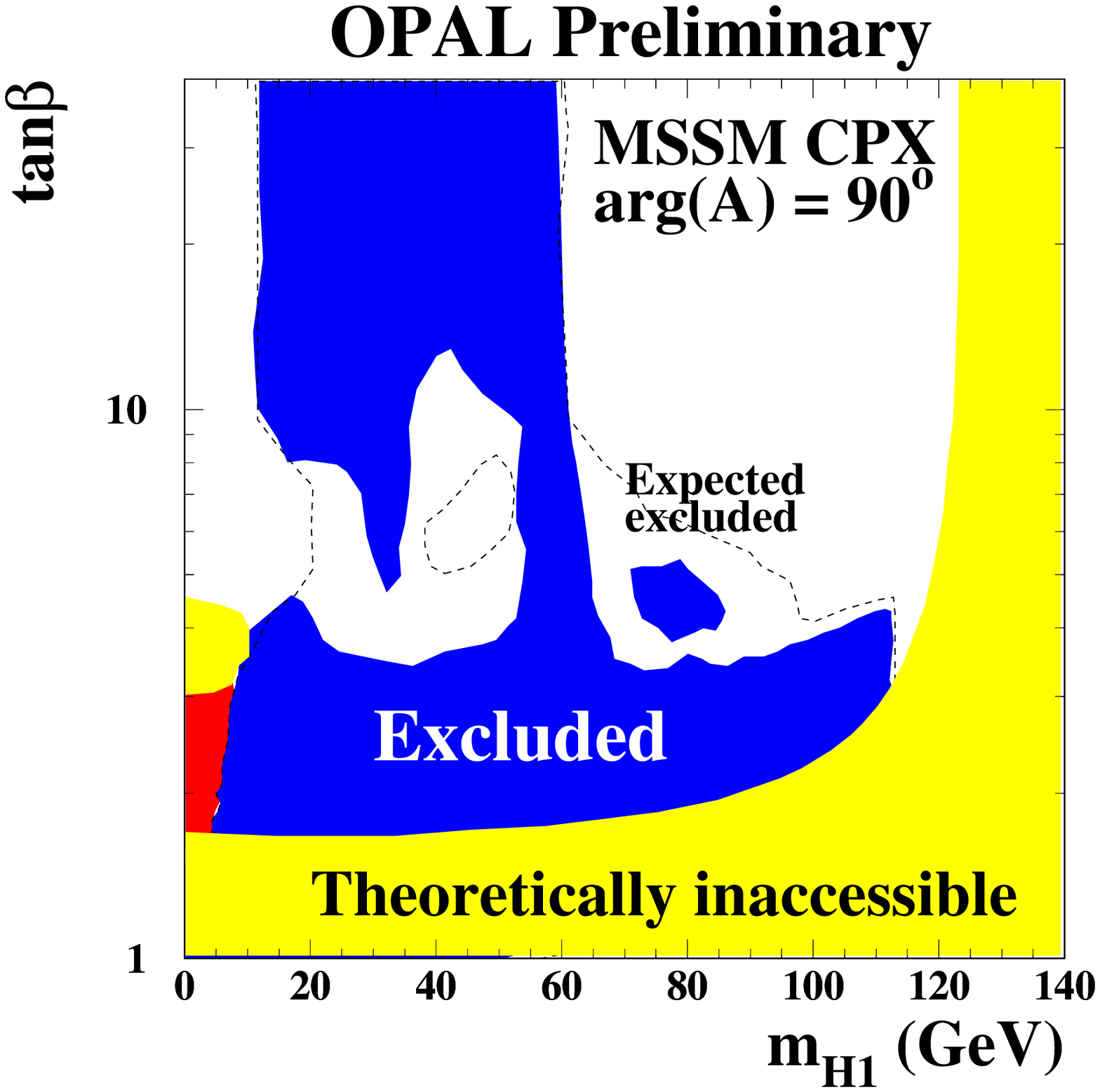}
\end{center}
\vspace*{-0.7cm}
\caption[]{CP-violation models.
Left: mass limits with no CP-mixing.
Right: mass limits with full CP-mixing.
\label{fig:cpmix}}
\vspace*{-0.5cm}
\end{figure}

Instead of h, H and A, the Higgs bosons are named $\rm H_1, H_2$ and $\rm H_3$.
The reactions
$\rm \ee\ra H_2 Z \ra b\bar b \nu\bar \nu$ and 
$\rm \ee\ra H_2 Z \ra H_1 H_1 Z \ra b\bar b b\bar b \nu\bar \nu$ are searched for.
No indication of these processes is observed in the data as shown in Fig.~\ref{fig:cpvio}. 
Figure~\ref{fig:cpmix} from another study shows that a variation in CP-mixing reduces the MSSM mass limits 
significantly.

\clearpage

\section{Invisible Higgs Boson Decays}
\vspace*{-0.2cm}

\begin{figure}[htb]
\vspace*{-0.5cm}
\begin{center}
\includegraphics[scale=0.35]{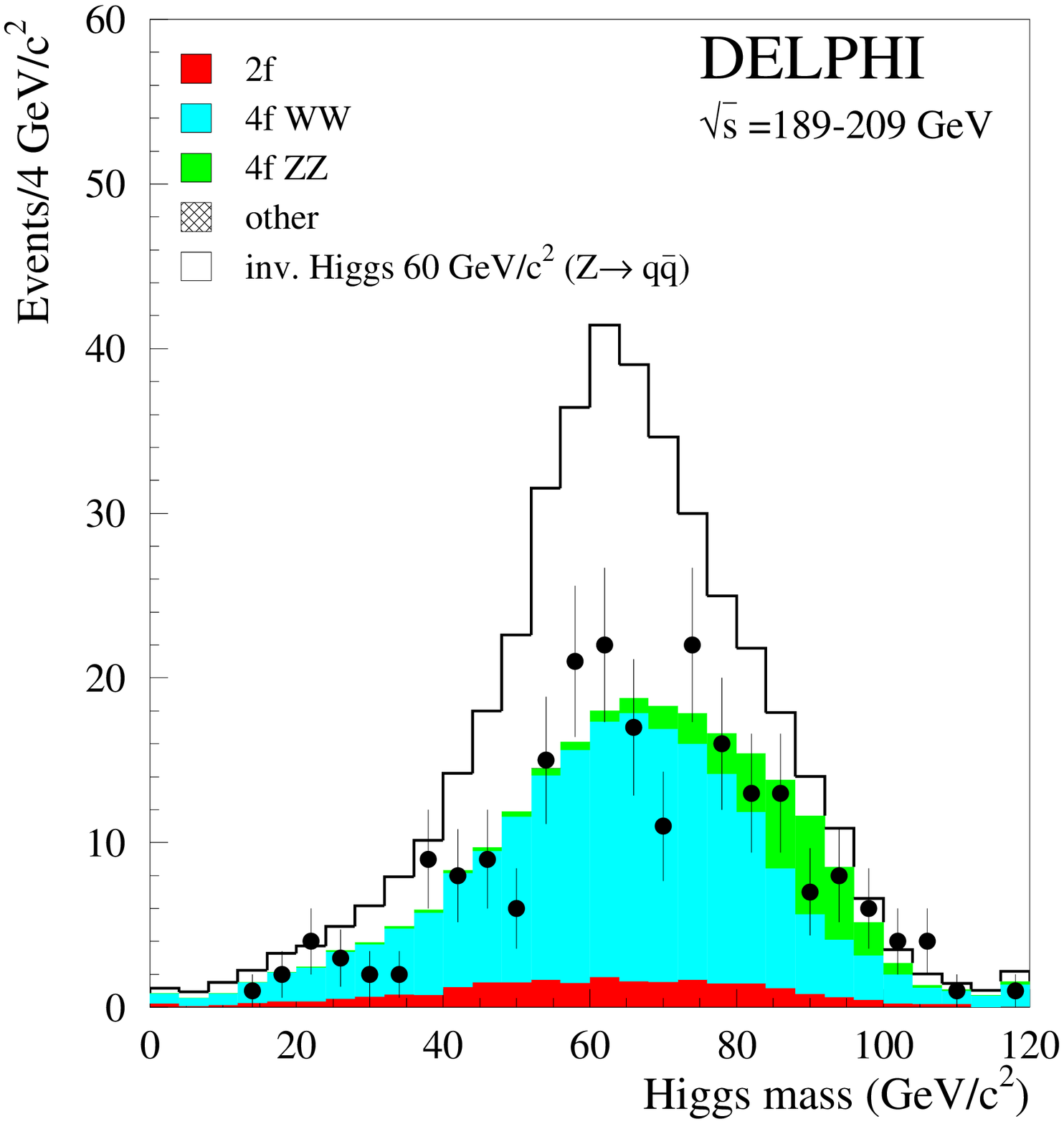}\hfill
\includegraphics[scale=0.35]{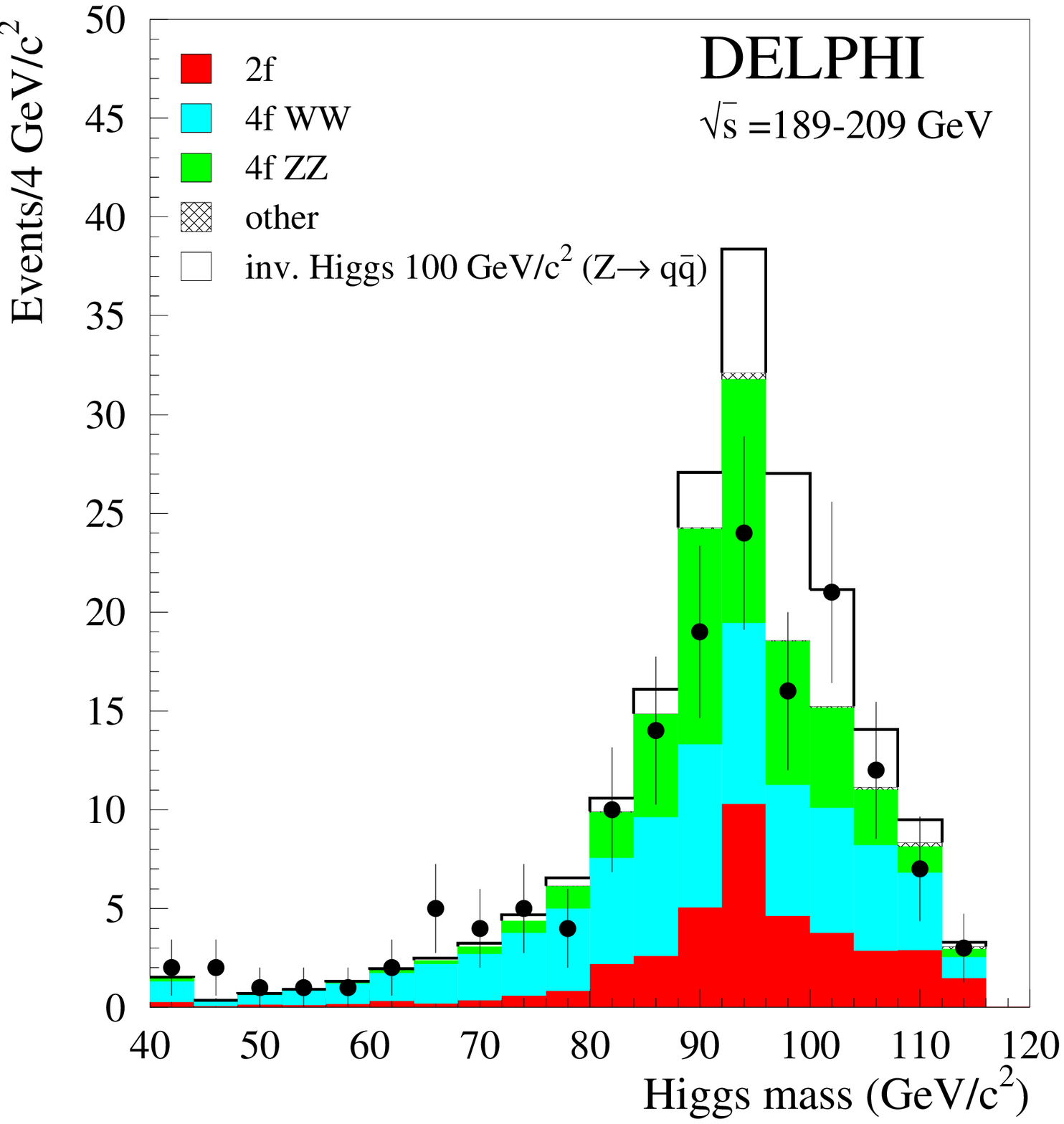}
\end{center}
\vspace*{-0.5cm}
\caption[]{No indication of invisibly-decaying Higgs bosons is observed 
           in searches optimized in low- and high-mass regions. 
\label{fig:invmass}}
\vspace*{-0.5cm}
\end{figure}

\begin{figure}[htb]
\vspace*{-0.2cm}
\begin{center}
\includegraphics[scale=0.35]{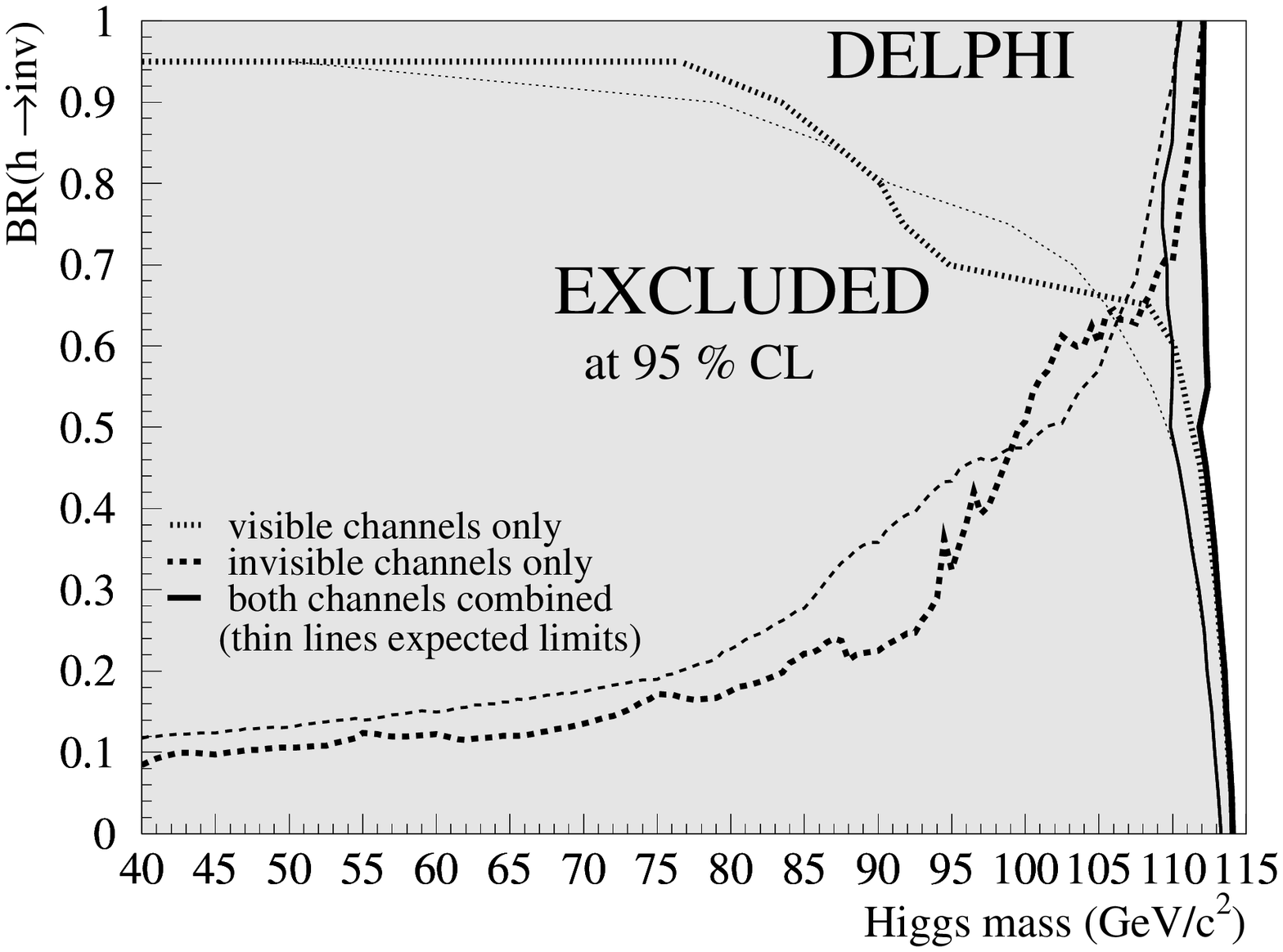}\hfill
\includegraphics[scale=0.3]{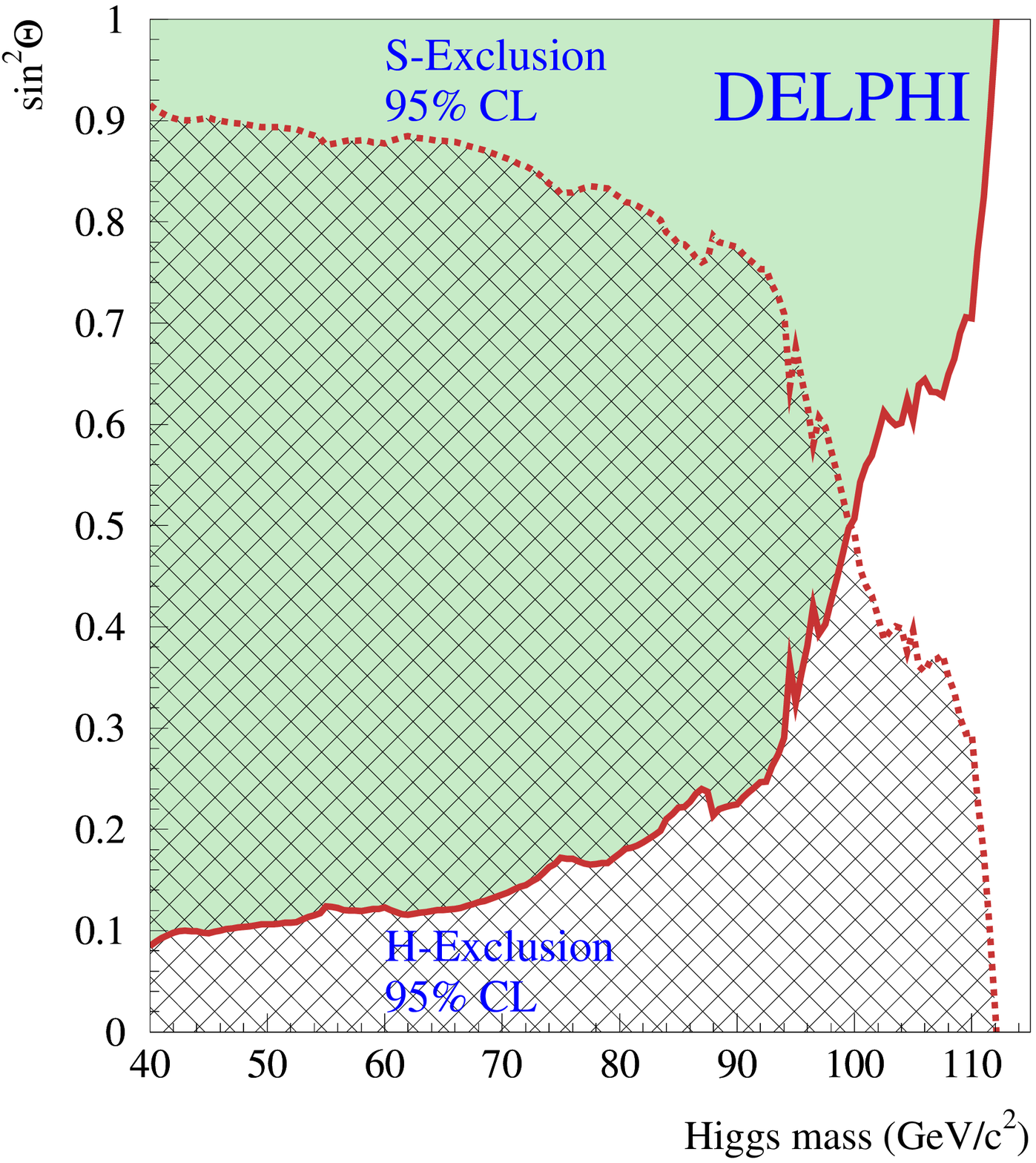}
\vspace*{-0.5cm}
\end{center}
\caption[]{Left: mass limits for SM and invisible Higgs boson decays combined.  
           Right: mass limits in Majoron models with an extra complex singlet,
           $\rm H/S\ra JJ$, where J escapes undetected. $\sin\theta$ is the H/S
           mixing angle.
\label{fig:invlimit}}
\vspace*{-2.3cm}
\end{figure}

No indication of invisibly-decaying Higgs bosons is observed as shown in
Fig.~\ref{fig:invmass}.
Figure~\ref{fig:invlimit} shows mass limits for SM and invisible Higgs boson decays combined, and  
                          in Majoron models with an extra complex singlet, $\rm H/S\ra JJ$,
                          where J escapes undetected. 

\clearpage
\section{Flavour-Independent Hadronic Higgs Boson Decays}
\vspace*{-0.2cm}

\begin{figure}[htb]
\vspace*{-0.2cm}
\begin{center}
\includegraphics[scale=0.3]{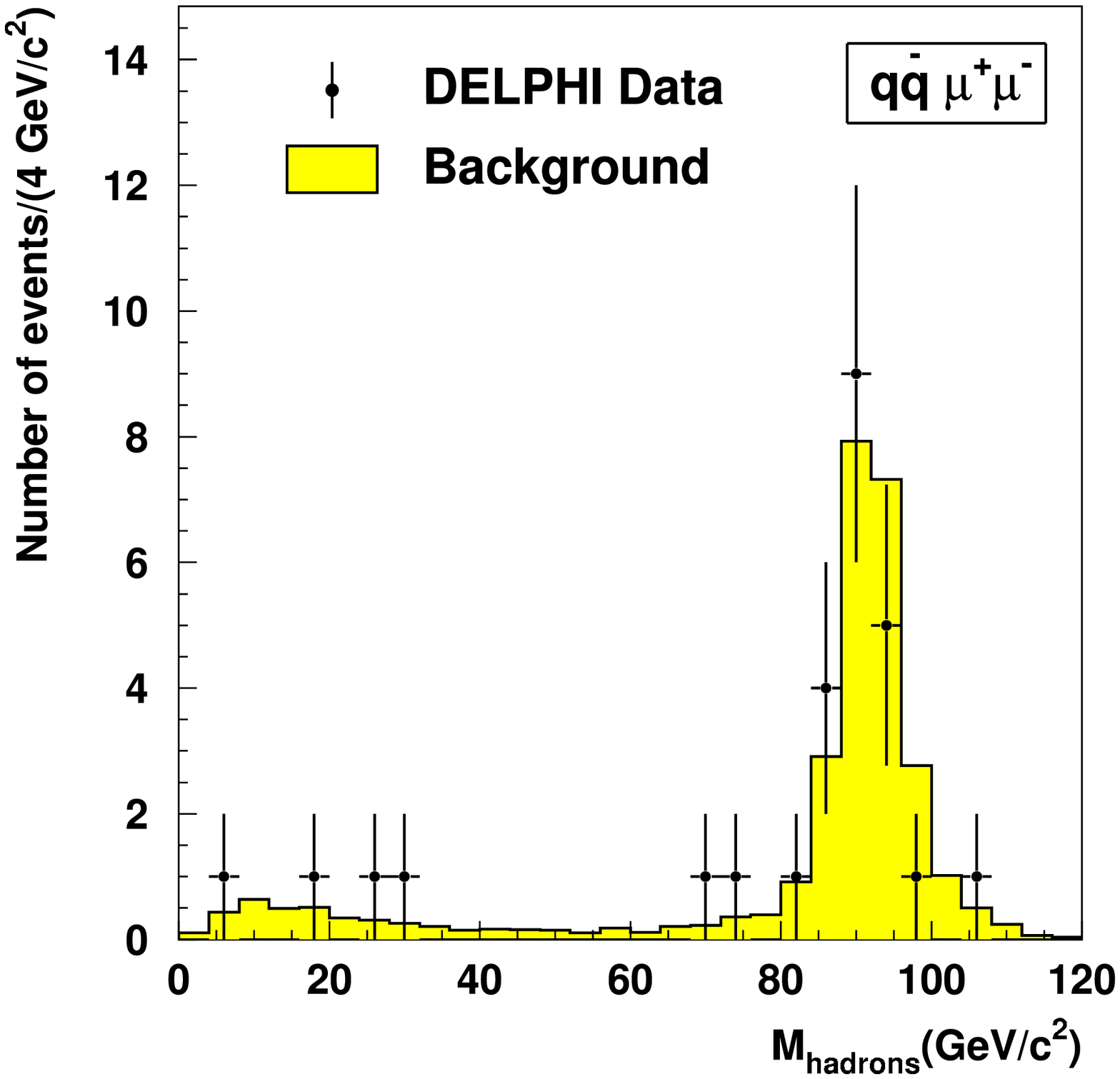}\hfill
\includegraphics[scale=0.3]{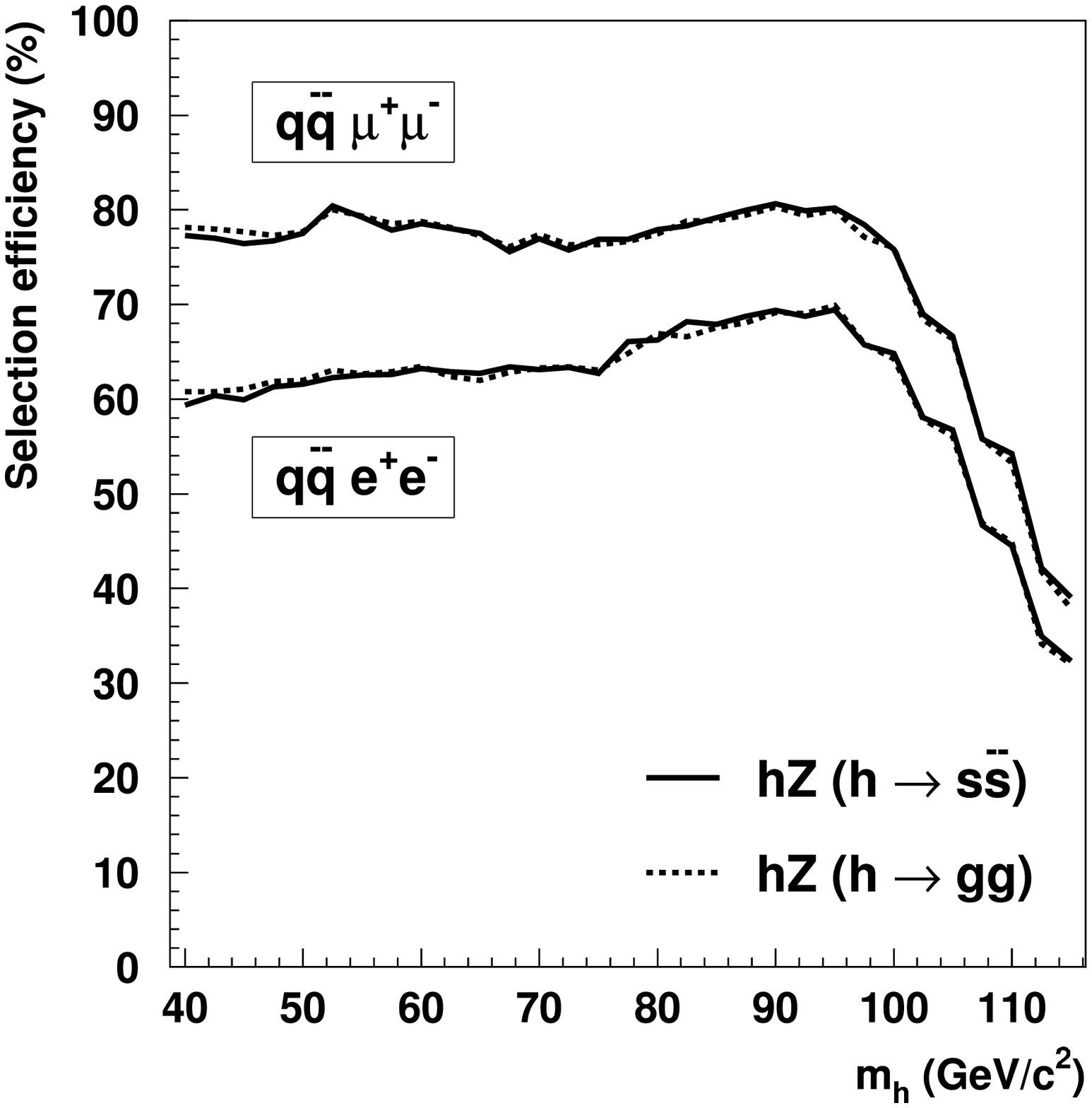}
\vspace*{-0.5cm}
\end{center}
\caption[]{Left: no indication of a signal for the process 
           $\rm hZ\ra q\bar q \ell^+\ell^-$ above the background 
           $\rm ZZ\ra q\bar q \ell^+\ell^-$.
           Right: expected signal efficiencies.
\label{fig:flind}}
\vspace*{-0.6cm}
\end{figure}

\begin{figure}[htb]
\vspace*{-0.2cm}
\begin{center}
\includegraphics[scale=0.27]{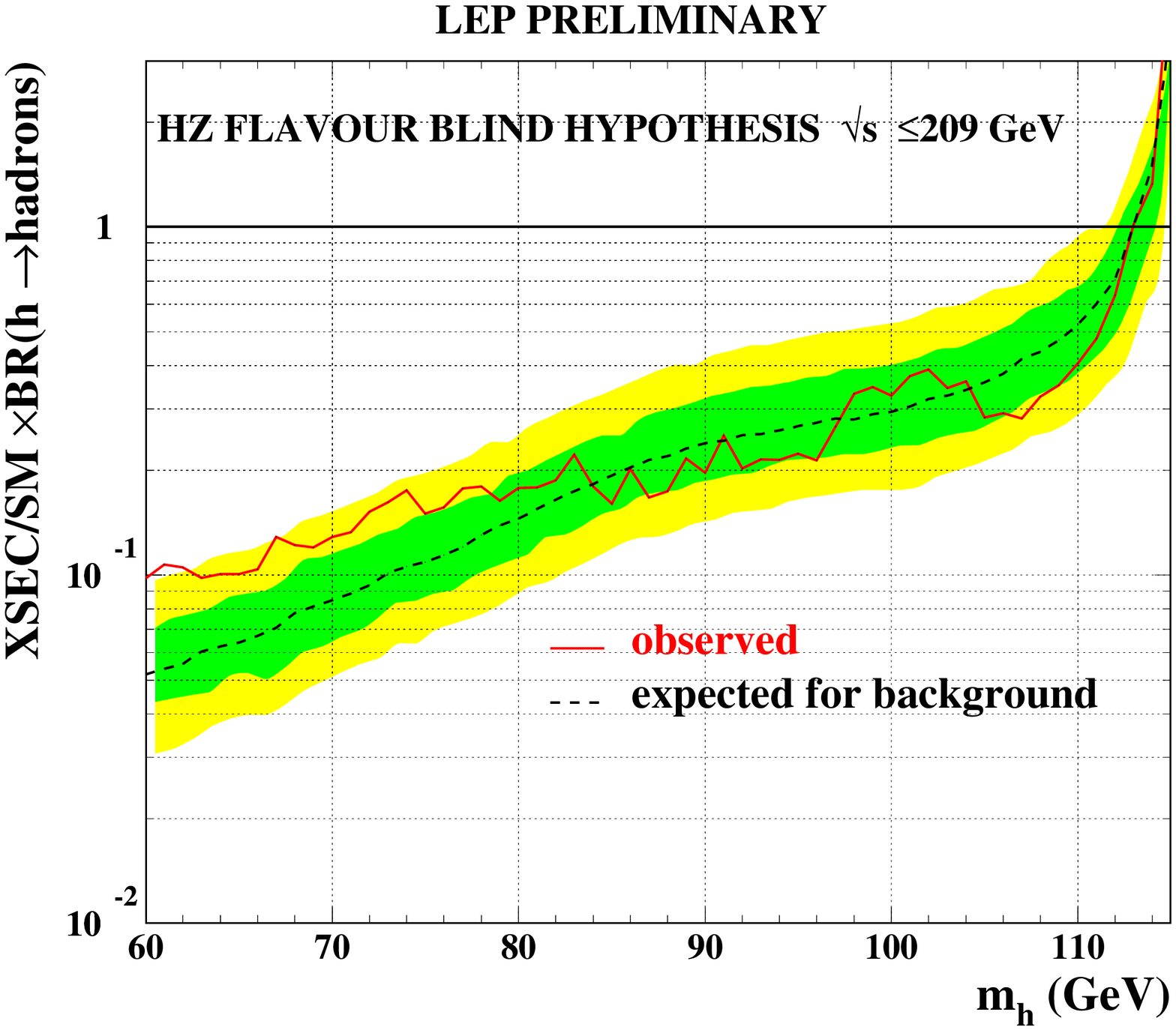}\hfill
\includegraphics[scale=0.4]{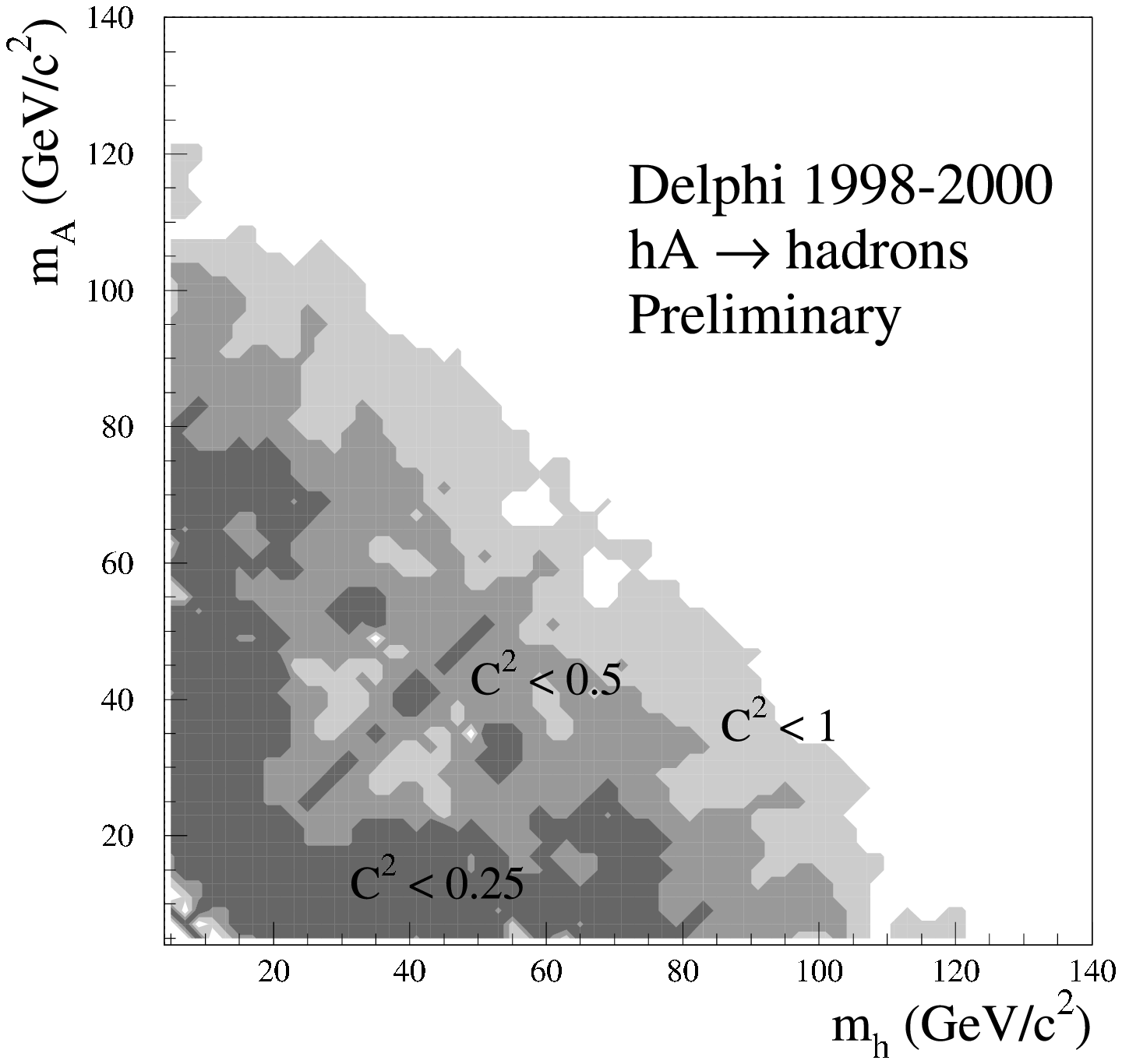}
\vspace*{-0.5cm}
\end{center}
\caption[]{Flavour-independent limits from searches for hadronic hZ and hA decays.
           No b-tagging requirement is applied. $C^2$ is the reduction factor on the maximum
           production cross section.
\label{fig:fblindlimit}}
\vspace*{-0.6cm}
\end{figure}

Figure~\ref{fig:flind} shows no indication of a signal for the process 
$\rm hZ\ra q\bar q \ell^+\ell^-$ above the background 
$\rm ZZ\ra q\bar q \ell^+\ell^-$.
In addition, the expected signal efficiencies are shown. Flavour-independent limits from 
searches for hadronic hZ and hA decays are shown in Fig.~\ref{fig:fblindlimit}.

\section{Neutral Higgs Bosons in the General 2-Doublet Higgs Model (2DHM)}
\vspace*{-0.2cm}

\begin{figure}[htb]
\vspace*{-0.3cm}
\begin{center}
\includegraphics[scale=0.32]{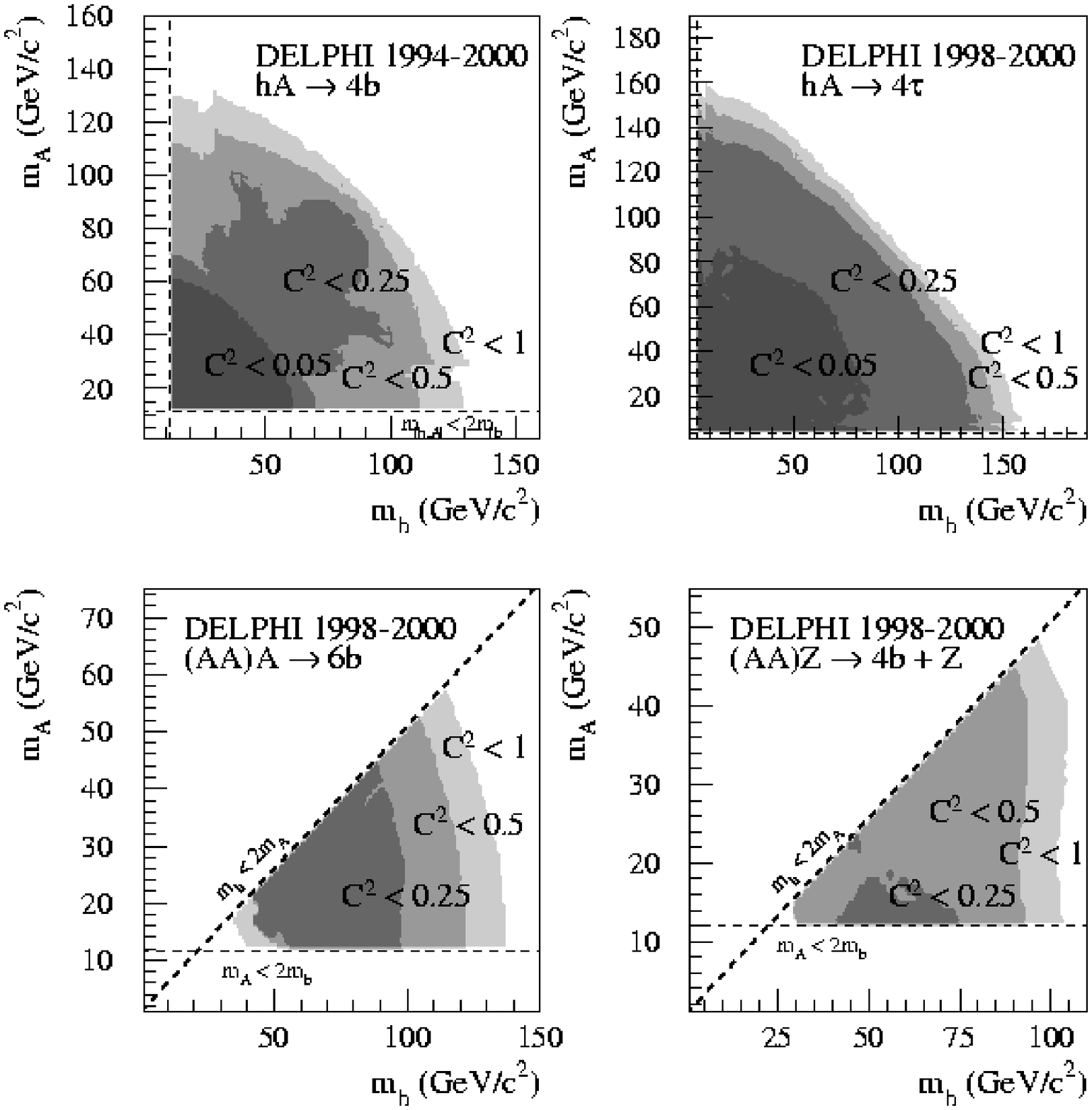}\hfill
\includegraphics[scale=0.35]{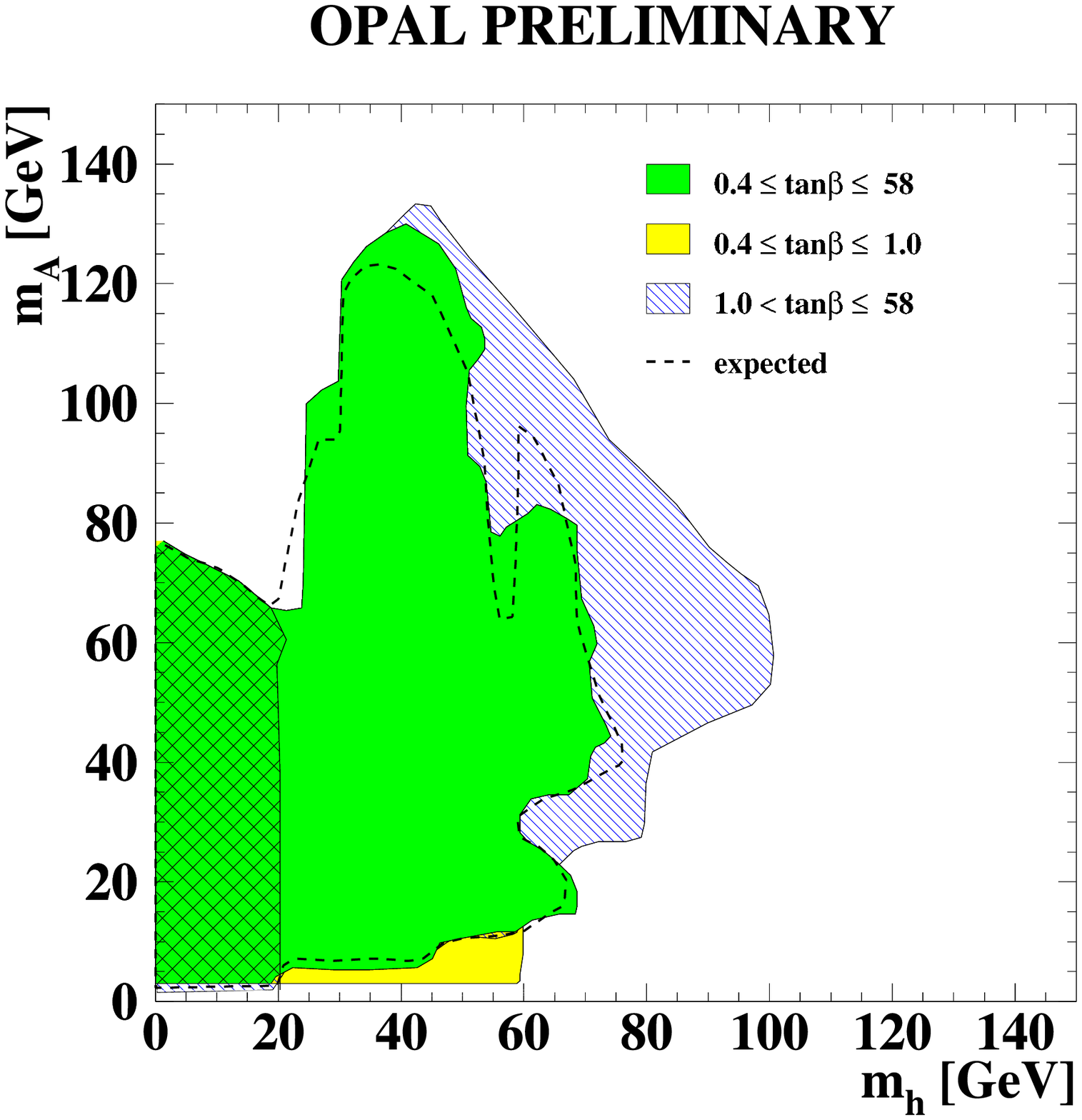}
\vspace*{-0.5cm}
\end{center}
\caption[]{2DHM. Left: mass limits from dedicated searches for hA production.
           $C^2$ is the reduction factor on the maximum production cross section. 
           \mbox{Right: mass limits from a general 2DHM parameter scan.}
\label{fig:2dhm}}
\vspace*{-1cm}
\end{figure}

Figure~\ref{fig:2dhm} shows mass limits from dedicated searches for
hA production and from a parameter scan.
The scan combines searches with b-tagging and flavour-independent searches.

\clearpage
\section{Yukawa Higgs Boson Processes 
\boldmath$\rm b\bar b h$ and $\rm b\bar b A$\unboldmath}
\vspace*{-0.2cm}

\begin{figure}[htb]
\vspace*{-0.2cm}
\begin{center}
\includegraphics[scale=0.5]{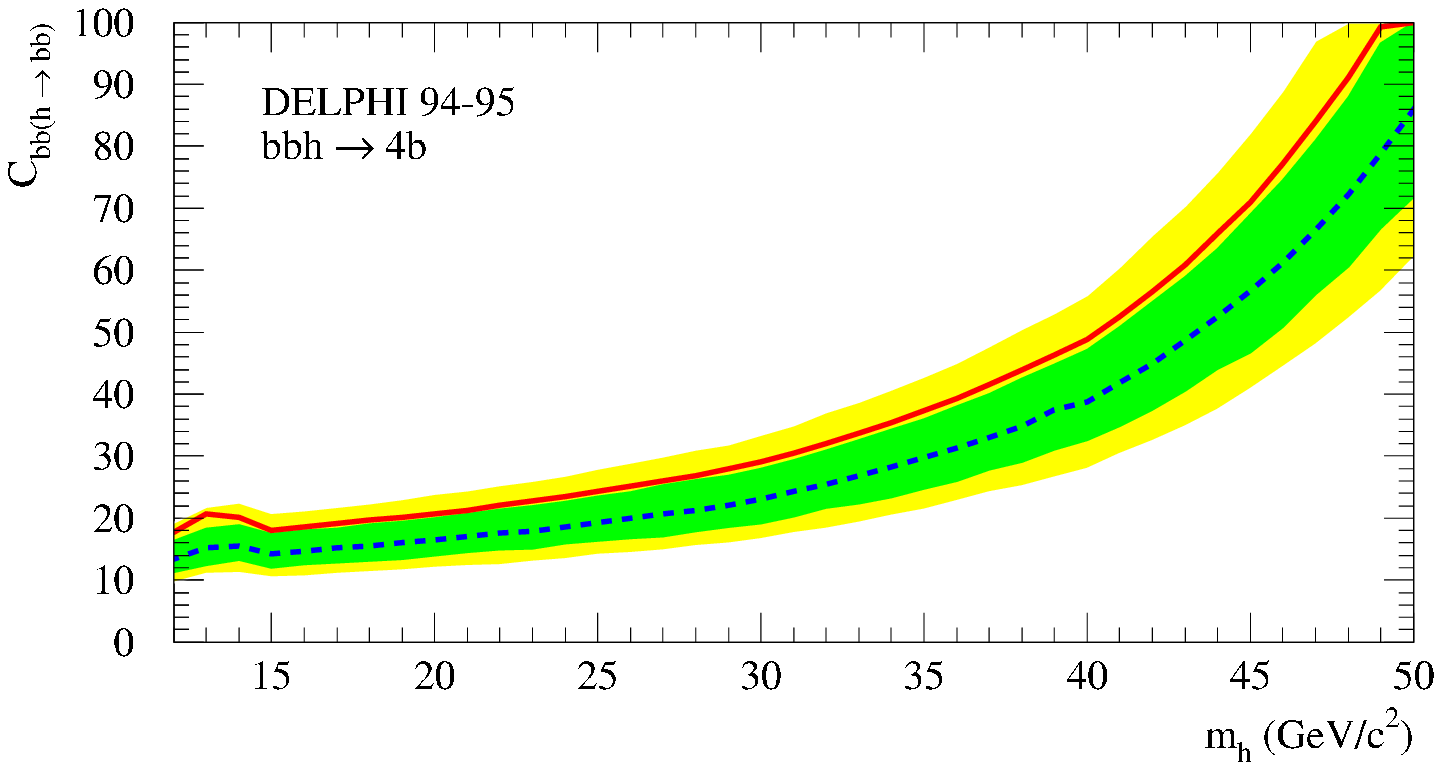}\hfill
\includegraphics[scale=0.5]{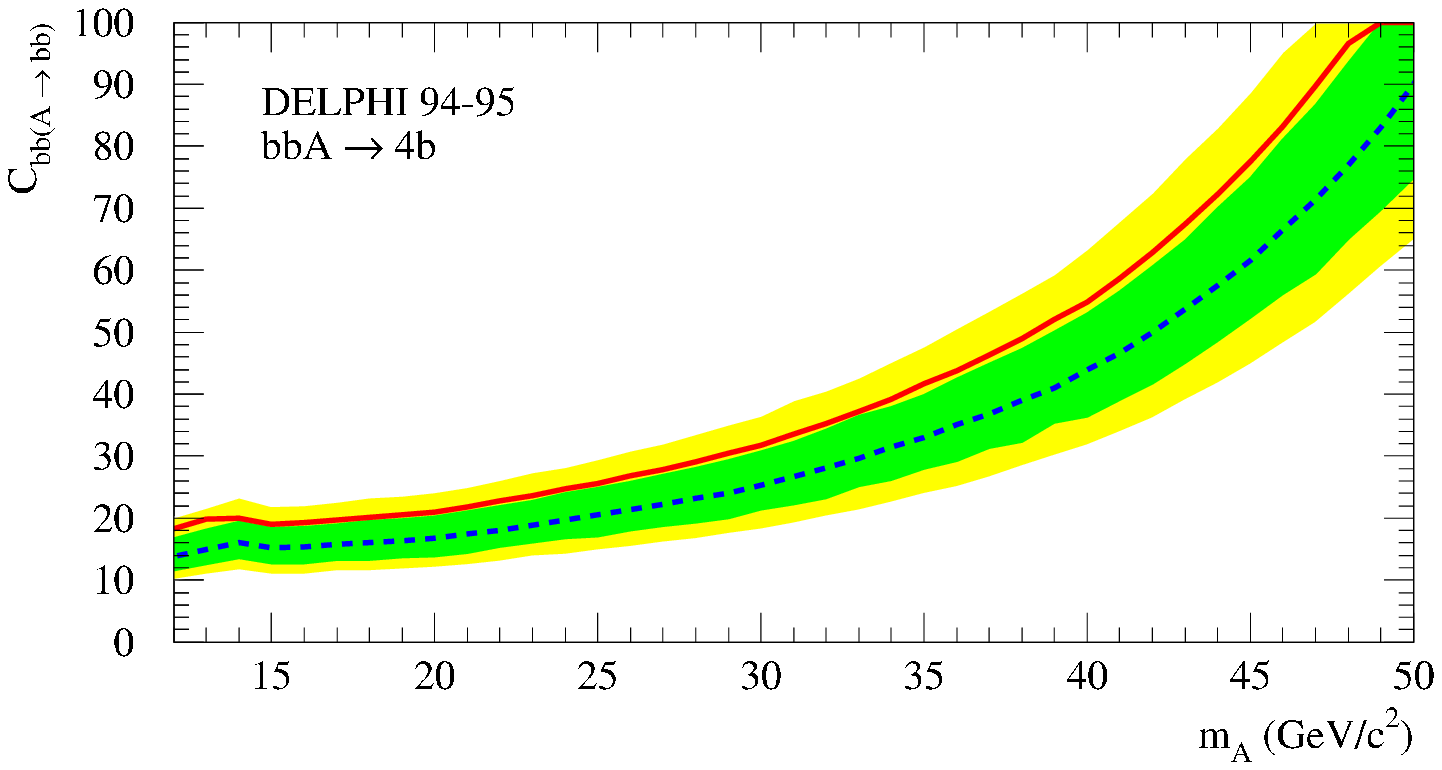}
\includegraphics[scale=0.5]{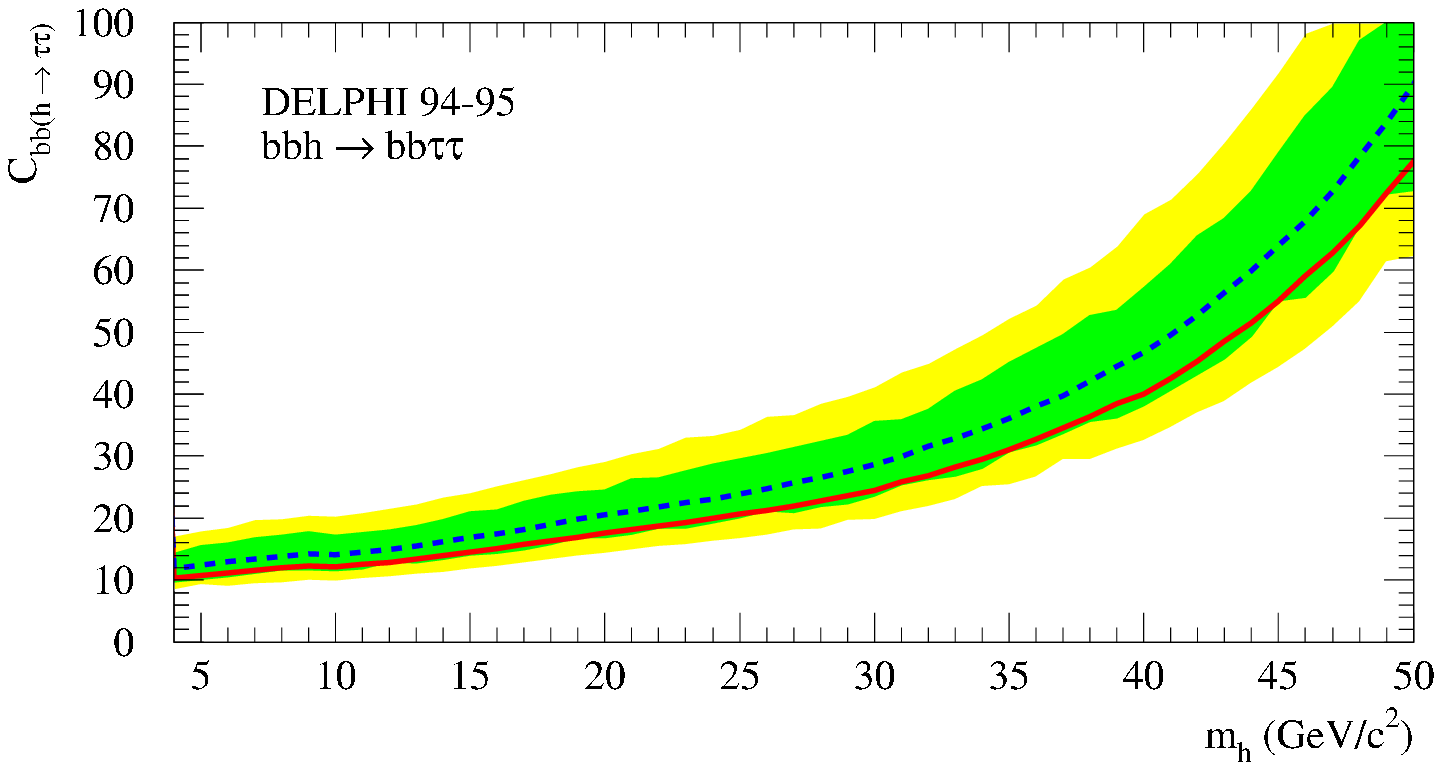}\hfill
\includegraphics[scale=0.5]{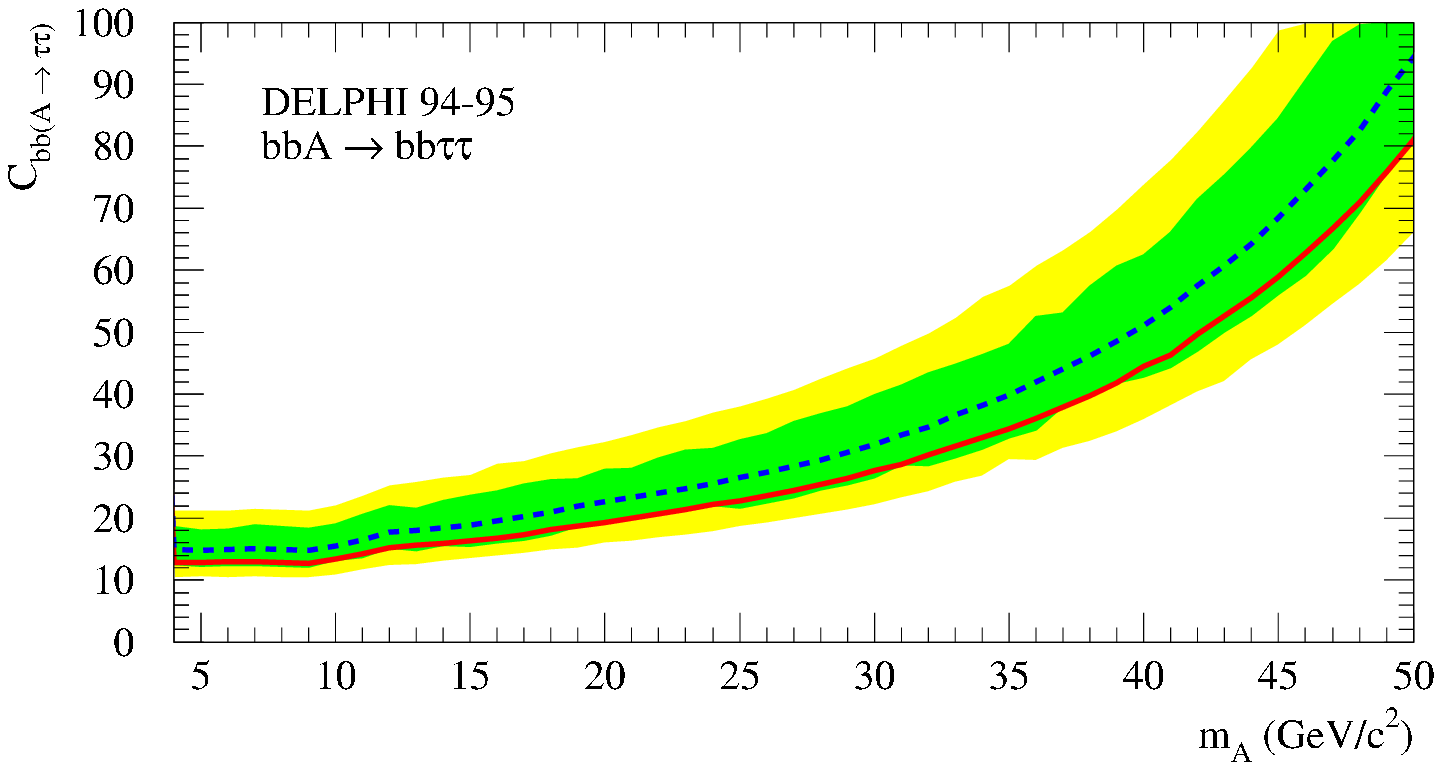}
\end{center}
\vspace*{-0.5cm}
\caption[]{Observed (solid line) and expected (dotted line) mass limits from searches 
           for the Yukawa processes 
           $\rm e^+e^-\rightarrow b\bar b\rightarrow b\bar bh,~b\bar bA$.
The $C$ factors include vertex enhancement factors and decay branching
fractions.
\label{fig:yukawa}}
\vspace*{-0.5cm}
\end{figure}

Figure~\ref{fig:yukawa}\,shows\,mass\,limits\,from\,searches\,for\,the\,Yukawa\,processes 
$\rm e^+e^-\rightarrow b\bar b \rightarrow b\bar bh,~b\bar bA$.

\section{Singly-Charged Higgs Bosons}
\vspace*{-0.2cm}

\begin{figure}[htb]
\hfill
\begin{minipage}{0.4\textwidth}
\vspace*{-0.5cm}
\begin{center}
\includegraphics[scale=0.3]{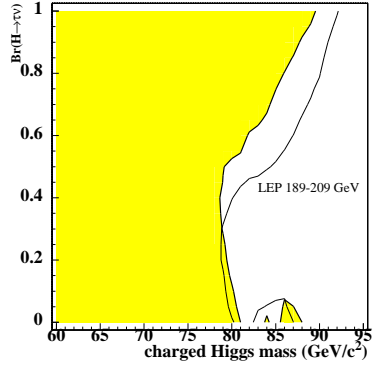}
\end{center}
\end{minipage}
\begin{minipage}{0.34\textwidth}
\caption[]{Excluded mass region (shaded area) from searches for
           $\rm \ee\ra H^+H^- \ra c\bar s\bar c s,$ $\rm cs\tau\nu$ and
           $\rm \tau^+\nu \tau^-\bar\nu$.
           The thin line shows the expected limit.
\label{fig:hpm}}
\end{minipage}\hfill
\vspace*{-0.2cm}
\end{figure}

\begin{figure}[htb]
\begin{center}
\vspace*{-0.5cm}
\includegraphics[scale=0.3]{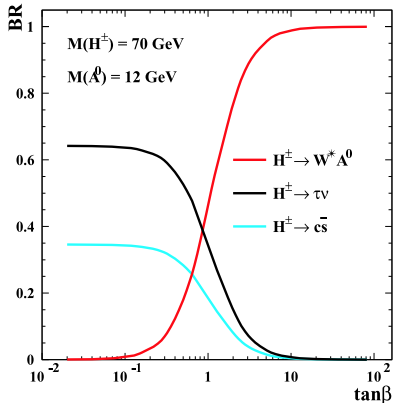}\hfill
\includegraphics[scale=0.3]{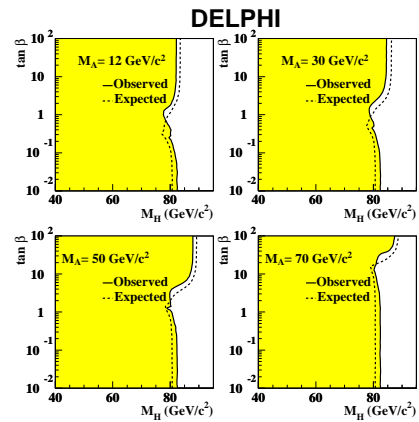}
\end{center}
\vspace*{-0.5cm}
\caption[]{Left: $\rm H^\pm \ra W^\pm A$ decays could be dominant for light A boson masses.
           Right: excluded mass region (shaded area) from searches for this process.
\label{fig:hwa}}
\vspace*{-1.8cm}
\end{figure}

Figure~\ref{fig:hpm} shows mass limits from searches for
$\rm \ee\ra H^+H^- \ra c\bar s\bar c s,~cs\tau\nu,~\tau^+\nu \tau^-\bar\nu$.
The decay $\rm H^\pm \ra W^\pm A$ could be dominant
and limits from searches for the process are shown in Fig.~\ref{fig:hwa}.

\clearpage
\section{Doubly-Charged Higgs Bosons}
\vspace*{-0.2cm}

\begin{figure}[htb]
\vspace*{-0.4cm}
\begin{center}
\includegraphics[scale=0.32]{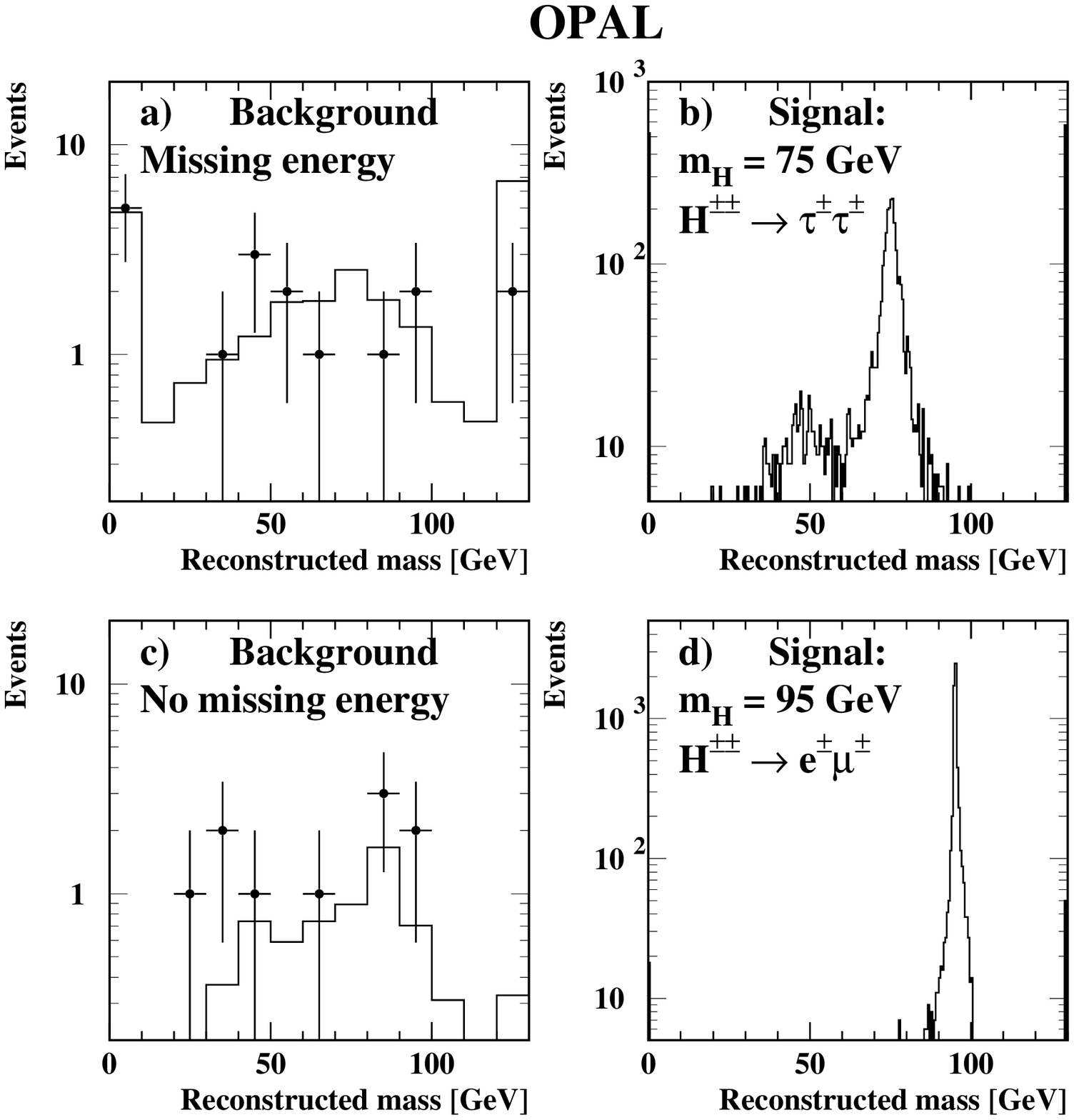}\hfill
\includegraphics[scale=0.35]{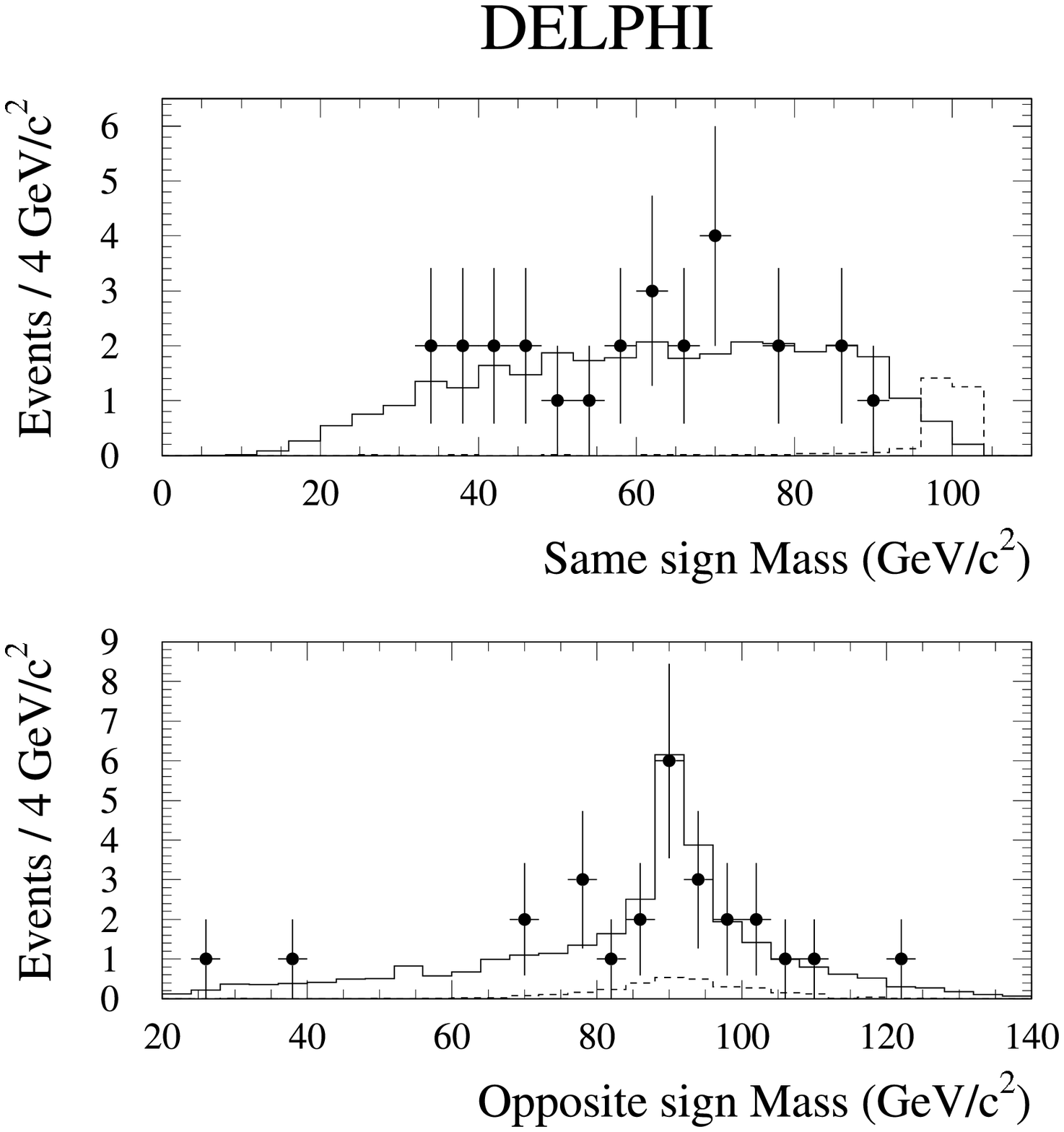}
\end{center}
\vspace*{-0.5cm}
\caption[]{No indication of 
$\rm \ee\ra H^{++} H^{--} \ra \tau^+\tau^+\tau^-\tau^-$ is observed in the data.
\label{fig:doubly}}
\vspace*{-0.3cm}
\end{figure}

\begin{figure}[htb]
\vspace*{-0.3cm}
\begin{center}
\includegraphics[scale=0.3]{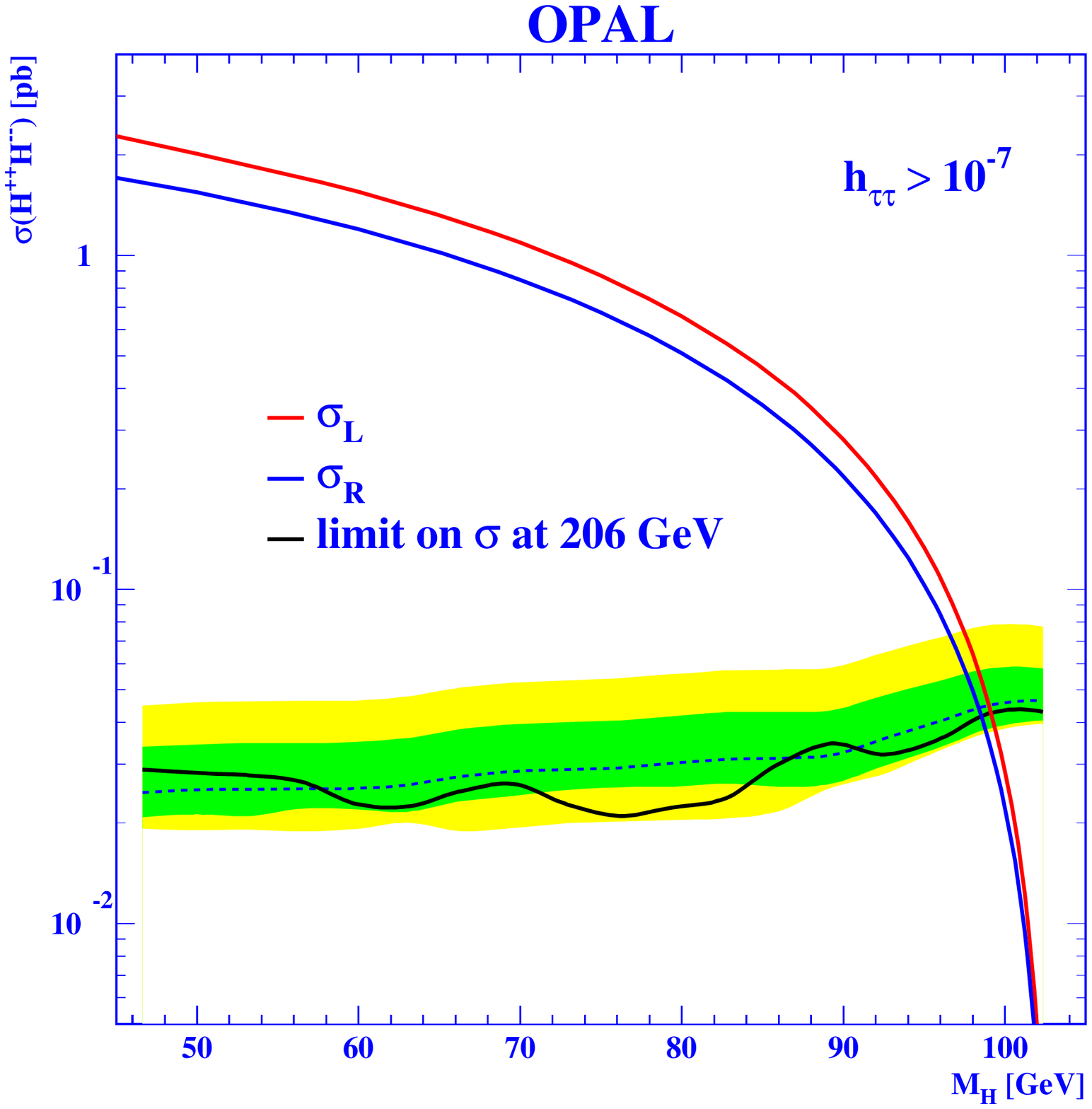}\hfill
\includegraphics[scale=0.4]{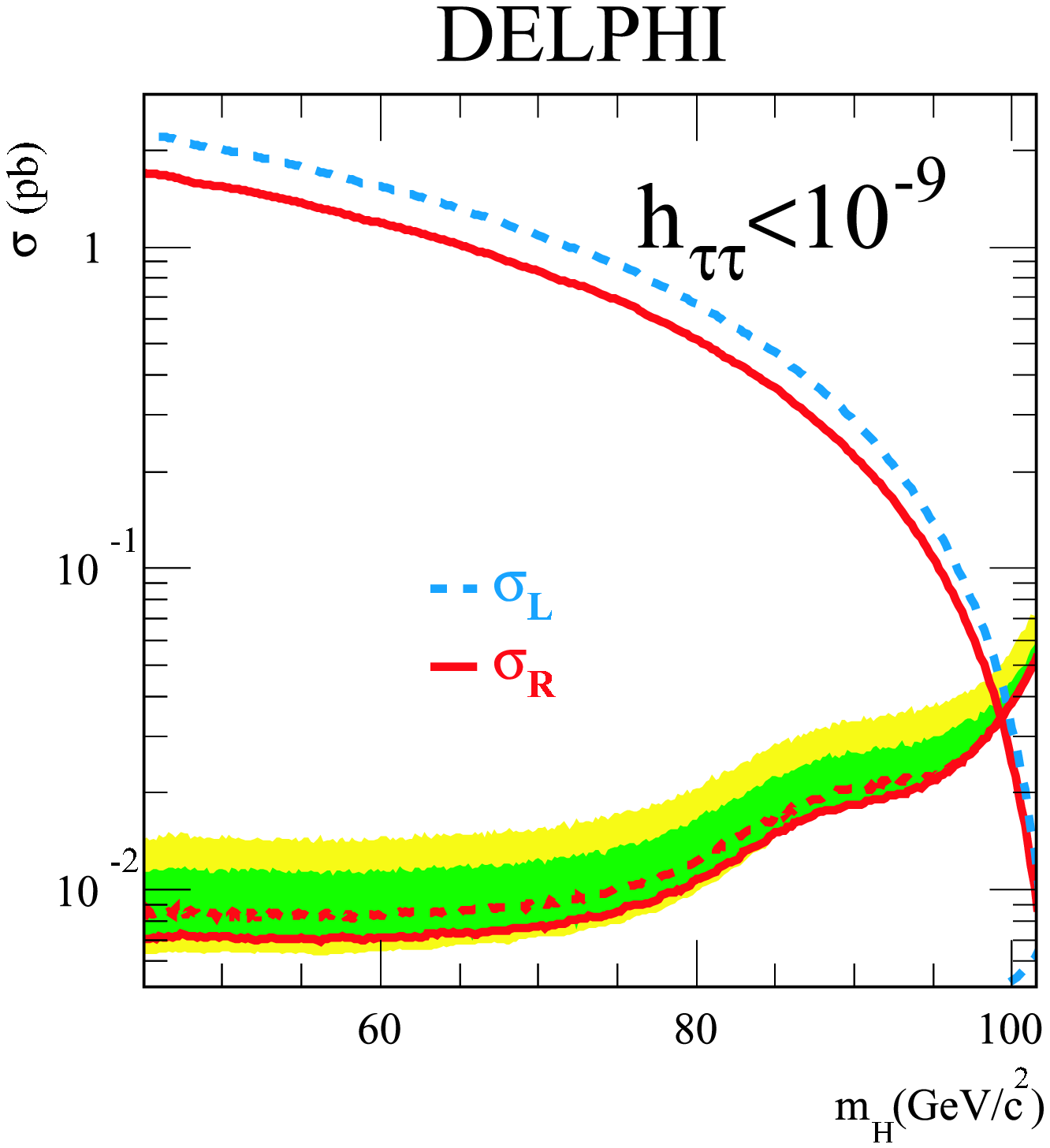}
\end{center}
\vspace*{-0.5cm}
\caption[]{Limits on the $\rm \ee\ra H^{++} H^{--}$ production cross section
           are set as a function of the doubly-charged Higgs boson mass.
\label{fig:doubly2}}
\vspace*{-0.5cm}
\end{figure}

The process $\rm \ee\ra H^{++} H^{--} \ra \tau^+\tau^+\tau^-\tau^-$ can lead
to decays at the primary interaction point 
($h_{\tau\tau}\geq 10^{-7}$)~\cite{d_doubly,o_doubly},
a secondary vertex, or stable massive particle signatures.
Figure~\ref{fig:doubly} shows no indication of a signal in the data.
Limits on the production cross section are given in Fig.~\ref{fig:doubly2}.

\section{Fermiophobic Higgs Boson Decays: 
\boldmath$\rm h\ra$ WW,~ZZ,~$\gamma\gamma$\unboldmath}
\vspace*{-0.2cm}

\begin{figure}[htb]
\vspace*{-0.5cm}
\begin{center}
\includegraphics[scale=0.3]{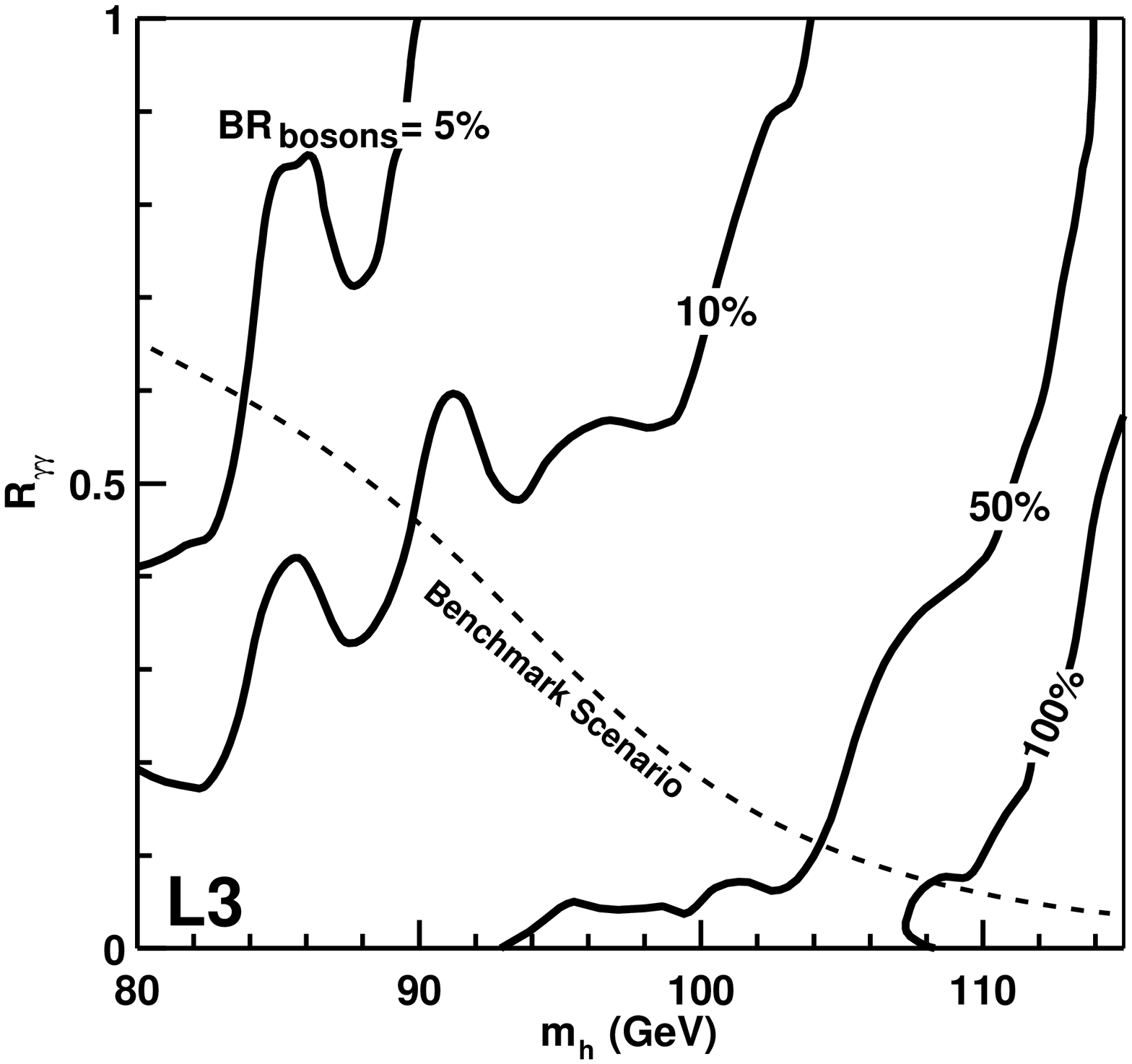}\hfill
\includegraphics[scale=0.31]{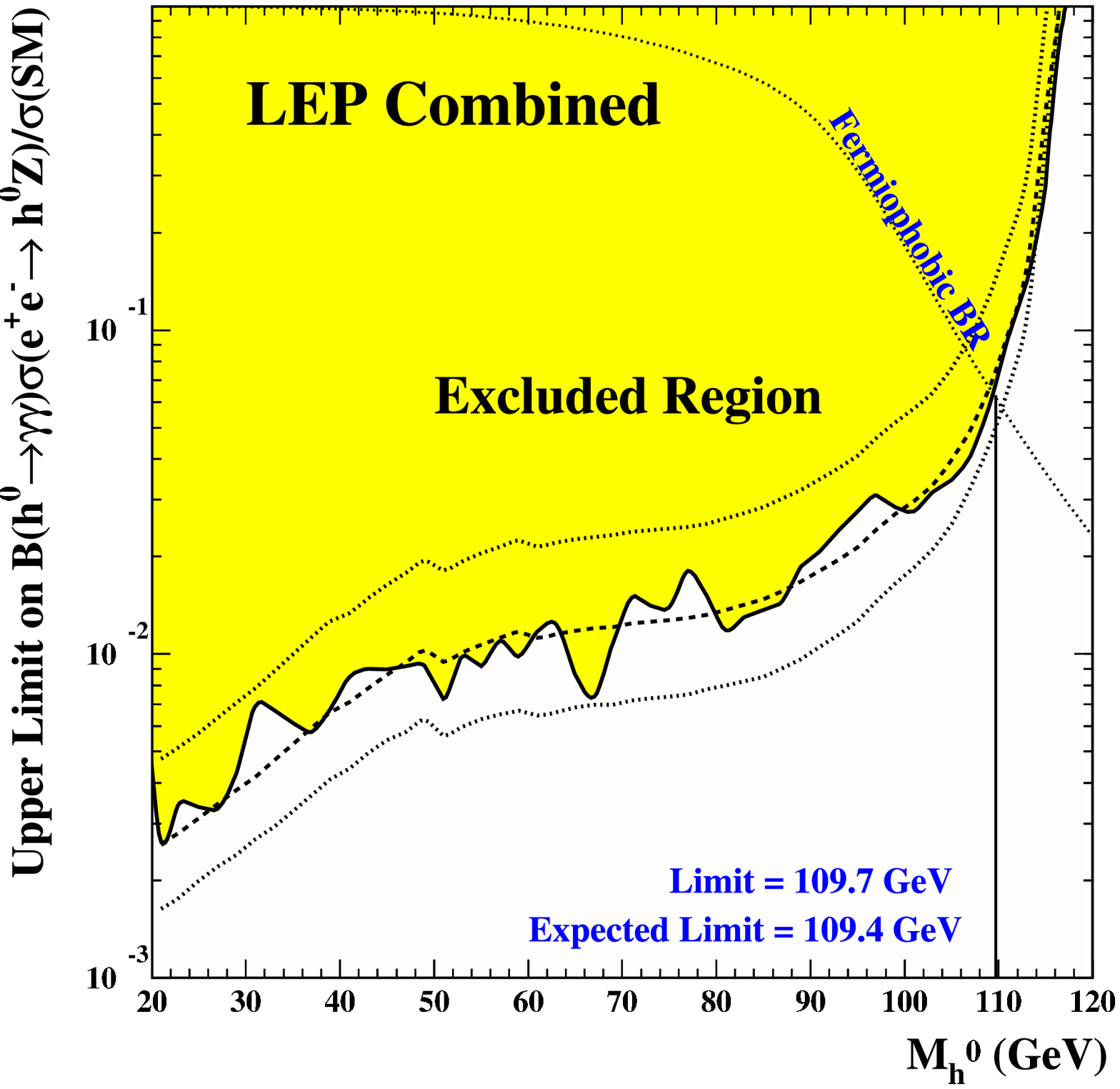}
\end{center}
\vspace*{-0.6cm}
\caption[]{Left: mass limits as defined in Ref.~\cite{l_bosons} 
           from $\rm h\ra$ WW,~ZZ,~$\gamma\gamma$ searches.
           Right: mass limits from $\rm h\ra \gamma\gamma$ combined results.
\label{fig:fermio}}
\vspace*{-0.9cm}
\end{figure}

If Higgs boson decays into fermions are suppressed, 
$\rm h\ra$ WW,~ZZ,~$\gamma\gamma$ decays could be dominant.
Mass limits from dedicated searches are shown in Fig.~\ref{fig:fermio}.

\clearpage
\section{Conclusions }
\vspace*{-0.2cm}

Immense progress over a period of 14 years has been made at LEP in 
searches for Higgs bosons and much knowledge has been gained in preparation 
for new searches.
No signal has been observed and various stringent limits are set as 
summarized in Table~\ref{tab:summary}. 

\begin{table}[htb]
\vspace*{-0.3cm}
\caption{Summary of Higgs boson mass limits at 95\% CL.
\label{tab:summary}}
\begin{center}
\begin{tabular}{c|c|r}
 Search                     & Experiment & Limit \\\hline 
Standard Model              &   LEP  
   & $m^{\rm SM}_{\rm H} > 114.4$ GeV \\ 
Reduced rate and SM decay &       
  & $\xi^2>0.05:$ $ m_{\rm H} > 85$ GeV \\
& & $\xi^2>0.3:$ $ m_{\rm H} > 110$ GeV \\
Reduced rate and $\rm b\bar b$ decay  &   
  & $\xi^2>0.04:$ $ m_{\rm H} > 80$ GeV \\
& & $\xi^2>0.25:$ $ m_{\rm H} >110$ GeV \\ 
Reduced rate and $\tau^+\tau^-$ decay & 
  & $\xi^2>0.2:$ $ m_{\rm H} > 113$ GeV \\ \hline
MSSM (no scalar top mixing) & LEP 
  & almost entirely excluded\\ 
General MSSM scan & DELPHI &  $m_{\rm h} > 87$ GeV, $m_{\rm A} >90$ GeV\\ \hline
CP-violating    & OPAL    &  strongly reduced limits  \\ \hline
Visible/invisible Higgs decays & DELPHI & $m_{\rm H} >111.8$ GeV\\ 
Majoron model (max. mixing) &  & $m_{\rm H,S} >112.1$ GeV\\ \hline
Flavour-ind. hadronic decay & LEP
  & $\rm hZ\ra q\bar q:$ $m_{\rm H} >112.9$ GeV\\ 
 (for $\sigma_{\rm max})$ &DELPHI  & $\rm hA\ra q\bar q q\bar q:$ 
      $m_{\rm h}+m_{\rm A} >  110$ GeV\\\hline
2DHM                      & DELPHI
  & $\rm b\bar b b\bar b:$
    $m_{\rm h}+m_{\rm A} >  150$ GeV\\
(for $\sigma_{\rm max}$) & 
  & $\tau^+\tau^-\tau^+\tau^-:$
    $m_{\rm h}+m_{\rm A} >  160$ GeV\\
& & $\rm (AA)A\ra 6b:$ $m_{\rm h}+m_{\rm A} >  150$ GeV\\
& & $\rm (AA)Z\ra 4b~Z:$ $m_{\rm h} >  90$ GeV\\
General 2DHM scan & OPAL
  & $\tan\beta > 1:$ $ m_{\rm h} \approx m_{\rm A} > 85$ GeV \\\hline 
Yukawa process & DELPHI & $C > 40:$ $m_{\rm h,A} > 40$ GeV \\\hline 
Singly-charged Higgs bosons & LEP 
  & $m_{\rm H^\pm} > 78.6$ GeV \\
$\rm W^\pm A$ decay mode & DELPHI& $m_{\rm H^\pm} > 76.7$ GeV \\ \hline
Doubly-charged Higgs bosons & DELPHI/OPAL
  & 
$m_{\rm H^{++}} > 99$ GeV \\\hline 
Fermiophobic $\rm H\ra WW, ZZ, \gamma\gamma$ & L3 
  &  $m_{\rm H} > 108.3$ GeV \\
$\rm H\ra \gamma\gamma$ &LEP &  $ m_{\rm H} > 109.7$ GeV \\
\end{tabular}
\end{center}
\end{table}

\section*{Acknowledgments}
\vspace*{-0.2cm}
I would like to thank the organizers of the NANP03 conference
for their kind hospitality, and Tom Junk and Bill Murray for comments 
on the manuscript.


\section*{References}
\vspace*{-0.2cm}
\begin{thebibliography}{99}

\bibitem{sm}
ALEPH, DELPHI, L3 and OPAL Collaborations
and the LEP working group for Higgs boson searches, 
Phys. Lett. {\bf B  565} (2003) 61.

\bibitem{summer2003}
ALEPH, DELPHI, L3 and OPAL Collaborations, contributed papers to the 
International Europhysics Conference on High Energy Physics
EPS, 17-23 July 2003, Aachen, Germany; and
XXI International Symposium on Lepton and Photon Interactions at High Energies,
11-16 August 2003, Fermi National Accelerator Laboratory, Batavia, Illinois, USA.

\bibitem{nanp01}
A.~Sopczak, Proc. NANP01, Phys. Atom. Nucl. {\bf 65} (2002) 2116.

\bibitem{as2000}
A.~Sopczak, Proc. DPF-2000, hep-ph/0011285.

\bibitem{d_doubly}
DELPHI Collaboration, J. Abdallah et al., Phys. Lett. {\bf B 552} (2003) 127.

\bibitem{o_doubly}
OPAL Collaboration, G. Abbiendi et al., Phys. Lett. {\bf B 577} (2003) 93.

\bibitem{l_bosons}
L3 Collaboration, P. Achard et al., Phys. Lett. {\bf B 568} (2003) 191.

\end{thebibliography}
\end{document}